# Orbital resonance and Solar cycles

by P.A.Semi

## *Abstract*

We show resonance cycles between most planets in Solar System, of differing quality. The most precise resonance - between Earth and Venus, which not only stabilizes orbits of both planets, locks planet Venus rotation in tidal locking, but also affects the Sun:

This resonance group (E+V) also influences Sunspot cycles - the position of syzygy between Earth and Venus, when the barycenter of the resonance group most closely approaches the Sun and stops for some time, relative to Jupiter planet, well matches the Sunspot cycle of 11 years, not only for the last 400 years of measured Sunspot cycles, but also in 1000 years of historical record of "severe winters". We show, how cycles in angular momentum of Earth and Venus planets match with the Sunspot cycle and how the main cycle in angular momentum of the whole Solar system (854-year cycle of Jupiter/Saturn) matches with climatologic data, assumed to show connection with Solar output power and insolation. We show the possible connections between E+V events and Solar global p-Mode frequency changes.

We futher show angular momentum tables and charts for individual planets, as encoded in DE405 and DE406 ephemerides. We show, that inner planets orbit on heliocentric trajectories whereas outer planets orbit on barycentric trajectories.

We further TRY to show, how planet positions influence individual Sunspot groups,
in SOHO and GONG data... [This work is pending...]

## *Orbital resonance*

### Earth and Venus

The most precise orbital resonance is between Earth and Venus planets. These planets meet 5 times during 8 Earth years and during 13 Venus years. Then their resonance ratio is **13:8** ... After those 8 Earth-years, the planets meet on almost a same place, which differs by 2.586 Earth-days, in counter-orbital (retrograde) direction, which is only 2.55°, or 0.088% of the cycle. It takes 239.8 years to rotate the resonance shape by 1/5 of the circle (which is 30x the 8-year cycle), then 1199 years to rotate one full round (which is 150x the 8-year cycle minus 1 year).

This difference makes the **higher-order wave** of the resonance. Only in the case of Earth/Venus, this higher-order wave among direct neighbours is in retrograde (counter-orbital) direction. All outer planets have got the higher-order wave in prograde direction among direct neighbours.
It takes 243 years (30x the 8-year cycle + 2 meetings) between the planets meet on an almost-same place.

The resonance shape gets little modified occasionally, mostly due to perturbations by the Jupiter planet, but it always restores briefly. Hence, the resonance stabilizes planet orbits, because if one planet gets a little late to the meet-point, it is attracted by the other planet. Since the orbital time determines the distance of the planet from the Sun, forcing planets to regular meet-points stabilizes orbit trajectory to a large degree.

Note the start of Fibonacci sequence of 5,8,13, that approaches the *Golden section* ratio. The actual difference of the resonance rate from a Golden section is only **0.46%**...

The shape, traced by the bary-center of the two planets (by this resonance sub-system), look like this:

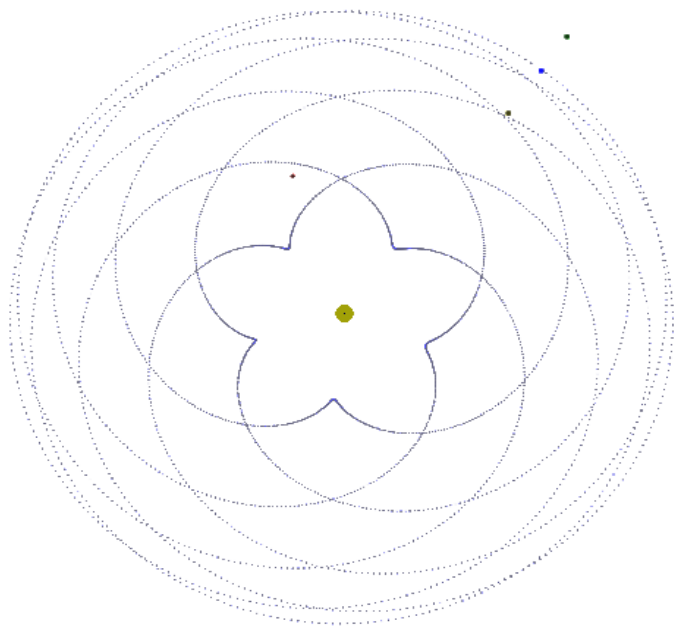

11.2.1910 15:40:09 (UTC)

Figure 1 - Barycenter trace between Earth and Venus planets, from 1902 to 1910.

Figure 1 also shows Sun, Mercury, Venus and Earth planets, and the barycenter between Earth and Venus, which traced the shape during 8 years. Planets and Sun are not to scale with their distances (their scale in the image is rather logarithmic and rounded to nearest pixel size, barycenter size is arbitrary).

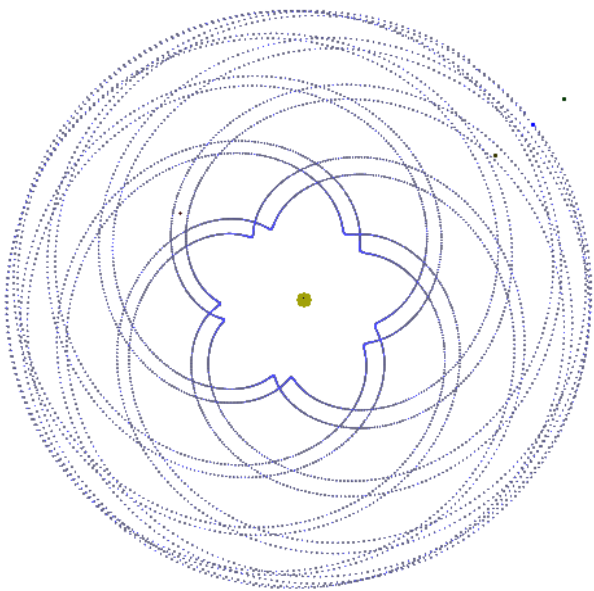

27.1.1958 22:36:15 (UTC)

Figure 2 - Two versions compared, 5 interleaved...

Figure 2 shows two versions of the "resonance shape", separated by roughly 40 years, and shows the rotation of the shape.

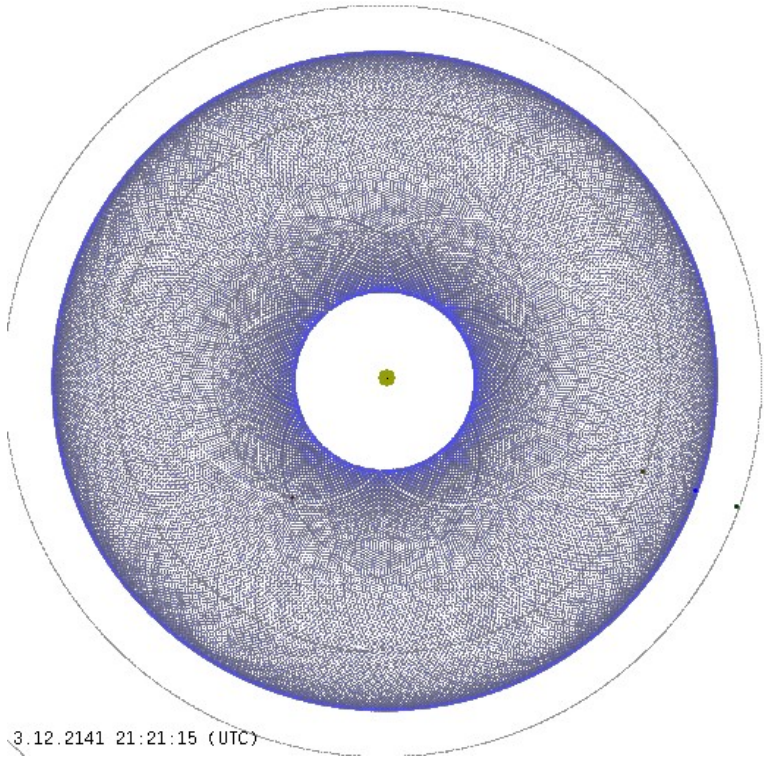

Figure 3 - After 240 years, (1/5 of the round)

The figure 3 shows the barytrace during 240 years. Repeated and rotated "shapes" fill the whole circle and almost connect with the preceeding version.

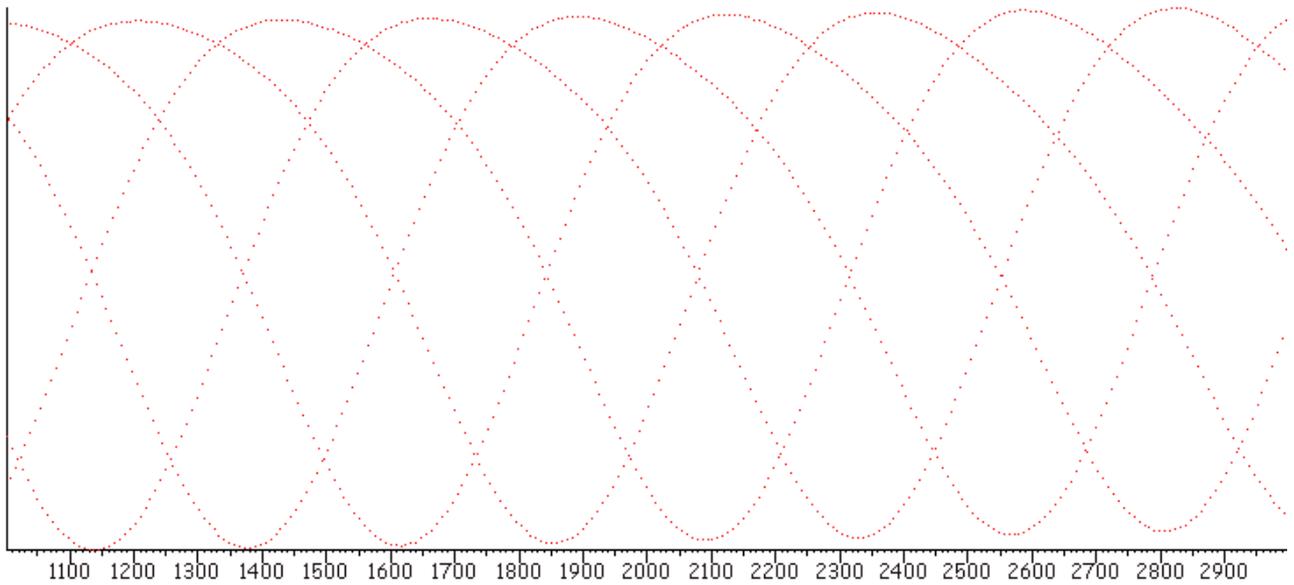

Figure 4 - Earth-Venus Meet-point distance chart, from 1000 to 3000

The distances during meet-points (most close approaches) between Earth and Venus planets (fig. 4) vary due to the distance of meet-points from Earth perihelion and probably also due to the different inclinations of the orbits (the longer trend (growing distance) is also caused by the change of excentricities of the orbits).

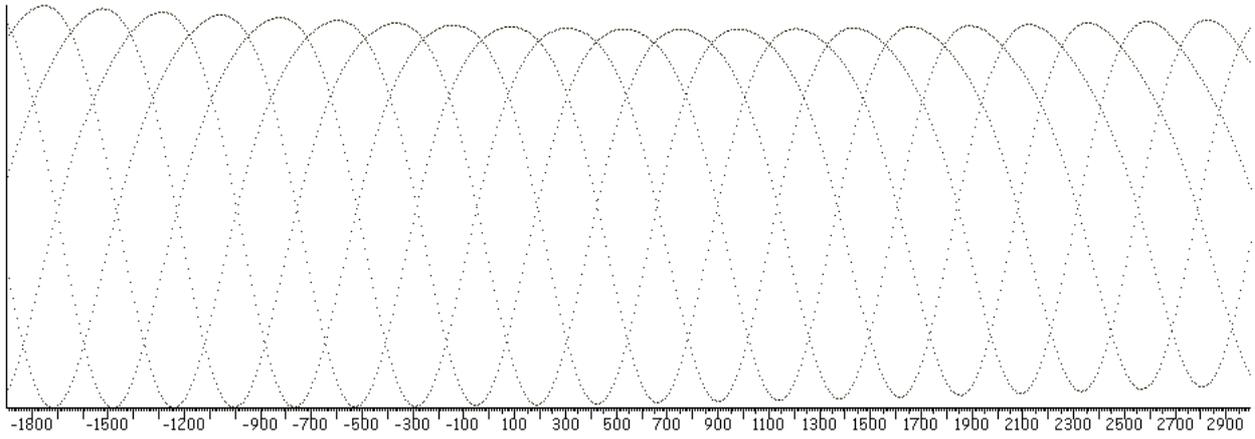

Figure 4b - EMB-Venus meet-point distance chart, longer tendency, from -1900 to 3000.

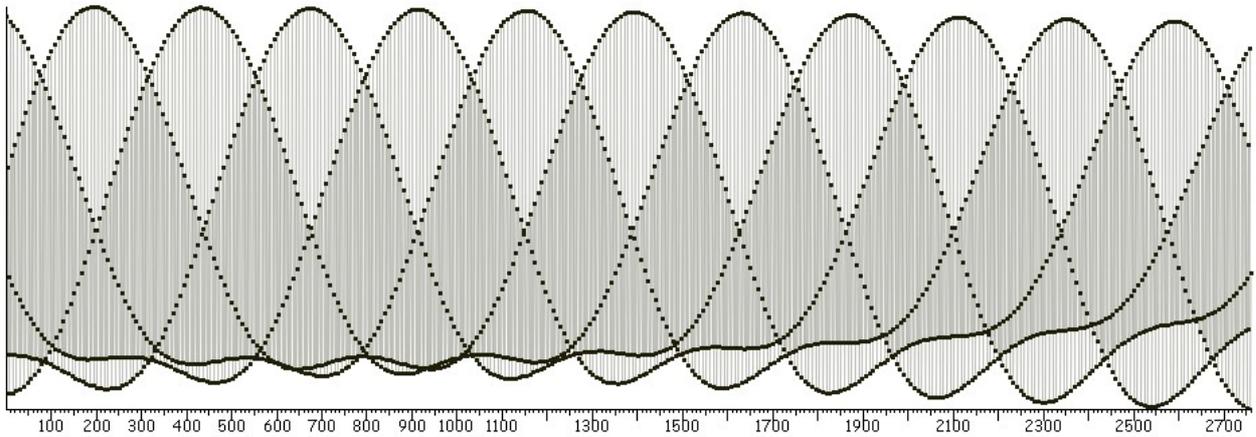

Figure 4c - Frequency of corresponding EMB-Venus meet-points (after 238 years), during 3 millennia, in nHz

Frequency of corresponding meet-points between EMB and Venus planets (fig. 4c) is on average 238.205 years, which is 0.13302878 nano Hertz, or *almost* the tone **D**:

Minimum: 87001.9d (238.198y, F=0.13303238 nHz, tone=C# +93.18%)
Maximum: 87006.0d (238.209y, F=0.13302622 nHz, tone=C# +93.10%)
Average: 87004.3d (238.205y, F=0.13302878 nHz, tone=C# +93.14%)

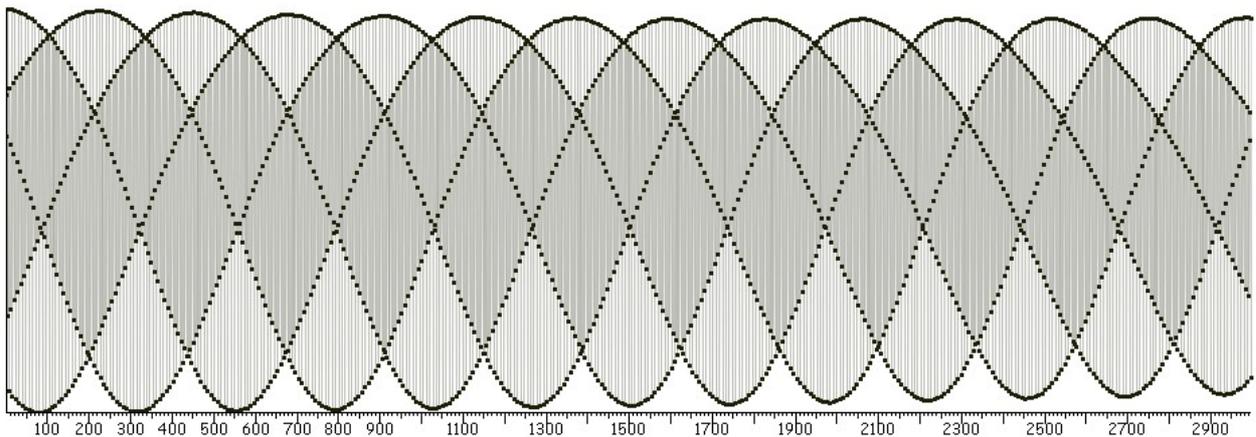

Figure 4d - Frequency of all meet-points between EMB and Venus (584 days) in nHz

Minimum: 581.2d (1.591y, F=19.91382016 nHz, tone=E +63.72%)
Maximum: 587.3d (1.608y, F=19.70695948 nHz, tone=E +45.59%)
Average: 583.9d (1.599y, F=19.82133173 nHz, tone=E +55.61%)

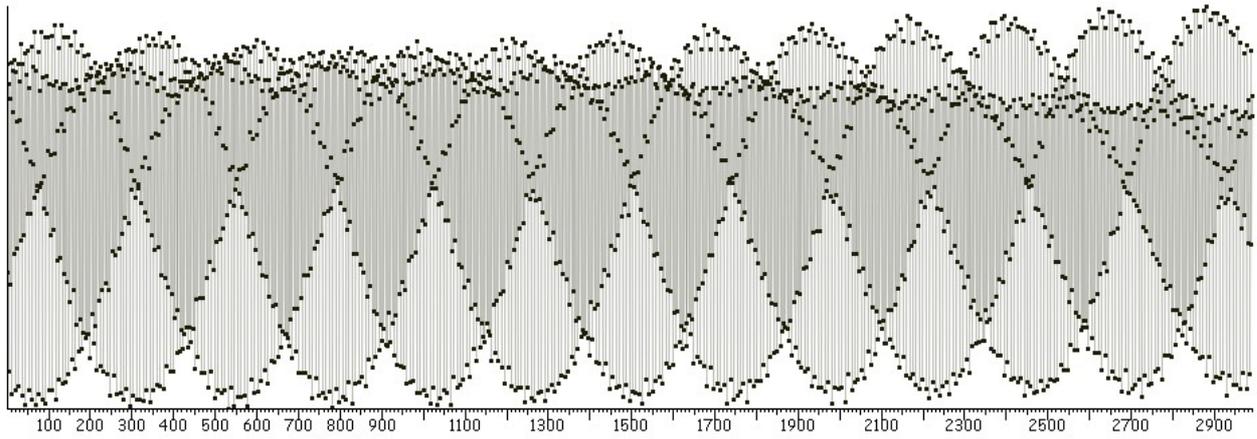

Figure 4e - Frequency of corresponding meet-points between EMB and Venus (8 years) in nHz

Note many irregularities in EMB-Venus corresponding 8-year meet-point frequency (fig. 4e, calculated from time between meet-points separated by 8 years). Yet the planet orbits are always re-stabilized by the resonance. (We just hope the irregularities are not our error, we checked this result twice... Figures 4c, 4d and 4e were all calculated from a same dataset, only the frequency filter has changed...)

Minimum: 2919.5d (7.993y, F=3.96435194 nHz, tone=C +69.44%)
Maximum: 2919.7d (7.994y, F=3.96413548 nHz, tone=C +69.34%)
Average: 2919.6d (7.993y, F=3.96425779 nHz, tone=C +69.40%)

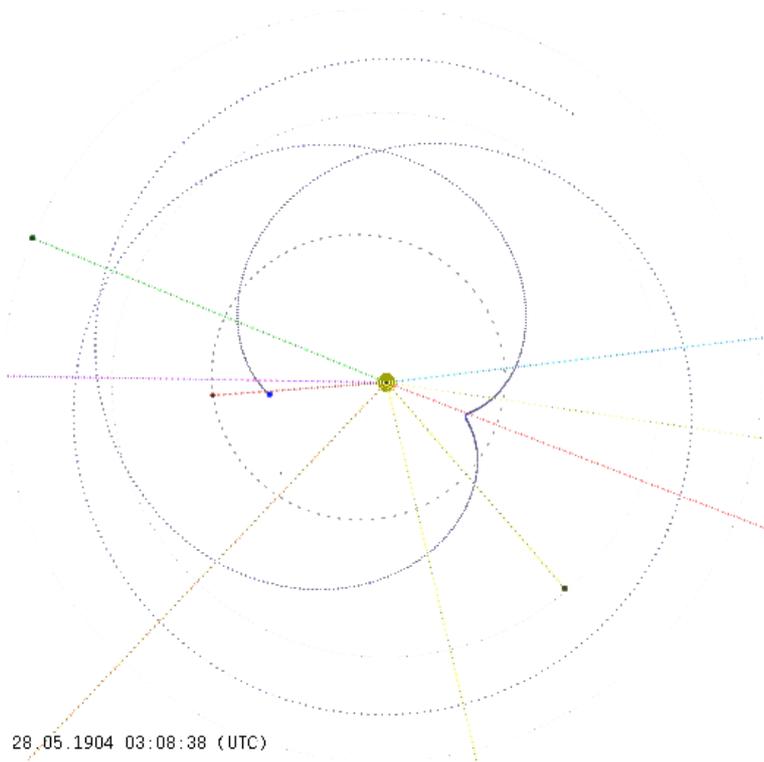

Figure 5 - Barycenter trace in progress, compared with planet trajectories.

The plots at figures 5 and 6 are 1 dot per 1 earthday. From the dot density, you can see the speed of the barycenter at various parts of the trajectory. The point moves fast when far from the Sun, and almost stops when it is nearest. This event is very dynamical: the barycenter suddenly starts approaching the Sun in orbital direction, then it stops almost on one place for 2 weeks, then it starts receding from Sun and moving again in the orbital direction (fig. 6)...

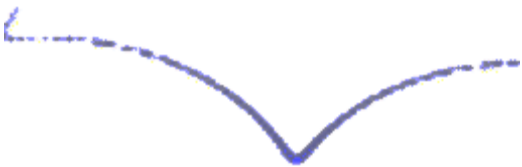

Figure 6 - Barycenter (E+V) trace in detail, position of Sun is marked by a single dot.

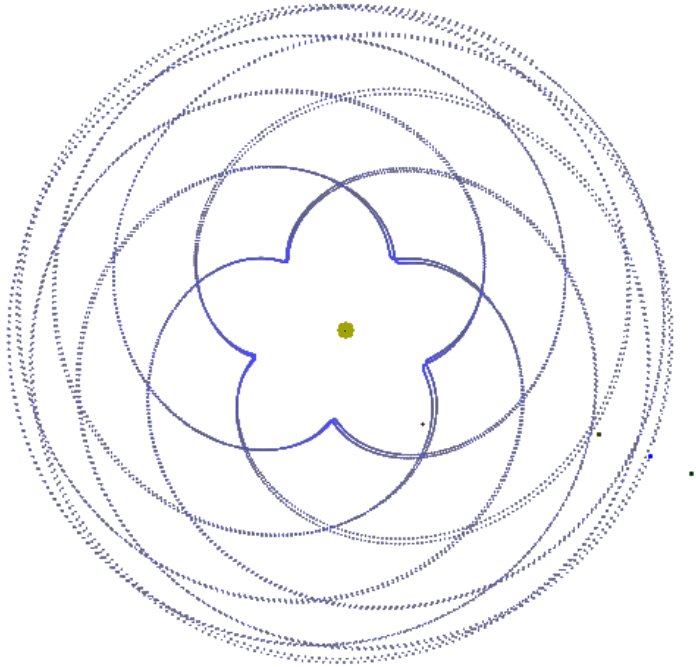

1.12.2149 11:13:38 (UTC)

Figure 7 - Compared two rounds of the resonance shape between Earth and Venus.

When compared with 1/5 round turn after 240 years (fig. 7), the shape almost closes onto its previous version, but not completelly - probably due to disturbances by larger planets, mostly by Jupiter... Yet despite these disturbances, the resonance keeps the orbits more stable (so that the resonance shape restores briefly (within few years) after the perturbations), and on a larger scale, there is no chaos.

## *Tidal locking of Venus planet*

The Venus spin/rotation seems to be tidaly locked by the Earth:
The Venus sidereal day (when compared to fixed star background) is -243.0185 Earth-days in counter-orbital (retrograde) direction. After the 13 orbits of Venus, when it meets Earth for 5th time during the resonance cycle, it makes almost exactly **12** day spins relative to the star background. Since this, the ratio between Earth year and Venus day is almost 8/12, which is **2/3**, more exactly 0.66533.
The apparent Solar day on Venus is 116.75 days, and it takes almost exactly **5** apparent solar days between consecutive meetings of Earth and Venus planets (25 during the whole resonance cycle), so that the Venus planet shows always *almost* the same face to the Earth planet during each meeting, and shows that same face to both Earth and Sun during heliocentric opposition of Earth and Venus planets.
One reason for the counter-orbital (retrograde) spin direction seems, that the tidal force is highest, when the planets meet (and are nearest to each other), and in these times, the Earth moves *retrograde* from Venus point of view, same as if Venus meets Mercury or is nearest to Jupiter, they also move retrograde from Venus point of view. The other reason (of ill spin/rotation) may be the absence of any moon on Venus?

## *Jupiter - Saturn resonance*

Jupiter and Saturn planets resonate in **5:2** rate, they meet 3 times during 5 Jupiter years and 2 Saturn years, with a difference of 242 Earth-days in orbital (prograde) direction (which is 9.7309°, or 1.125% of the cycle). The difference makes the shape rotate, and close almost after 854 Earth-years, which is almost 29 Saturn-years and 72 Jupiter-years. This (854 earth-years) is also the cycle in their relative orbital inclination and the main angular momentum cycle of the Solar planet system. (See figures 8-13)

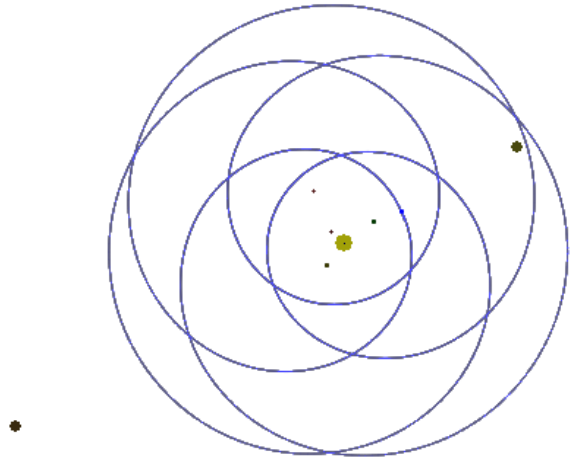

26.1.2050 10:45:56 (UTC)

Figure 8 - Barycenter trace between Jupiter and Saturn planets.

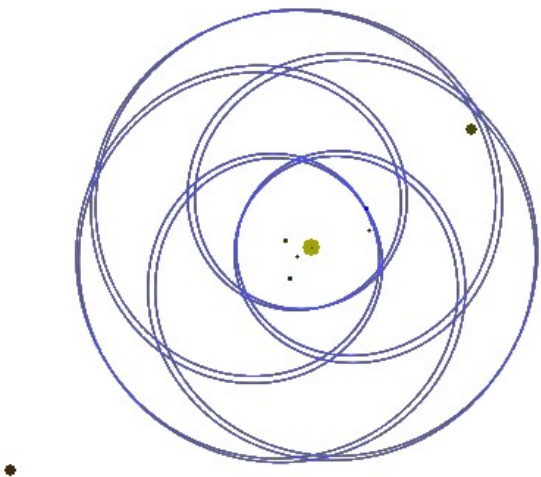

19.8.2109 10:45:56 (UTC)

Figure 9 - Compared two versions of Barycenter trace between Jupiter and Saturn planets

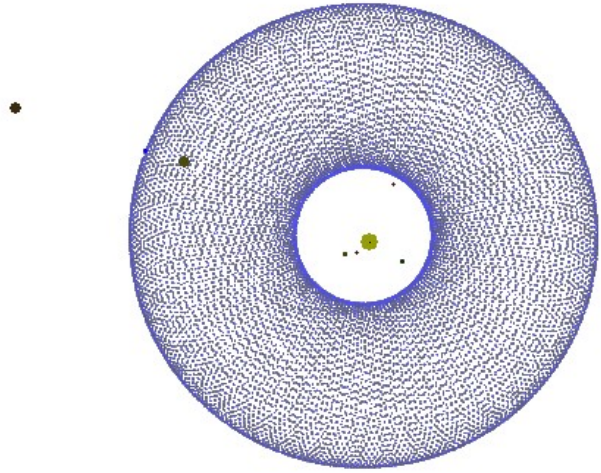

28.11.2457 6:12:16 (UTC)

Figure 10 - Barycenter trace between Jupiter and Saturn after 854 Earth-years

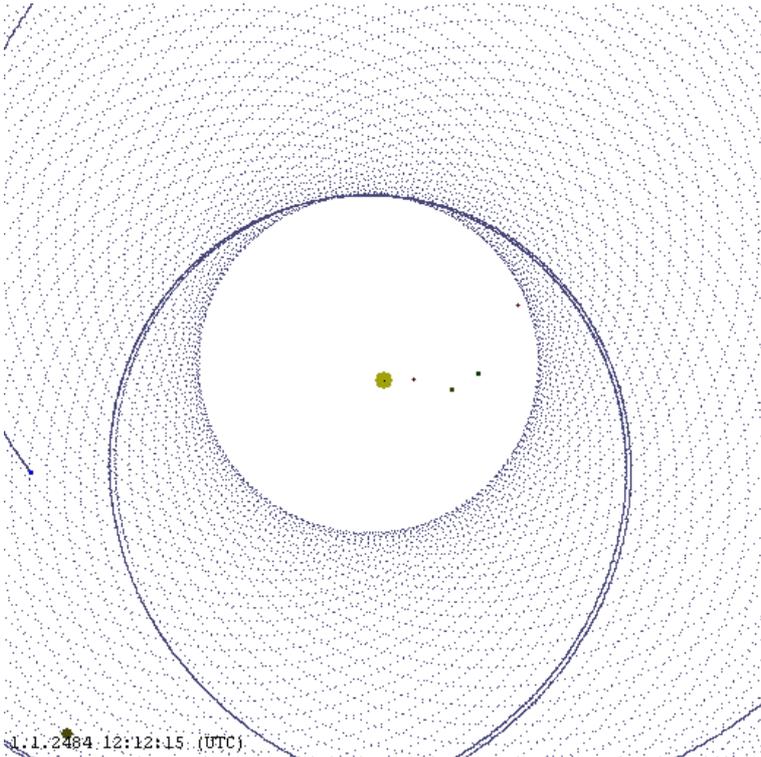

1.1.2484 12:12:15 (UTC)

Figure 11 - Barycenter trace between Jupiter and Saturn planets, compared with the next cycle, in detail.

Note (fig. 11), that the barycenter trace between Jupiter and Saturn planets does not close exactly with the previous round of the cycle.

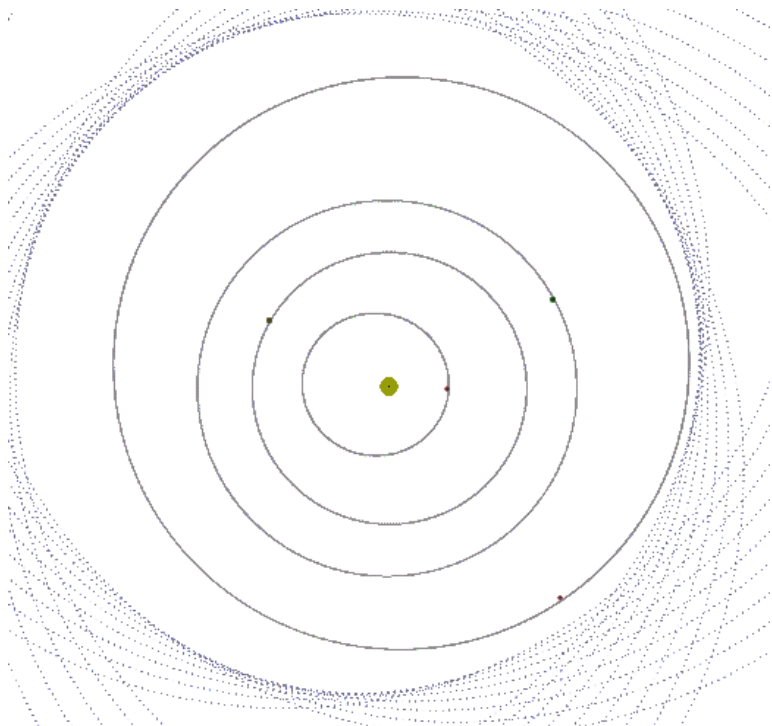

Figure 12 - Barycenter between Jupiter and Saturn planets, compared with innet-planet trajectories.

Note (fig. 11,12), that Mars just does not pass through the bary-center trajectory of its outer neighbours. The Asteroid belt is in the place, where a planet could have been, but it either could not form or it was destroyed, since it passed through the bary-center of Jupiter and Saturn...

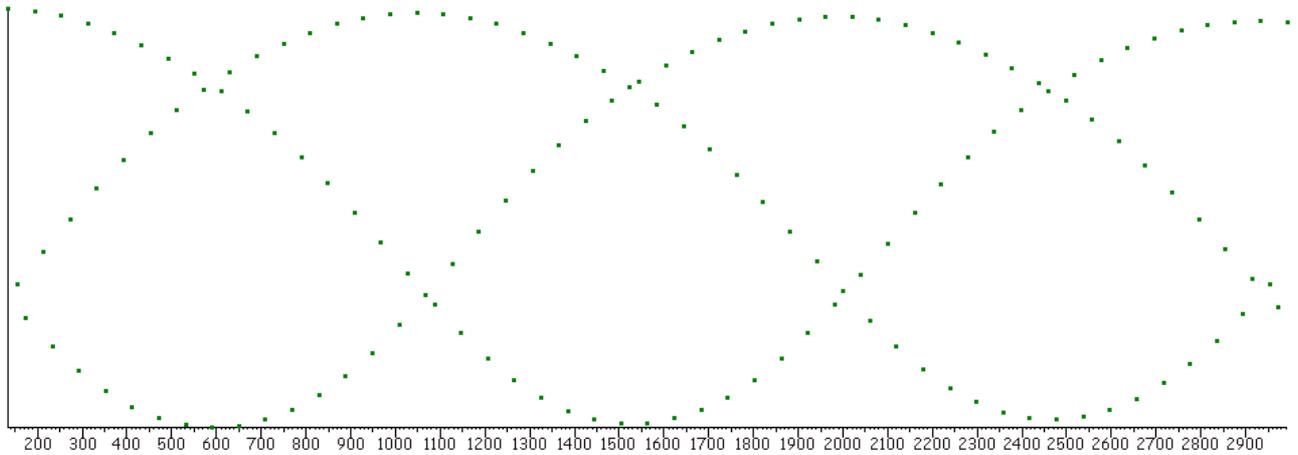

Figure 13 - Meet-point distance chart between Jupiter and Saturn planets

Distance of Jupiter and Saturn planets at meet-points (fig. 13) is influenced by the orbital excentricity (distance of meet-points from the perihelions of planets) and shows the 854-year cycle of Jupiter/Saturn and a main cycle of angular momentum (absolute value) change in the Solar system (see later chapters on Angular momentum).

## *Saturn - Uranus resonance*

Saturn and Uranus resonate in **20:7** rate (which is close to 3:1), with a difference of 7.75° (after 588.8 years of the cycle) in prograde direction.

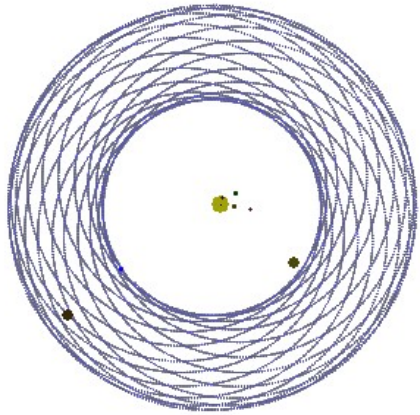

Figure 14 - Barycenter trace between Saturn and Uranus planets

Note (fig. 14), that the Jupiter planet just does not pass through the bary-center trace of its outer neighbours...

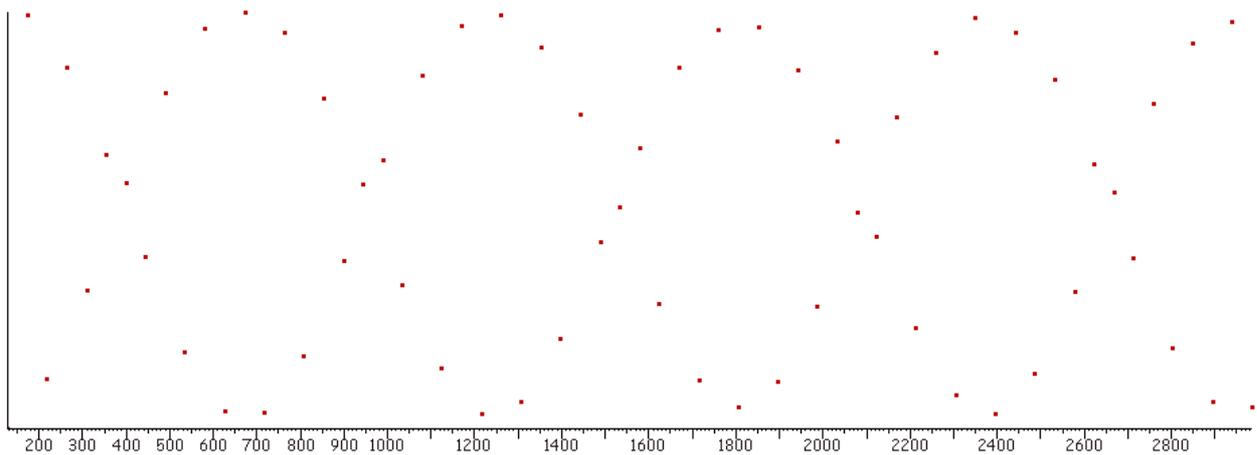

Figure 15 - Meet-point distance chart between Saturn and Uranus planets.

## *Uranus-Neptune resonance*

Uranus and Neptune planets resonate probably in **51:26** rate (which is close to 2:1), with a difference of 4.278° (after 4286.8 years of the cycle) in prograde direction.

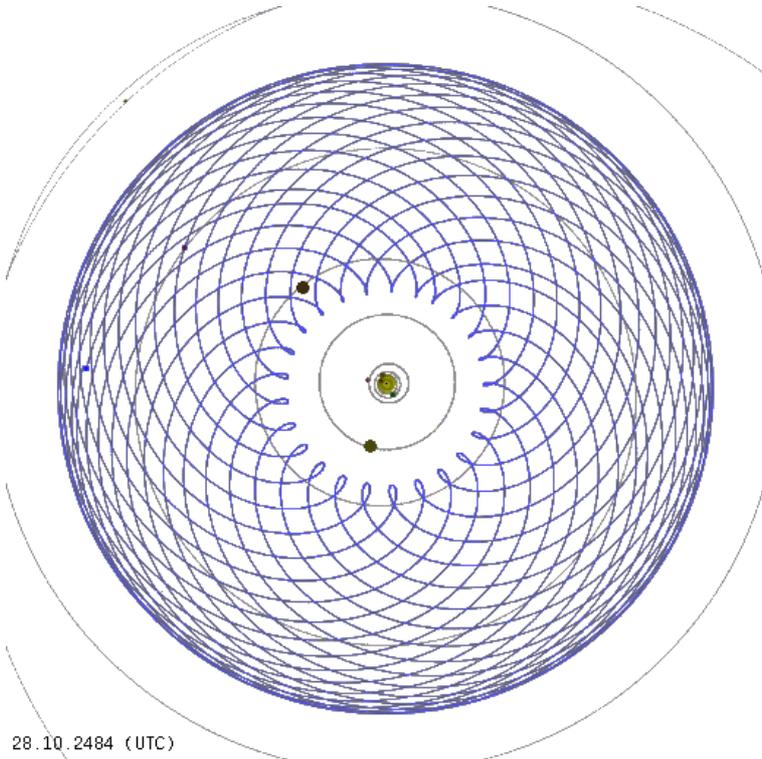

Figure 16 - Bary-center trace between Uranus and Neptune planets

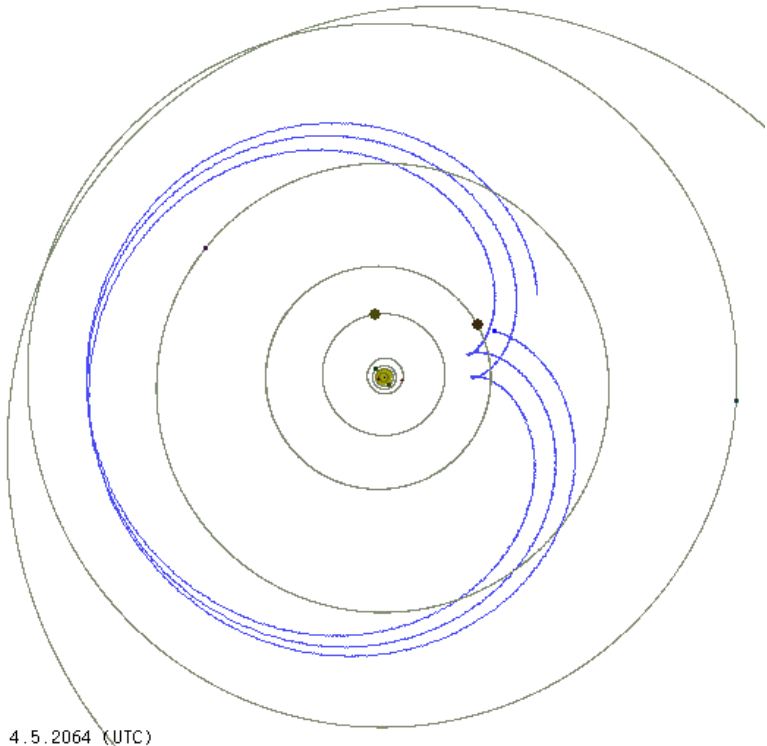

Figure 17 - Saturn planet meets the bary-center of Uranus/Neptune planets.

Note (fig. 17), that Saturn passes through the barycenter trajectory of its outer neighbours, and seems to meet that bary-center regularly at present time, but according to these ephemerides it was not meeting that barycenter in past centuries...

## *Neptune - Pluto resonance*

Neptune/Pluto resonate in **3:2** rate, with a difference of 2.7° - 3.1° in retrograde direction during most meetings in present era. Anyhow (according to DE406), before some 4 millennia the drift has occured sometimes in prograde direction also. Their resonance (beside Earth-Mars one) seems to be one of the least stable of the resonances.
No bary-trace is available, since their weight is incomparable and the barycenter between Neptune and Pluto planets is almost inside the Neptune planet, same as barycenter between Jupiter and Mars planets is almost inside the Jupiter planet...

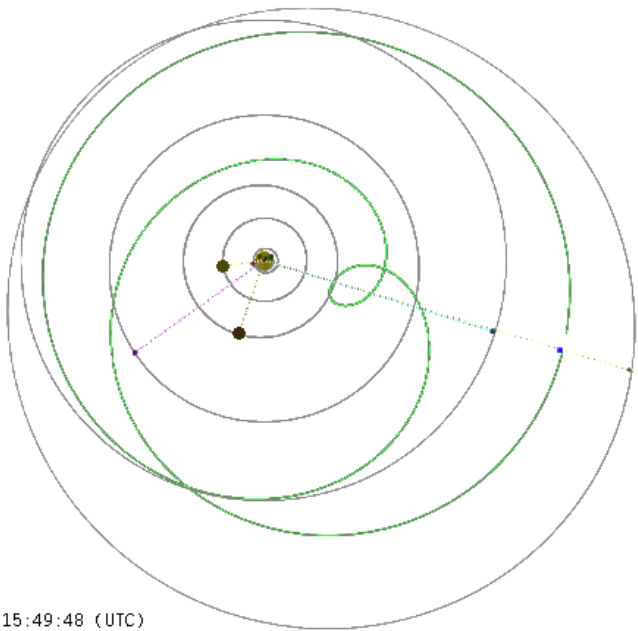

Figure 18 - Mid-point trace between Neptune and Pluto planets, compared with planet trajectories.

The mid-point trace (fig. 18) has got no physical meaning and is included just for an illustration...

## *Venus - Mercury resonance*

**Mercury/Venus** planets weakly resonate in mean **23:9** rate.

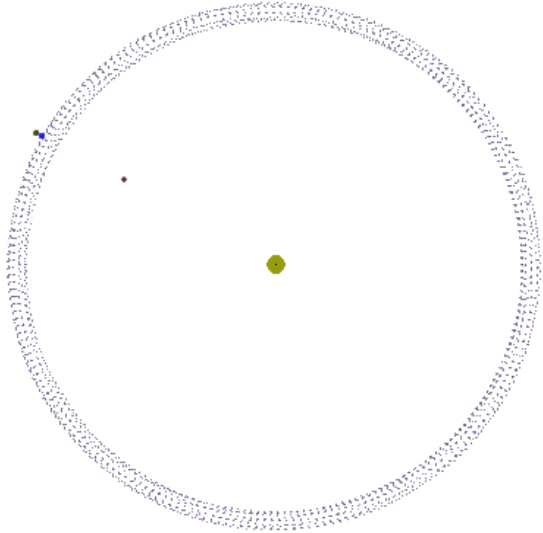

16.10.2011 04:52:31 (UTC)

Figure 19 - Barycenter trace between Venus and Mercury planets

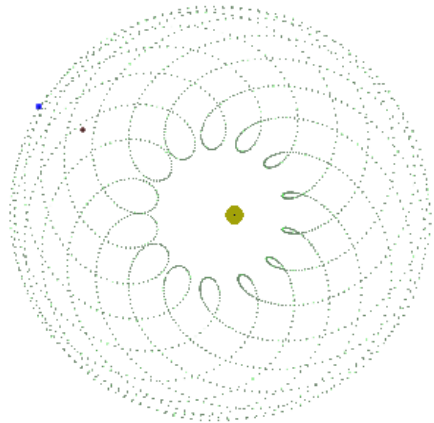

16.10.2011 04:52:31 (UTC)

Figure 20 - Mid-point trace between Venus and Mercury planets

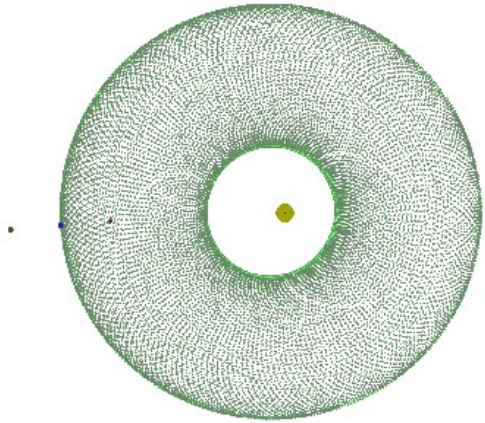

`04.09.2061 10:48:22 (UTC)`

Figure 21 - Ten succesive Mid-point traces between Venus and Mercury planets

The weight of Venus and Mercury planets is little comparable, so that bary-center trajectory trace (fig. 19) little differs from Venus orbital trajectory.
When drawing middle-point traces between Mercury and Venus planets (figures 20 and 21) (*which has got no physical meaning*), it reveals the excentric shape of 14 petals, which rotates in *prograde* direction, by one petal after 10 these shapes.

Mercury orbit is highly excentrical. Mercury is also one of a few planets, that pass through the trajectory of barycenter of its outer neighbours (the other one is Saturn), and seems to meet it regularly (only) at some times... Consequence of this bary-center meeting on the orbit regularity (of planet Mercury) needs further investigation...

## *Earth - Mars resonance*

The resonance of **Earth/Mars** is little close to **15:8** (which is close to 2:1), but it seems the least perfect (or least stable, most perturbed) of the planetary resonances...

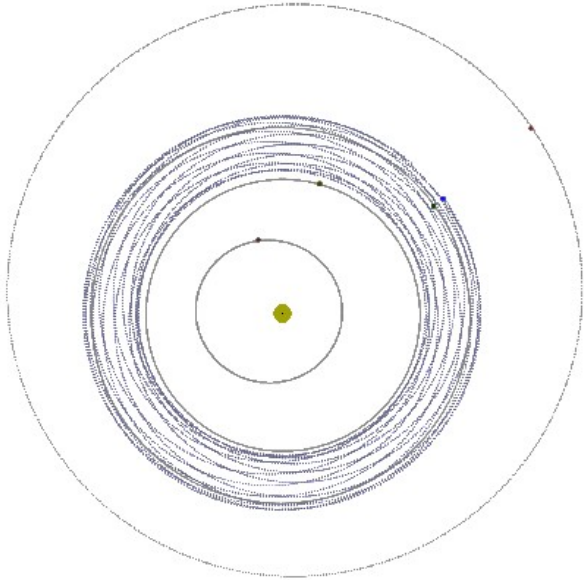

25.01.1931 15:01:28 (UTC)

Figure 22 - Bary-center trace between Earth and Mars planets, compared with planet trajectories

The barycenter trajectory between Earth and Mars planets (fig. 22) seems erratic, probably due to disturbations of the Mars planet by the asteroids and by the Jupiter planet...
It takes 14.96 Earth-years to complete this figure (fig. 22).

Note the Venus planet trajectory, as it just does not touch the Earth/Mars barycenter trace...

## *Venus - Mars resonance*

The resonance of Venus/Mars is **3:1**. Venus meets Mars (heliocentric conjunction) two times during this cycle, with an asymetric lengths (353.9 Earth-days and 314.8 Earth-days at some times), which mostly depends on the Mars excentricity. The total time between 2 meetings (3 Venus years, 1 Mars year) is 668.7825 Earth-days, with a difference of -8.544° (in retrograde direction).
The whole shape rotates after 32.9 years by 1/3 of circle, and after 98.7 years by 1 whole circle...

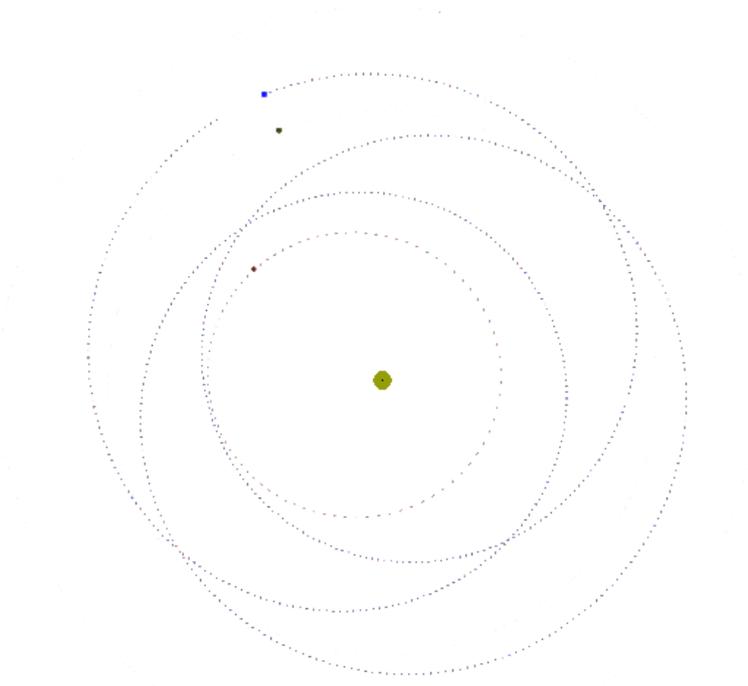

Figure 23 - Barycenter trace between Venus and Mars planets

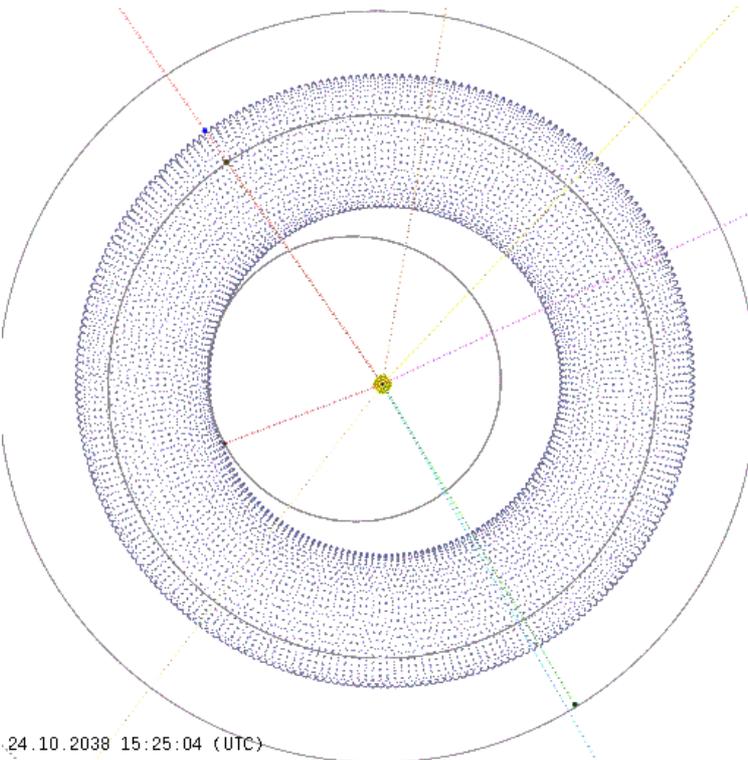

Figure 24 - Barycenter trace between Venus and Mars planets after rotating 1/3 round.

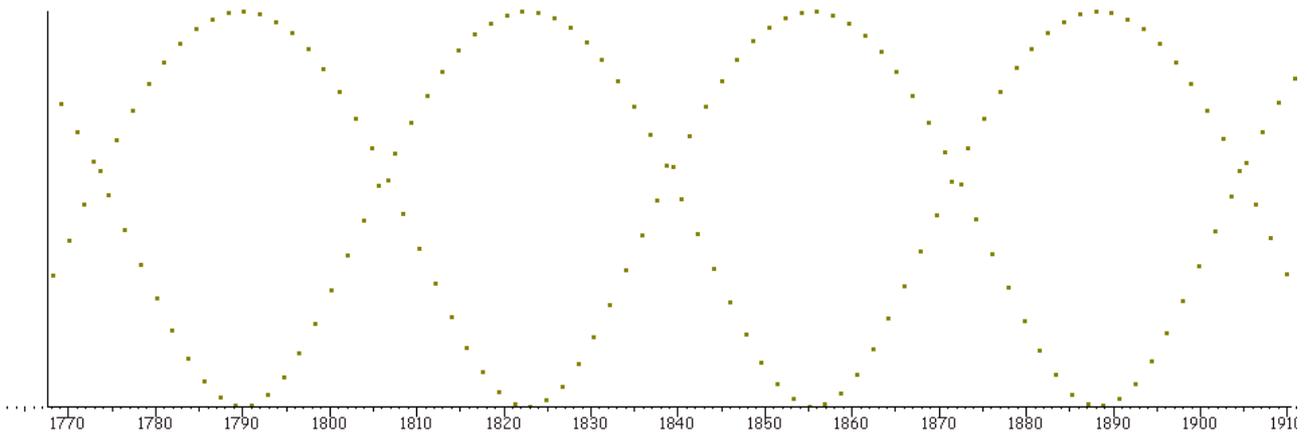
Figure 25 - Meet-point distance chart between Venus and Mars planets

The Mercury planet sometimes meets the barycenter of Venus and Mars planets (fig. 24), but even at closest approaches (during present 5 millennia) it is separated vertically (Z ecliptic coordinate) by at least 3,318,844 km (at 12.9.414 AD)

## Mercury - Earth resonance

The resonance of **Mercury/Earth** is **29:7**, with 22 nodes drifting in retrograde direction after 46 years by 1/7 of the circle...

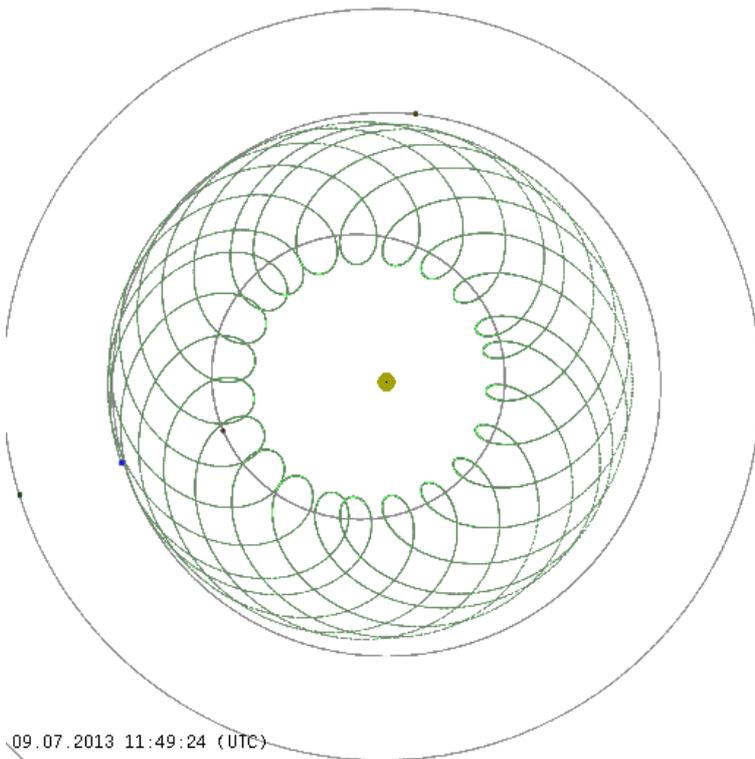
Figure 26 - Mid-point trajectory between Emb and Mercury planets

The weight of Earth and Mercury planets is not much comparable, so the bary-center trajectory runs almost in Earth center. The mid-point trajectory (*which has got no physical meaning*), looks like this (figure 26).

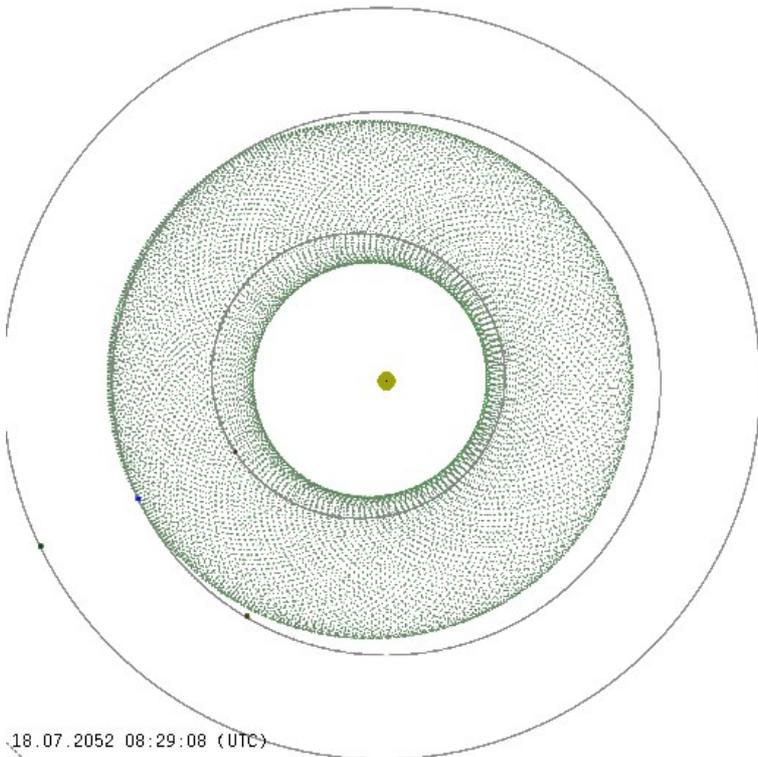
Figure 27 - Mid-point trajectory between Earth and Mercury planets after 46 Earth-years

## *Non-neighbour resonances in the outer group*

Resonance can also be found between non-neighbouring planets in the outer group:

Ratio between **Jupiter/Uranus** (fig. 28) is **7:1**, with difference of -4.48475° in retrograde direction.
Ratio between **Jupiter/Neptune** (fig. 29) is **14:1**, with difference of +3.0176° in prograde direction.
Ratio between **Saturn/Neptune** (fig. 30) is **28:5**, with difference of +2.16647° in prograde direction.

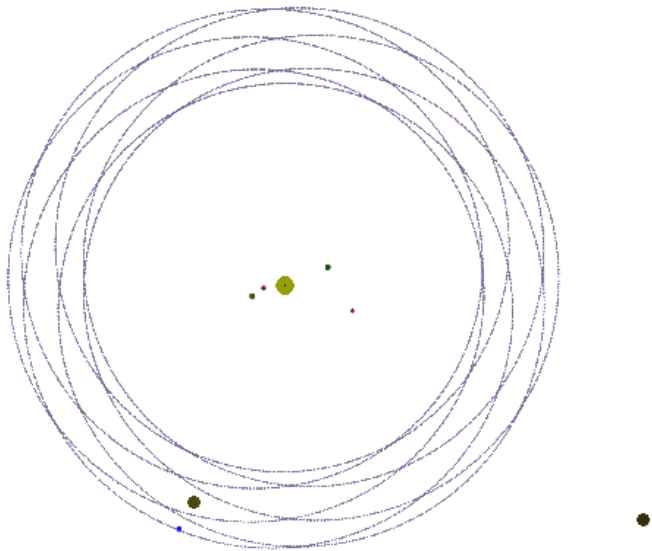
Figure 28 - Barycenter trace between Jupiter and Uranus planets

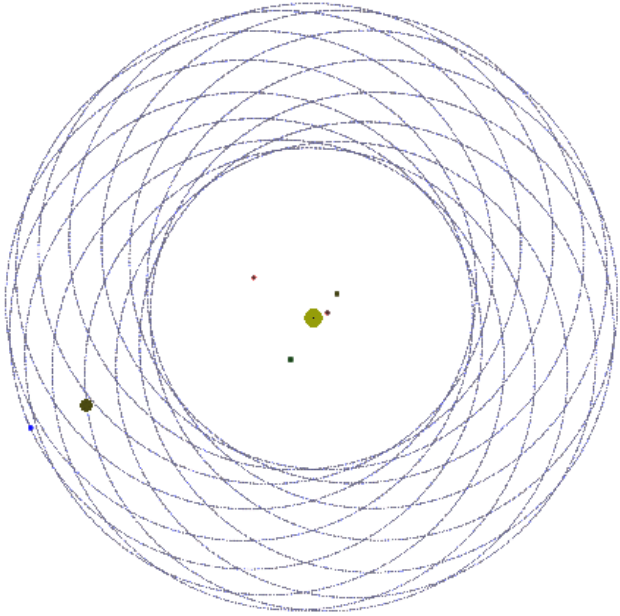

18.08.1664 23:09:17 (UTC)

Figure 29 - Barycenter trace between Jupiter and Neptune planets

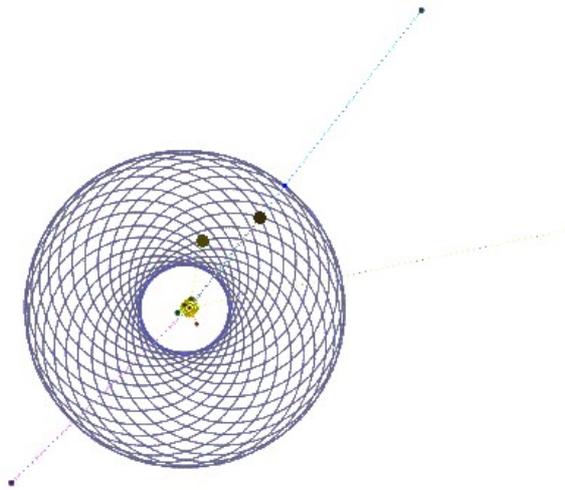

20.07.2419 19:43:33 (UTC)

Figure 30 - Barycenter trace between Saturn and Neptune planets

## *Resonance groups*

There is no apparent resonance between Mars and Jupiter (ratio of orbital periods 6.30792 which is close to 82/13) and the resonance between Jupiter and Earth (12:1) is very far from perfect, so this is the place, that separates two resonating groups of planets:
- Mercury/Venus/Earth/Mars
- Jupiter/Saturn/Uranus/Neptune/Pluto

Within the group, the middle planets resonate on both sides and even with non-neighbour peers.

The planets in the two groups also differ by the important trajectory characteristic - the inner small planets orbit on heliocentric trajectories, whereas the outer large planets orbit on barycentric trajectories (see later the chapter about Angular momentum).

## *Conclusion of orbital resonance*

There is no chaos in planetary motions, but instead the cycles are more complex then was previously thought. The perturbations (see Angular momentum chapter below) average in longer-scale cycles, and the planetary motions are stabilized by the resonance.

## *Angular momentum*

### Model

This is my analysis of planet moves, encoded in DE405 and DE406 ephemerides, which were calculated by E.M.Standish, JPL NASA.

### Center

Inner (small) planets orbit arround the Sun (heliocentric trajectories).
Outer (large) planets, on the other hand, orbit arround the Solar system barycenter (SSB), as a counter-weight of the Sun.

### Angular momentum table

In a simple Keplerian model, the angular momentum of a planet, with respect to the center of its trajectory, is constant. The planet moves faster, when near to center, and slower, when far from center, which conserves the angular momentum.

Anyhow, in real world, angular momentum of individual planets is not constant, due to tugs (perturbations) by all other planets and compensations by different inclinations to the invariant plane...

Relative values of orbital angular momentum of individual planets, in arbitrary units (probably kg*m$^2$ * 1.4959787E+29, metric units divided by AU and scaled decimally), as determined from *preliminary* ephemerides versions DE414 & DE415. (See Table 1)

Table 1 - relative values of orbital angular momentum of individual planets with respect to specified center.

| Planet | Center Sun | | | | Center Ssb | | | | Mass (kg) (G=6.6742*10$^{-11}$) |
|---|---|---|---|---|---|---|---|---|---|
| | Minimum | Maximum | Scatter | Tendency | Minimum | Maximum | Scatter | Tendency | |
| Sun | | | | | 5.9000 | 290.0000 | 284.1000 | oscilating | 1.98843966345011E+30 |
| Mercury | 5.9865 | 5.9866 | 0.0001 | reducing | 5.8308 | 6.1763 | 0.3455 | oscilating with Sun | 3.30108185047165E+23 |
| Venus | 123.2947 | 123.2971 | 0.0024 | growing | 121.5867 | 125.0140 | 3.4273 | oscilating with Sun | 4.867381346361E+24 |
| Emb (*) | 180.0427 | 180.0474 | 0.0047 | growing | 178.2222 | 181.9053 | 3.6831 | oscilating with Sun | 6.04571684215467E+24 |
| Mars | 23.4882 | 23.4906 | 0.0024 | reducing | 23.3269 | 23.6571 | 0.3302 | oscilating with Sun | 6.41699179158458E+23 |
| Jupiter | 128 875.4927 | 128 923.2083 | 47.7155 | narrow oscilating, reducing | 128 519.1412 | 128 815.7483 | 296.6071 | smooth oscilating contrary to Sun | 1.89854360313697E+27 |
| Saturn | 52 156.0531 | 52 345.9642 | 189.9111 | wide oscilating | 52 193.5867 | 52 244.4528 | 50.8661 | narrow oscilating, growing | 5.68450446981147E+26 |
| Uranus | 11 265.6617 | 11 326.0695 | 60.4079 | wide oscilating | 11 292.6605 | 11 296.5708 | 3.9103 | long cycle | 8.66045149559652E+25 |
| Neptune | 16 773.9927 | 16 791.7912 | 101.7985 | wide oscilating | 16 824.2558 | 16 825.3984 | 1.1426 | long cycle | 1.02953279430512E+26 |
| Pluto | 2.7635 | 2.7864 | 0.0229 | wide oscilating | 2.7742 | 2.7745 | 0.0003 | slowly growing | 1.52956896907633E+22 |
| | | | | | | | | | |
| sum (*1) | | | | | 209470.9488 | 209471.1088 (*2) 209481.96 | 0.1600 11 (*2) | 854y cycle | |

These minimum and maximum values are during 20th century, for Pluto during 3 centruries, and for the Sum during 1.5 millennia.

Notes to the Table 1:
(*1) - In the sum there are included 9 planets and the Sun, all with respect to Ssb. The sum is scalar. The planets are Mercury, Venus, Emb (Earth-Moon system), Mars, Jupiter, Saturn, Uranus, Neptune, Pluto. It does not include asteroids and spin momentum of bodies, namely of Sun.
The whole sum is almost constant, but there is a small difference of $8.11*10^{-7}$ of the whole. The difference from a constant value is mainly caused by performing the scalar sum instead of vector sum, and also it is divided between orbital angular momentum of asteroids, trans-neptunians and spin angular momentum of Sun. (but of these only asteroids were included in ephemerides calculation)
(*2) - These high swings are at the times, when the Sun is approaching the solar-system barycenter too closely, and then the space curvature (not regarded by me), plays a significant role...? No, the Sun is moving retrograde at these times and its vector angular momentum is actually negative. There is no negative scalar angular momentum, which causes this difference.

On the scale of millennia, the Earth and Venus angular momentum (relative to Sun) grows, these planets move in concentric spirals(?) and are speeding up (as determined by huge Gaussian filtering), or rather by the change in orbit excentricities. The average approach to Sun after 5 millennia is still much smaller than annual approaching and receding due to excentricity... On this scale, the Mercury and Mars angular momentum (relative to Sun) shrinks.

### *EMB*

Planet Earth, when considering orbital characteristics and when interacting gravitally with other planets, is actually a system of 2 bodies, Earth and Moon. Their weight is comparable (1:81.300587). Here we call this system EMB, the **Earth-Moon Barycenter**...
Of this two, only the Earth interacts magnetically, and it oscilates during its orbital motion by the counter-weight of Moon by +-4900 km, with a main frequency of **29.53** days and an envelope of 1.13y (fig. 31). The oscilation does not seem important, and is negligible for ex. when compared with Solar wind velocities etc...

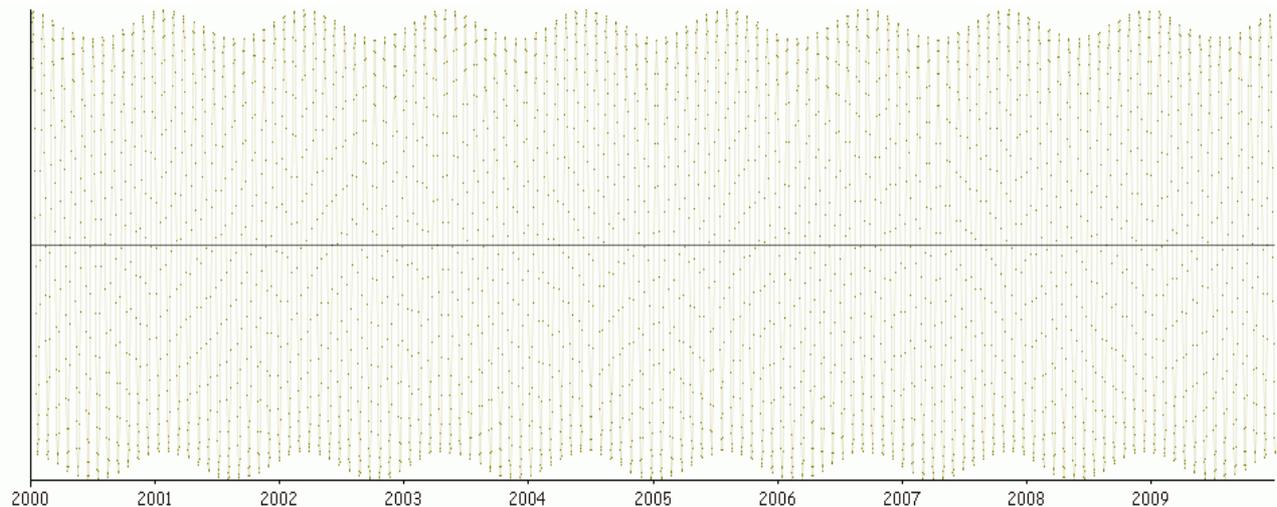

Figure 31 - Distance of Earth from Sun, compared with distance of Emb from Sun

Angular momentum of EMB with respect to Solar-system barycenter:

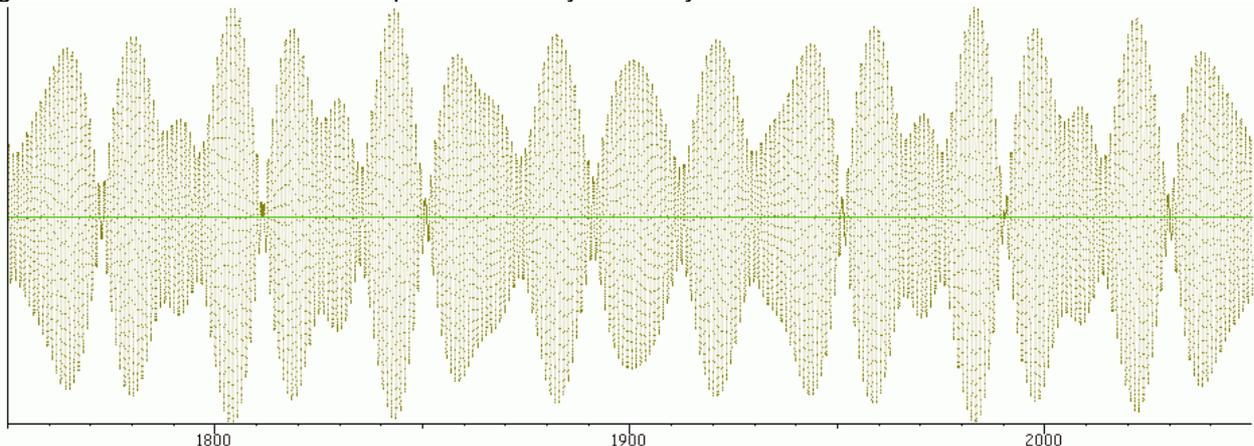

Figure 32 - Angular momentum of EMB with respect to SSB.

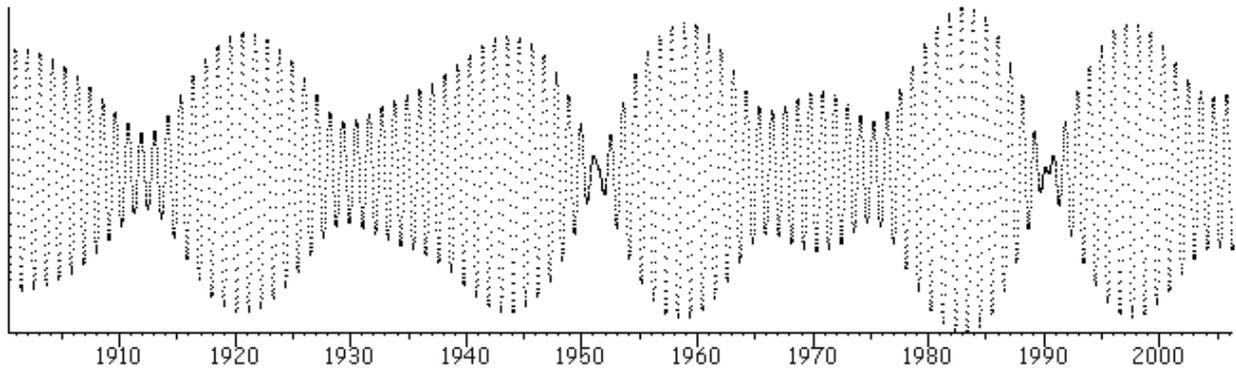
Figure 33 - Angular momentum of EMB with respect to SSB in detail.

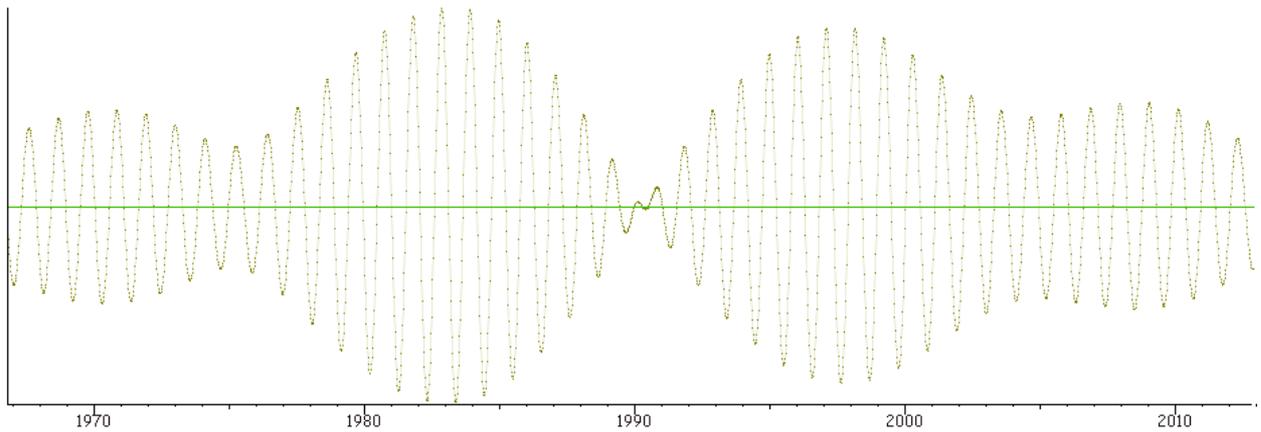
Figure 34 - Angular momentum of EMB with respect to SSB in more detail.

Angular momentum of EMB with respect to Solar system barycenter (olive), compared with angular momentum of EMB with respect to Sun (green) (in fig. 32). It is evident, that angular momentum of Emb is much more constant with the center in Sun, whereas with the center in SSB it oscilates with the Sun-SSB oscilation. Thereby, the center of EMB's orbit is the Sun, not the SSB.
(*The charts are self-relative, with minimum to maximum stretched to fill the image. Actually, if they displayed 0 also, the data would be one straight line at the top of the image, since the oscilation is incomparably smaller than the absolute, almost-constant value...*)

Main frequency is here **1.09** year (which is a frequency of angle between Earth motion vector and vector from Earth to Jupiter), other frequencies are at 1.20y and 1.00y, and the envelope reflects the oscilation of Sun arround the Solar-system barycenter.

As the EMB orbits on a heliocentric trajectory (and not on barycentric), so the angular momentum relative to Sun is more constant and also more interesting...

Angular momentum of EMB with respect to Sun:

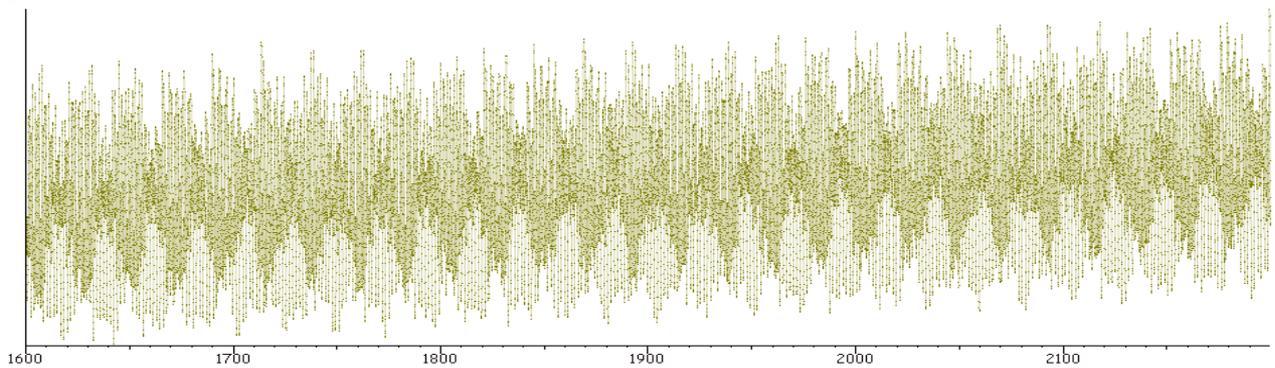
Figure 35 - Angular momentum of EMB relative to Sun

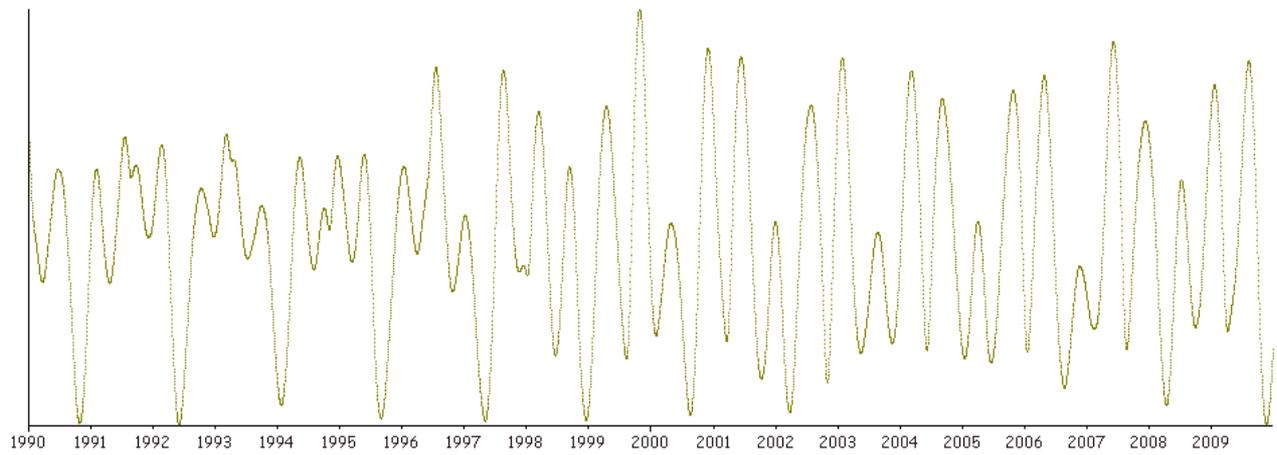
Figure 36 - Angular momentum of EMB relative to Sun, in detail.

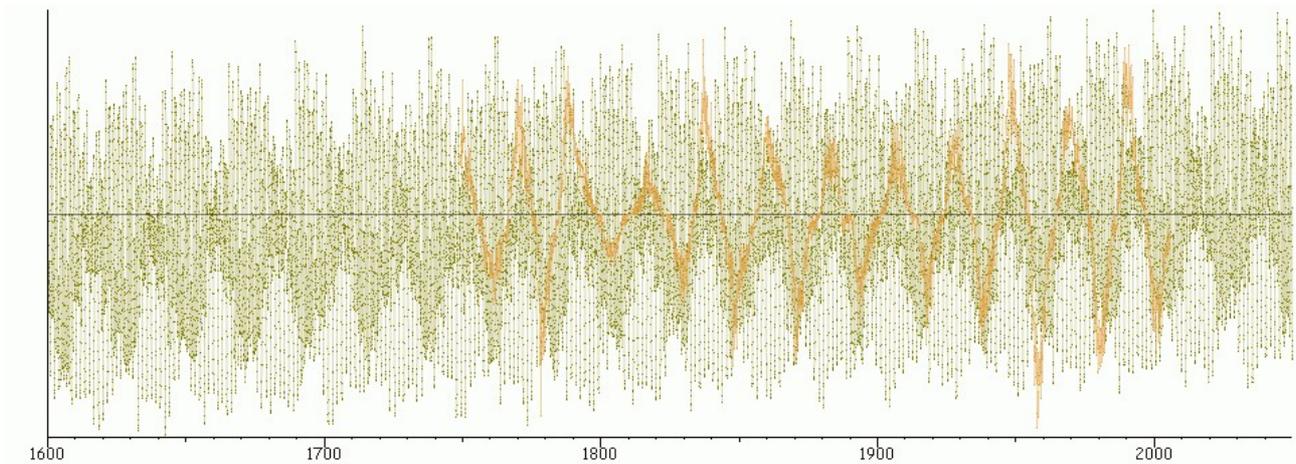
Figure 37 - Angular momentum of EMB relative to Sun compared with signed Sunspot cycle.

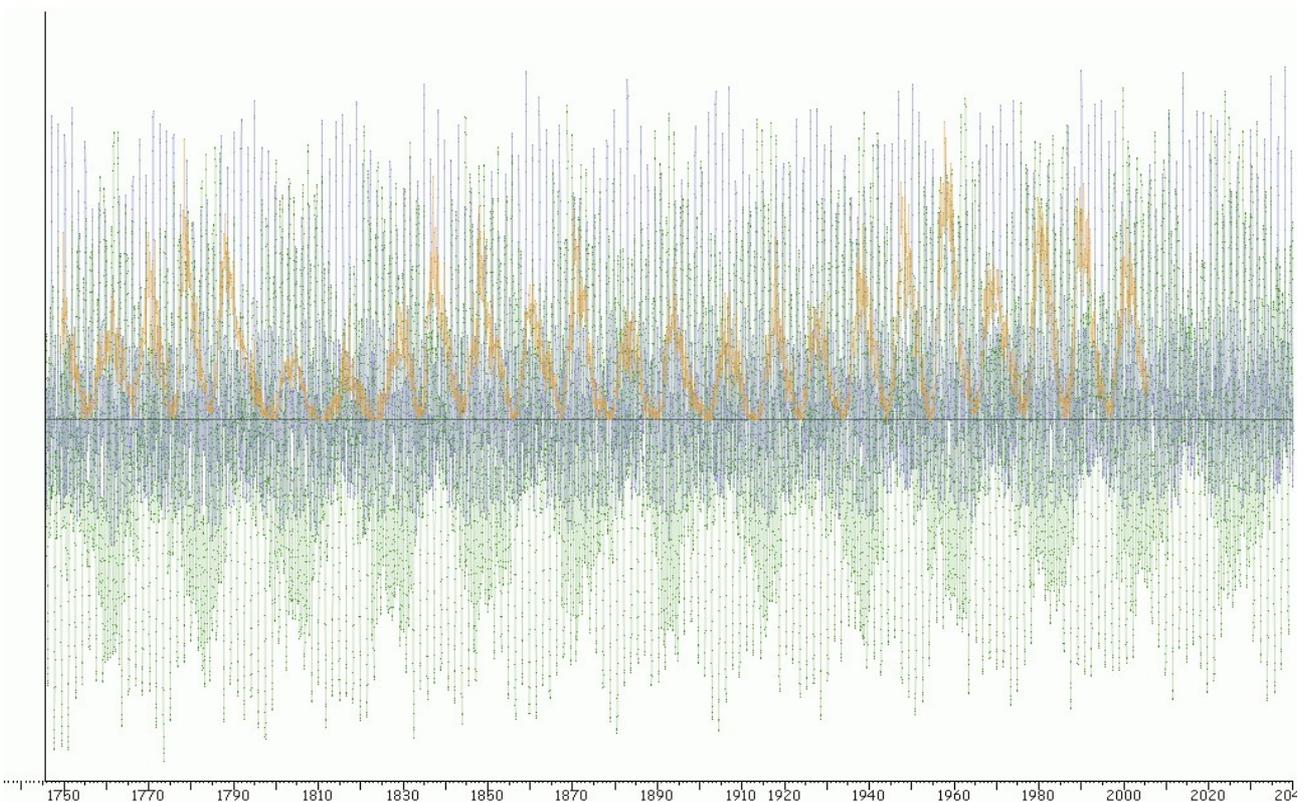
Figure 38 - Sum of Angular momentum of Emb and Venus relative to Sun compared with unsigned Sunspot cycle.

Main frequency is here **199.4** days, which is 1/2 of the synodic period of Jupiter as seen from Earth, then 584 days - the period between meetings of EMB and Venus (the resonance period), then 292 days, 133 days, 117 days, 1.09 years, 3.98 years, 11.84 years, 15.6 years and many other, as determined by FFT analysis...

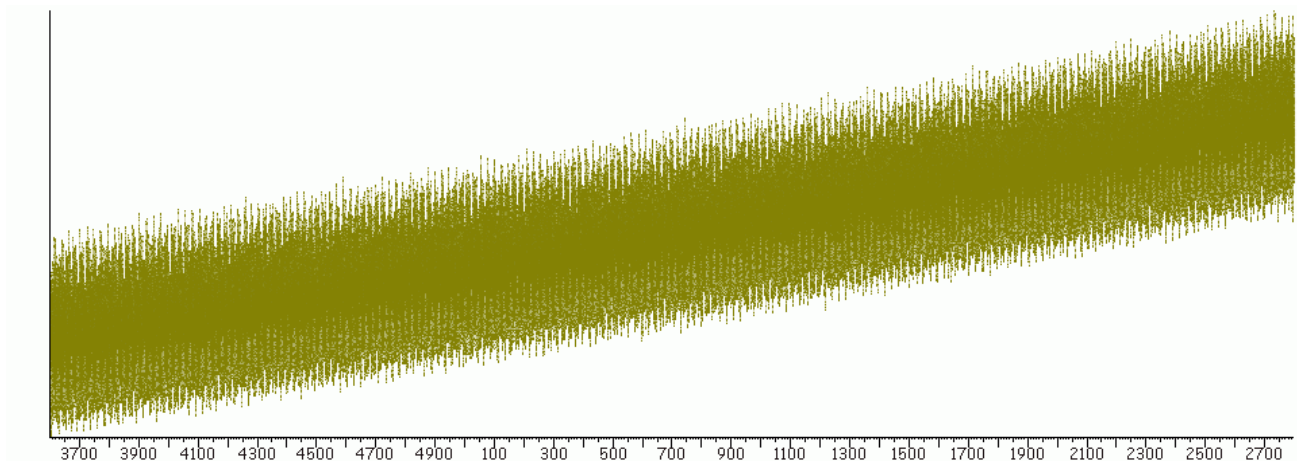

Figure 39 - Tendency of Angular momentum of EMB with respect to Sun, during 4000 years.

**Venus**

As with planet Earth, the Venus planet orbits on a heliocentric trajectory, not barycentric.

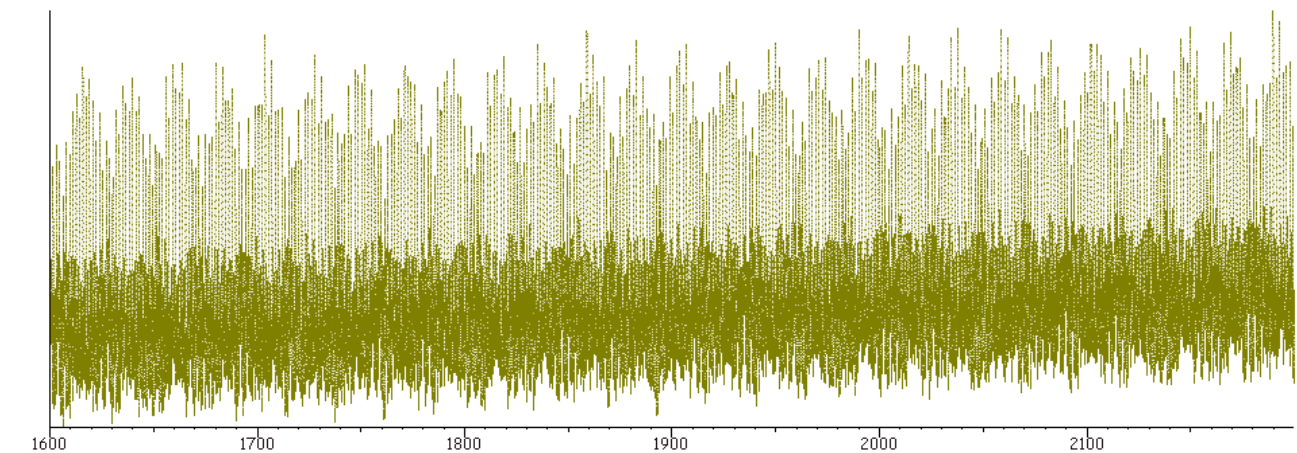

Figure 40 - Angular momentum of Venus with respect to Sun, 1600 - 2200.

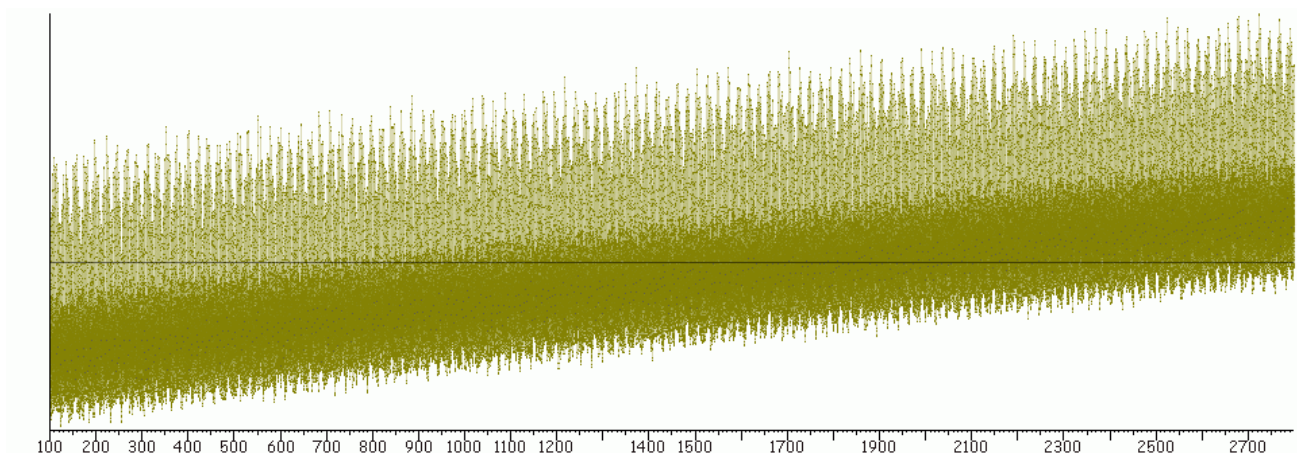

Figure 41 - Angular momentum of Venus with respect to Sun, longer trend (100 - 2800).

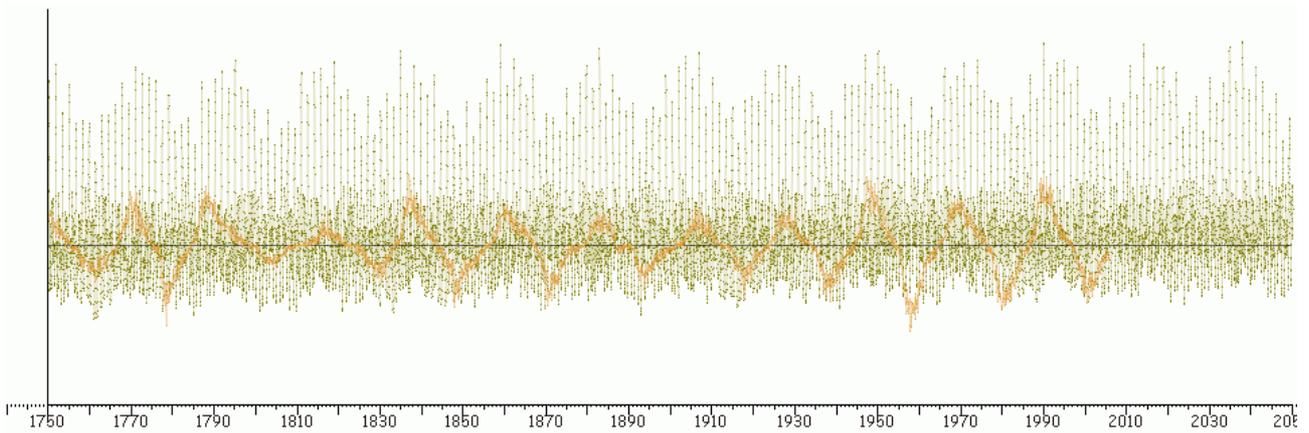
Figure 42 - Angular momentum of Venus with respect to Sun, compared with signed Sunspot cycle.

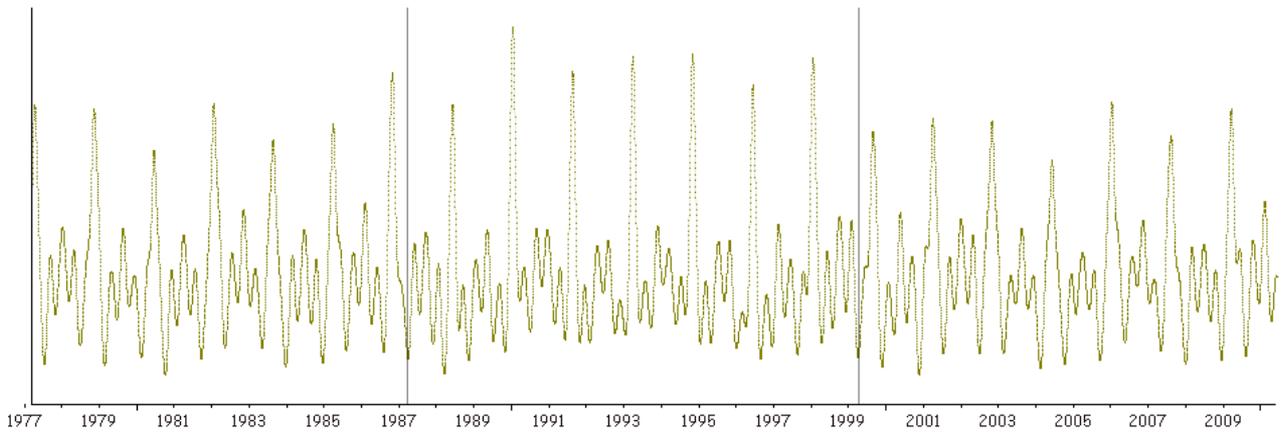
Figure 43 - Angular momentum of Venus with respect to Sun, in detail, with marked mode changes.

Main frequency is here **291.6** days (half of meet-period between Earth and Venus), then 118.4 days, 194.5 days (Jupiter?), 145.9 days (Mercury), 583.5 days (Earth), 116.7 days (1/5 Ea, Venus solar day), 3.96 years, 97 days, 83.4 days, 121.7 days and other, sorted by significance, as determined by FFT analysis...

On a detailed chart, there are remarkable two "modes" in the oscailation. One mode can be described as "1 large spike and 3 small spikes, of which the central one is larger than the other two". The second mode can be described as "1 large spike and 3 similar spikes or 1 large spike and 4 smaller spikes where the central two are larger than the outside two". (fig. 43)

Changes between these modes also approximatelly match the borders between Sunspot cycles (fig. 42, 91).

## Jupiter

It is counter-intuitive, that Jupiter angular momentum relative to Sun shows less swing than Jupiter angular momentum relative to SSB. Jupiter planet is the main counter-weight to the Sun, and so their distance is more constant than distance of Jupiter to Solar System Barycenter (SSB), that varies due to other planets. Also the SSB is always more near to Jupiter than the Sun to Jupiter (see fig. 44, 45).

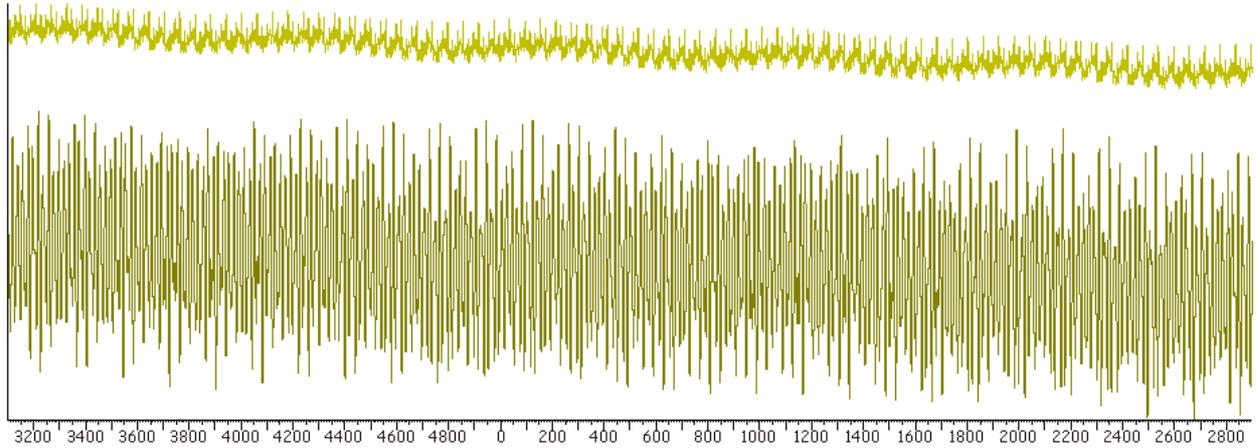

Figure 44 - Compared angular momentum of Jupiter planet with center in Sun (yellow) and SSB (olive)

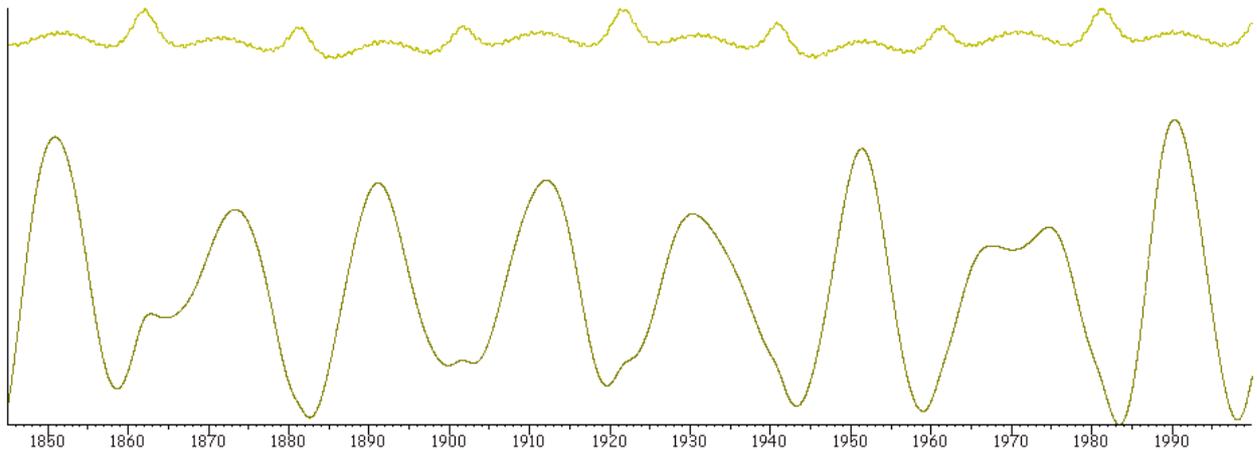

Figure 45 - Compared angular momentum of Jupiter planet with center in Sun (yellow) and SSB (olive), in detail.

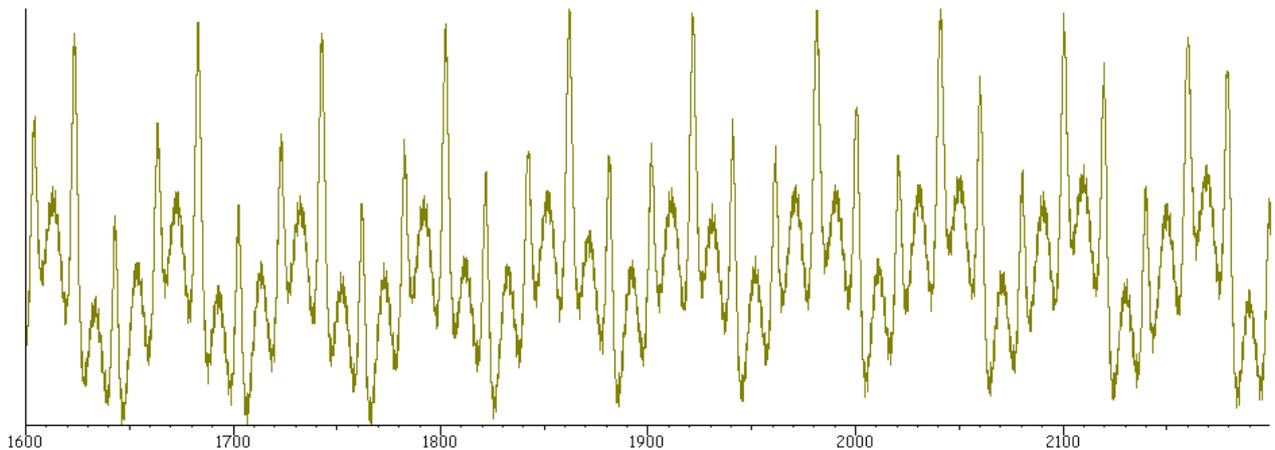

Figure 46 - Angular momentum of Jupiter relative to Sun (colors changed, now it is olive which was yellow above)

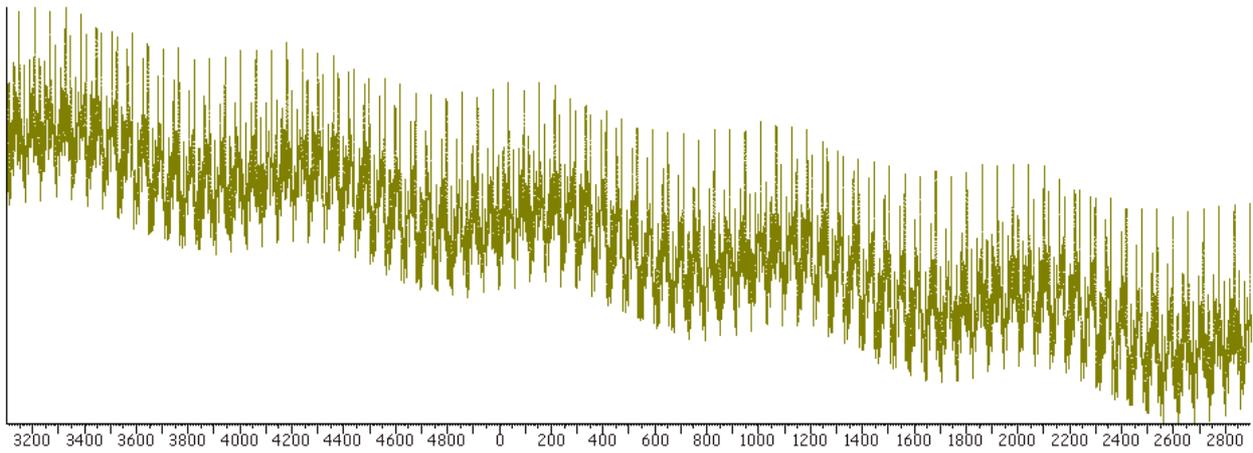
Figure 47 - Longer trend in angular momentum of Jupiter relative to Sun.

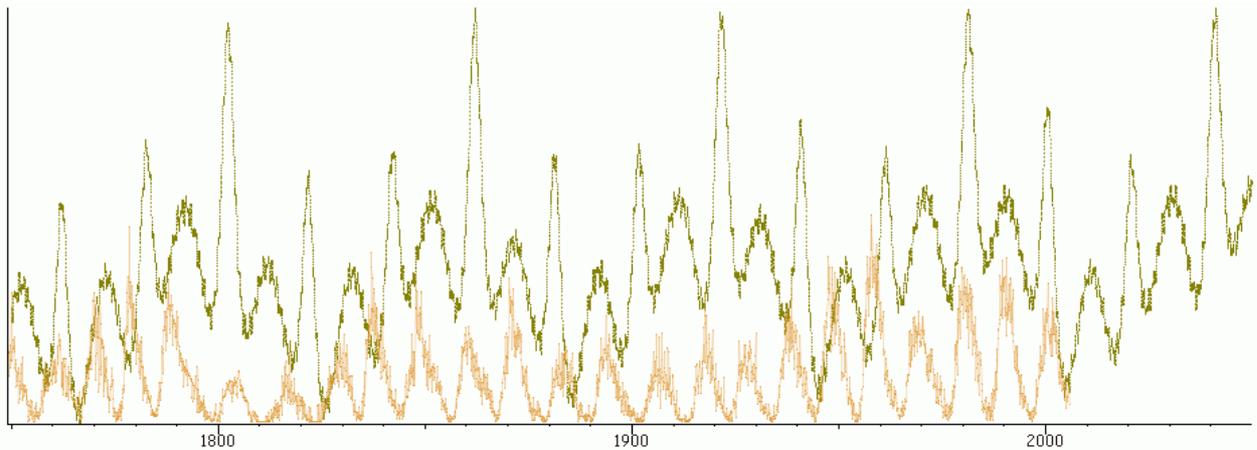
Figure 48 - Angular momentum of Jupiter relative to Sun, compared with Sunspot cycle.

It is evident, that Jupiter's angular momentum relative to Sun shows a different frequency than the Sunspot cycle. (fig. 48)

Angular momentum of Jupiter relative to Sun is roughly opposite of angular momentum of Saturn relative to SSB. (fig. 48b)
Angular momentum of Jupiter relative to SSB has 20-year frequency of Jupiter/Saturn cycle, similar (oposite) to angular momentum of Sun... (fig. 48c,70,71)

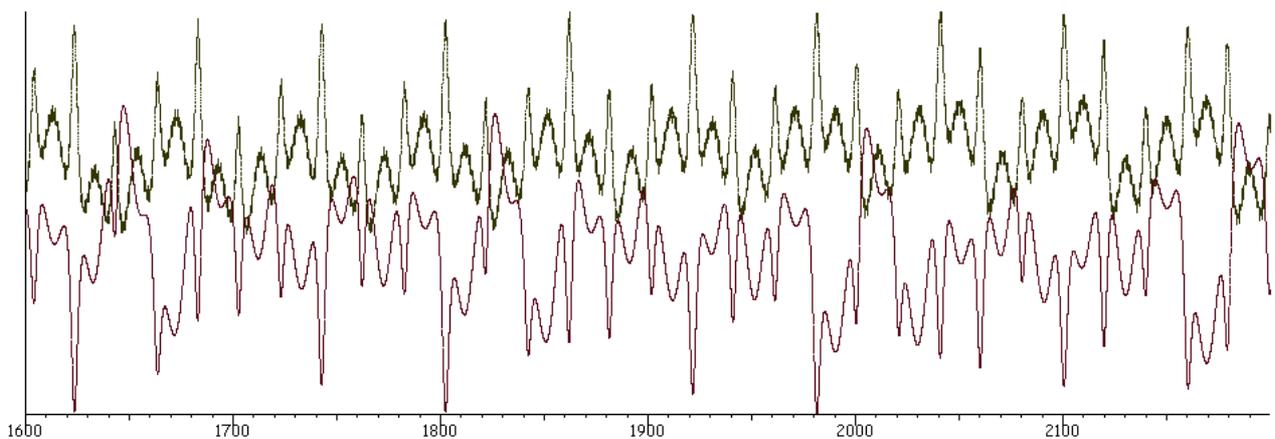
Figure 48b - Angular momentum of Jupiter relative to Sun (olive) compared with angular momentum of Saturn relative to SSB (maroon), offseted vertically without scaling to fit in one chart

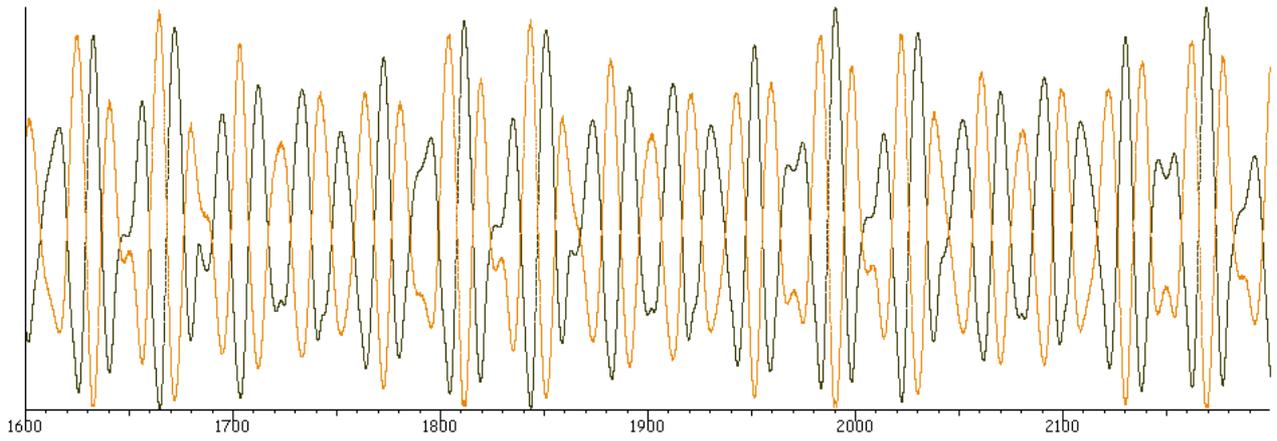

Figure 48c - Angular momentum of Jupiter relative to SSB (brown) compared with angular momentum of Sun relative to SSB (orange), offseted vertically without scaling to fit in one chart.

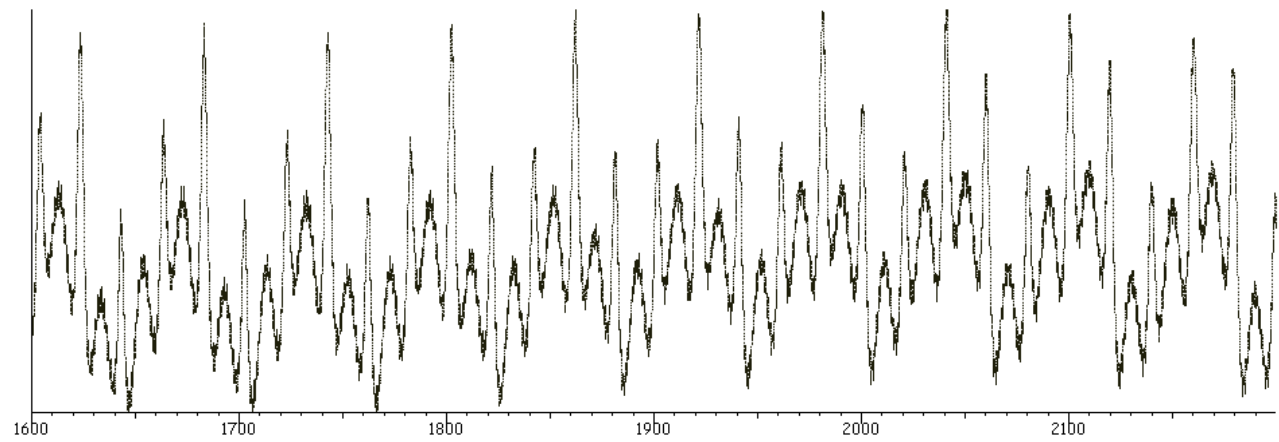

Figure 48d - Angular momentum of Jupiter relative to Sun.

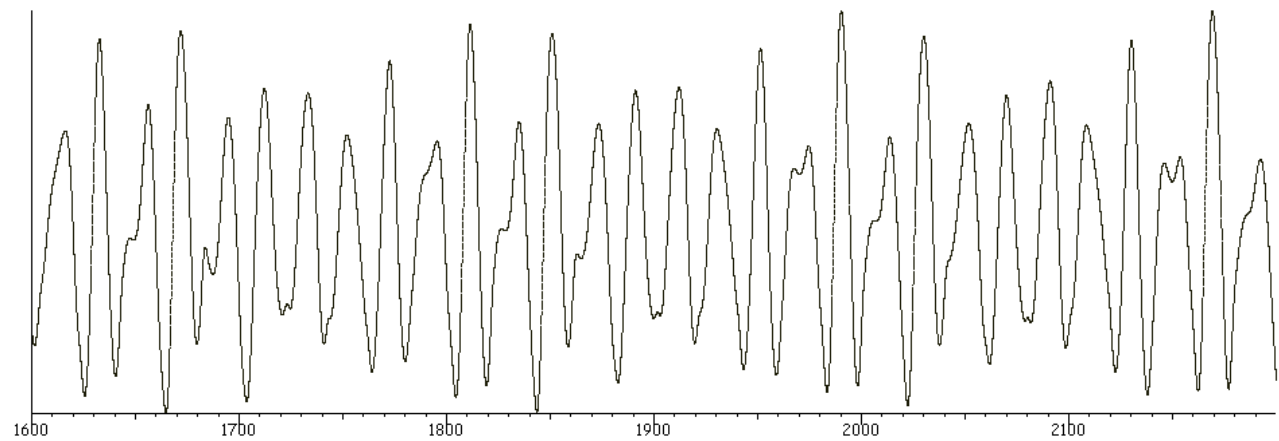

Figure 48e - Angular momentum of Jupiter relative to SSB.

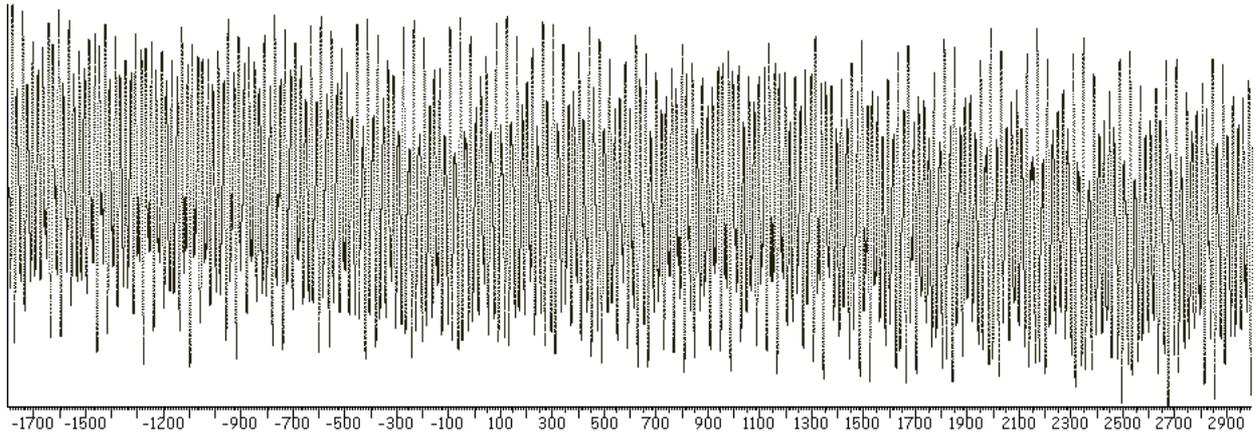

Figure 48f - Angular momentum of Jupiter relative to SSB, longer trend.

The oscilation of Jupiter's angular momentum at present era (during 5 millennia) is 355.593, whereas absolute values range from 128481.702 to 128837.295 (in arbitrary units, see Appendix 1), so the oscilation is only 0.276% of the absolute value. The absolute value is slowly shrinking.

Main frequency of Angular momentum of Jupiter relative to SSB is **19.68** years, then 12.7 years, 13.7 years, 60 years, 9.9 years, 11.8 years, 29.3 years, 7.4 years, 6.6, 4.97 years, 3.97 years, 3.3 years and some of their harmonics, sorted by importance, as determined by FFT analysis.

We have got no explanation for the longer wave of approximatelly 2600 years (fig. 48f, but not on fig. 48g), but its length is little similar to long PTC cycle of the Sun.

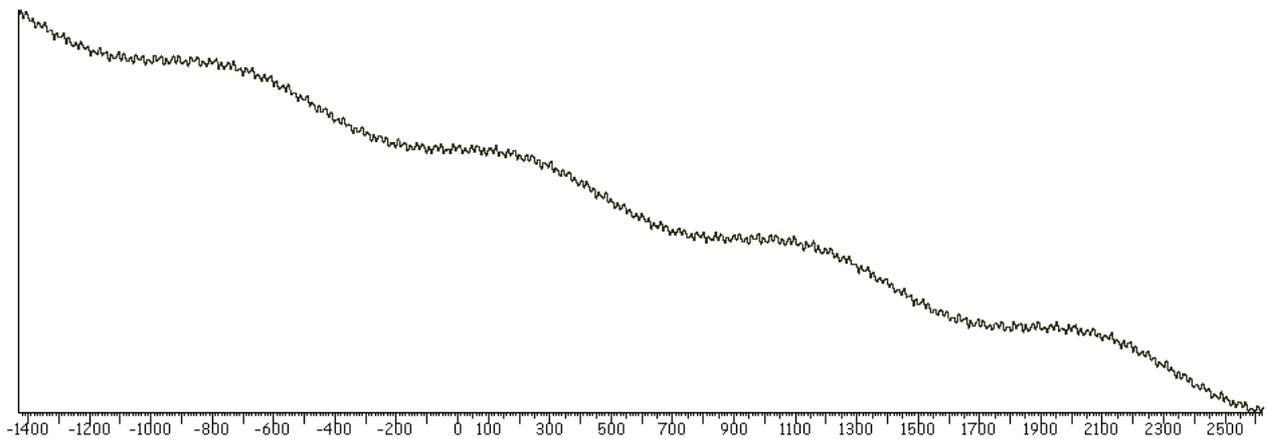

Figure 48g - Angular momentum of Jupiter relative to SSB, longer trend with Gaussian filtering with a window of 690 years.

Note the 854 year cycle on filtered (Gaussian filtering with a long window of 690 years) Angular momentum of Jupiter relative to SSB (fig. 48g).

## Saturn

Saturn is the first of the outer planets, that shows clearly higher swing relative to Sun than to SSB. It orbits on a barycentric trajectory and not a heliocentric one. (see fig. 49-51)

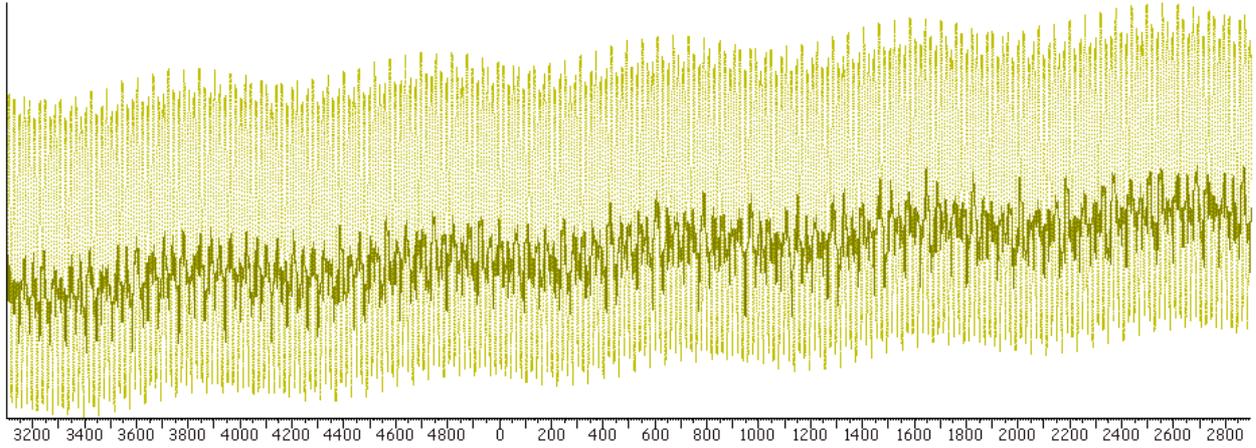

Figure 49 - Compared angluar momentum of Saturn planet relative to Sun (yellow) and SSB (olive, in middle)

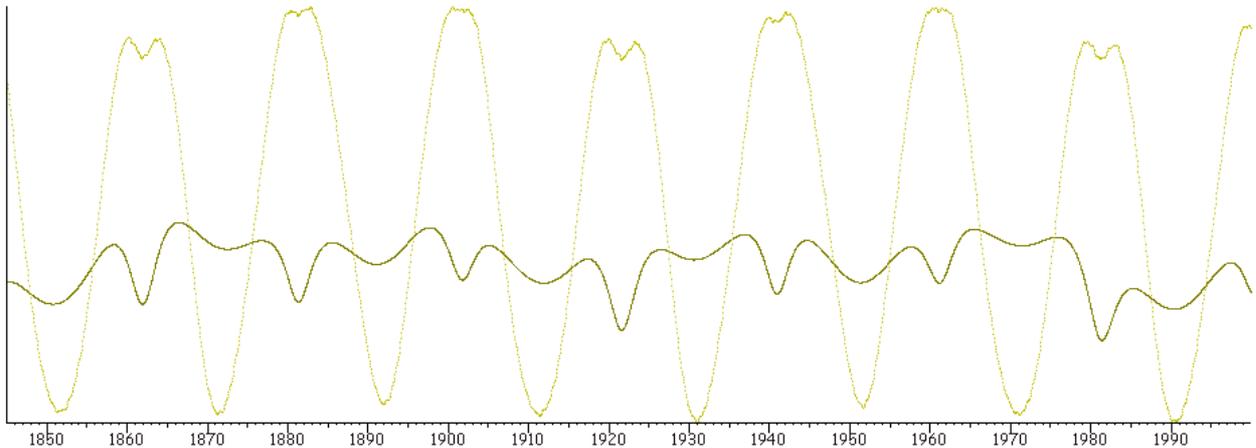

Figure 50 - Compared angular momentum of Saturn planet relative to Sun (yellow) and SSB (olive, in middle) in detail

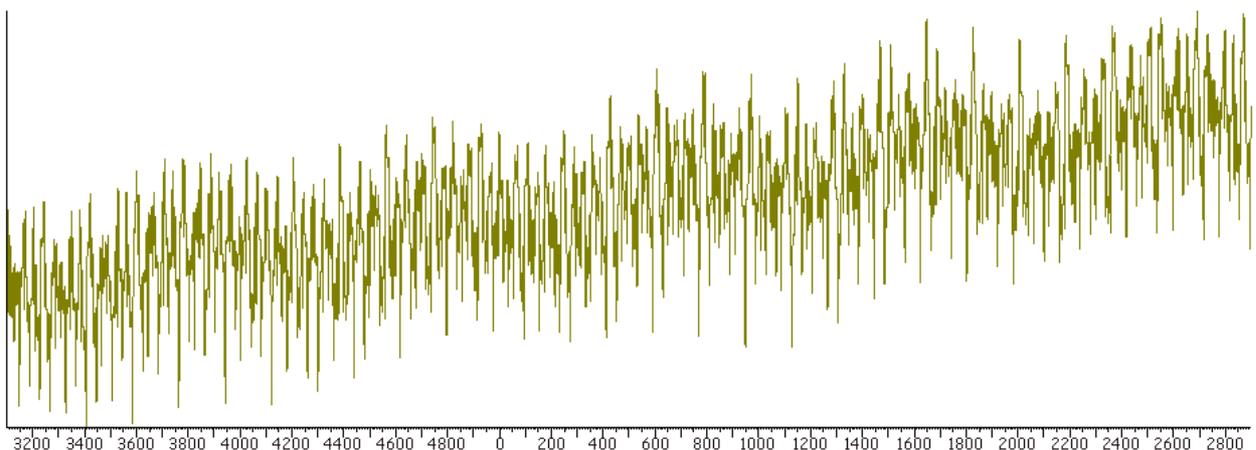

Figure 51 - Longer tendency in angular momentum of Saturn relative to SSB

Main frequency of perturbations in Angular momentum of Saturn relative to SSB is 35.4 years, then 9.9 years, 45.2 years, 60.4 years, 19.6 years, 14.9 years and many harmonics, sorted by significance, as determined by FFT analysis.

## Uranus

Uranus planet also orbits on a barycentric trajectory.

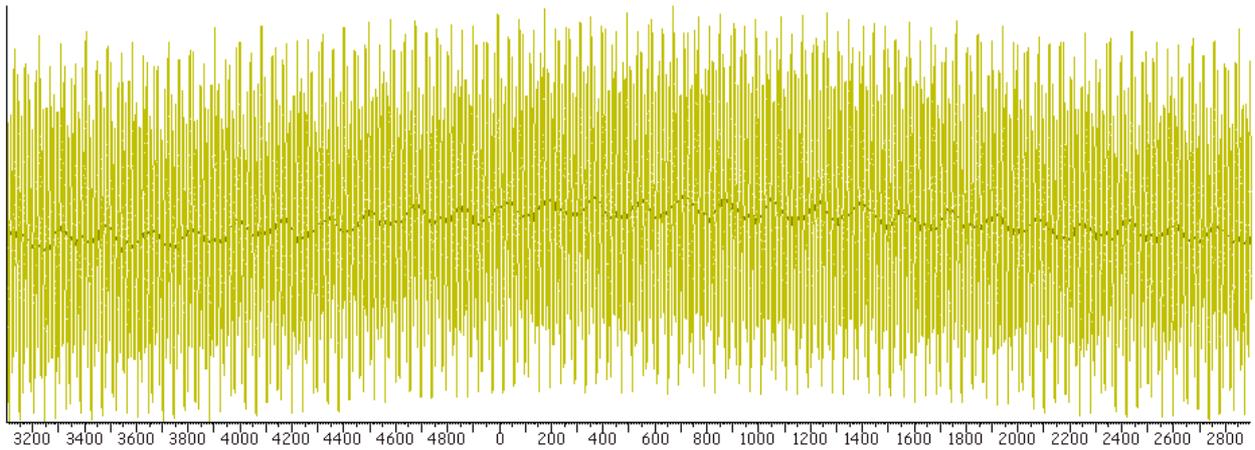
Figure 52 - Compared angular momentum of Uranus planet relative to Sun (yellow) and SSB (olive in middle)

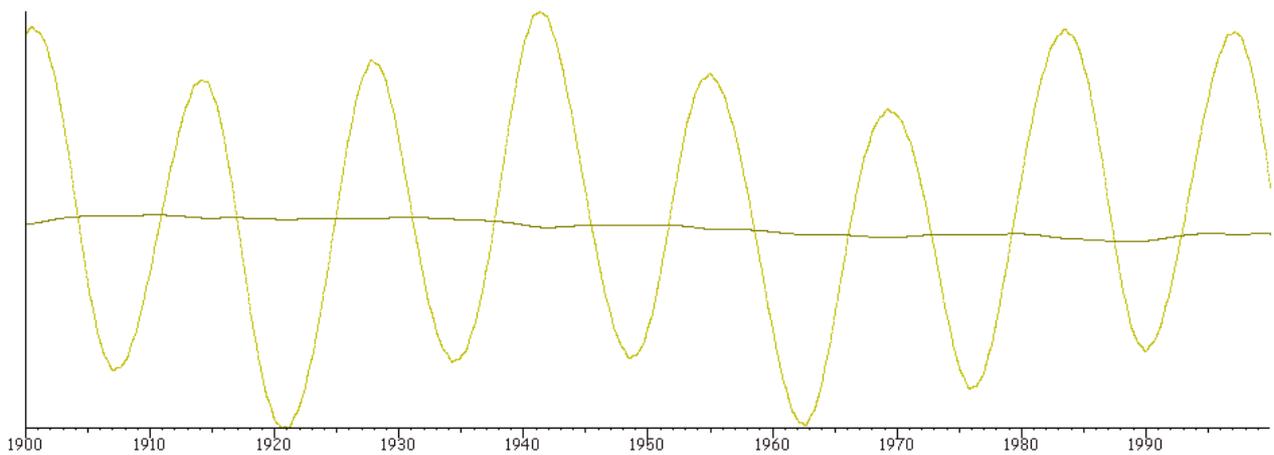
Figure 52 - Compared angular momentum of Uranus planet relative to Sun (yellow) and SSB (olive), in detail.

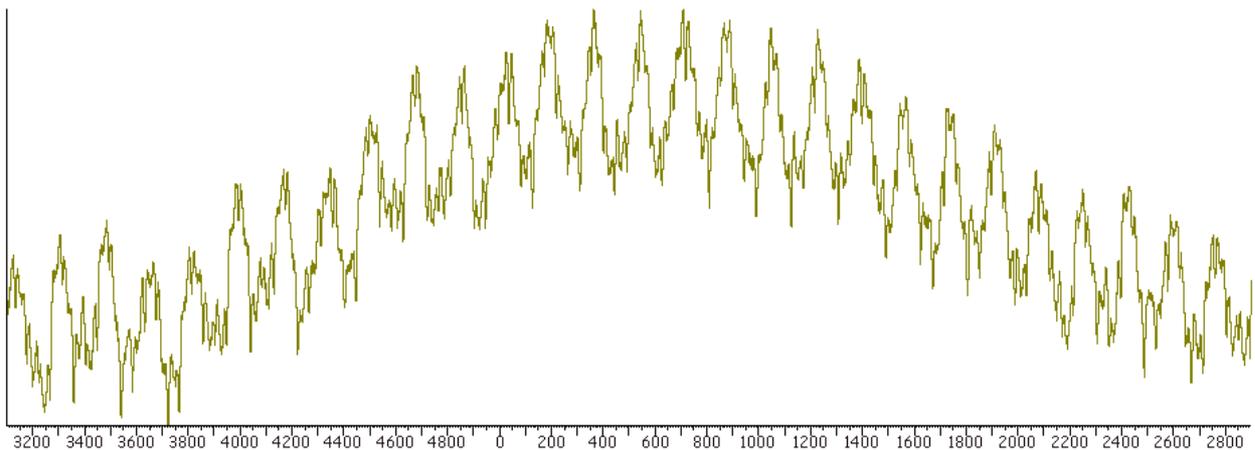
Figure 53 - Longer cycle of angular momentum of Uranus planet relative to SSB.

Main frequency of angular momentum of Uranus relative to the Sun is **13.74** years.

Main frequency of angular momentum of Uranus is 179-181 years (guessed value?), the resonance of Uranus/Neptune, with other frequencies at 164y, 22.4y, 15.03y, 6.9y and other...

## Neptune

Neptune planet also orbits on a barycentric trajectory, not on heliocentric one.

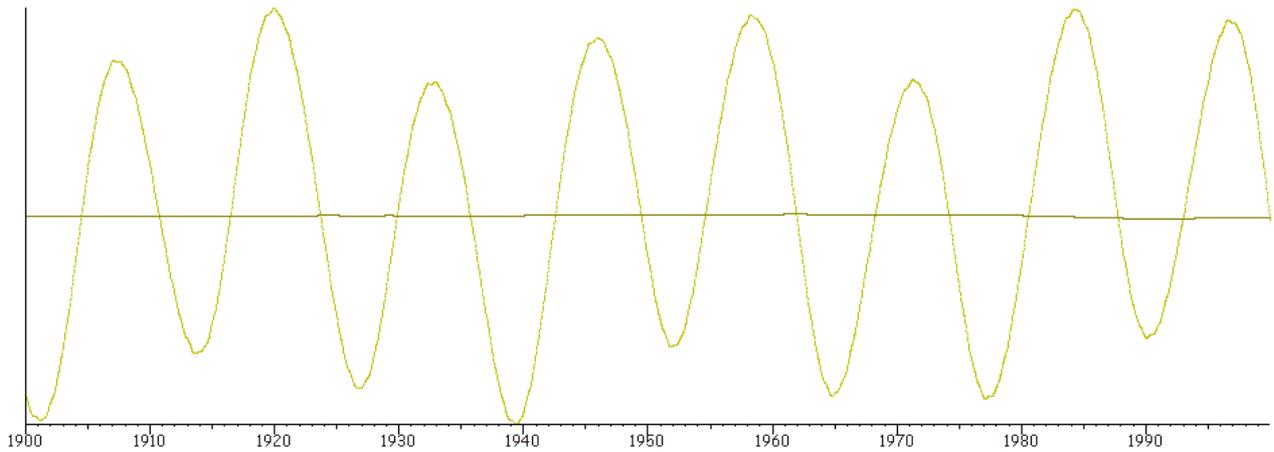

Figure 54 - Compared angular momentum of Neptune planet relative to Sun (yellow) and relative to SSB (olive, in middle)

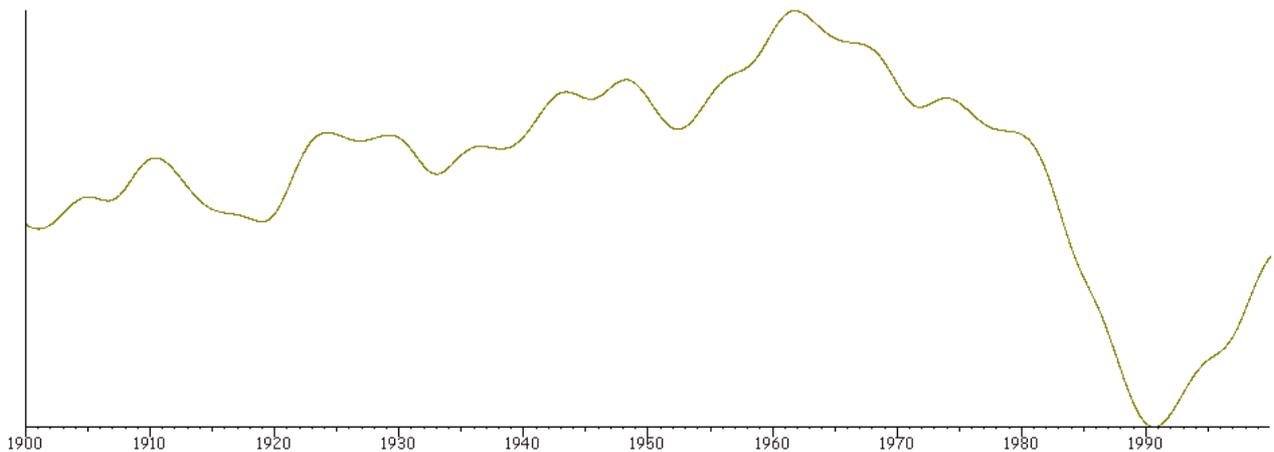

Figure 55 - Detail of angular momentum of Neptune relative to SSB, during 20th century.

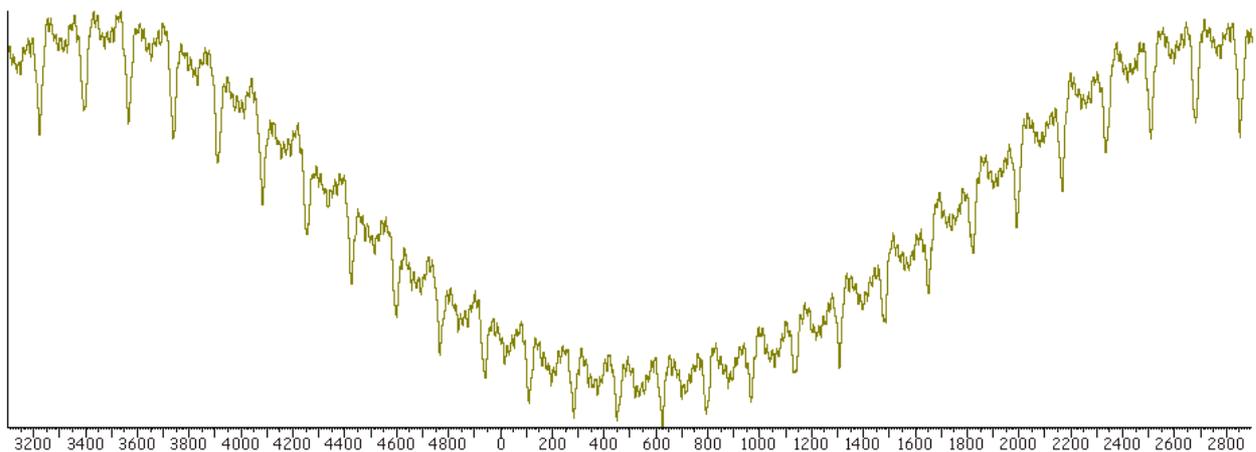

Figure 56 - Longer cycle in angular momentum of Neptune relative to SSB.

Main frequency of angular momentum of Neptune relative to Sun is **12.70** years.
Main frequency of angular momentum of Neptune is 168-171 years (guessed value?!), with other at 82.0y, 56.3y, 42.5y, 17.9y, 6.4y and other...

## Pluto

Pluto (dwarf) planet is rather chaotic (see fig. 58) and its resonance with Neptune is of a little quality (seems instable, see fig. 18).

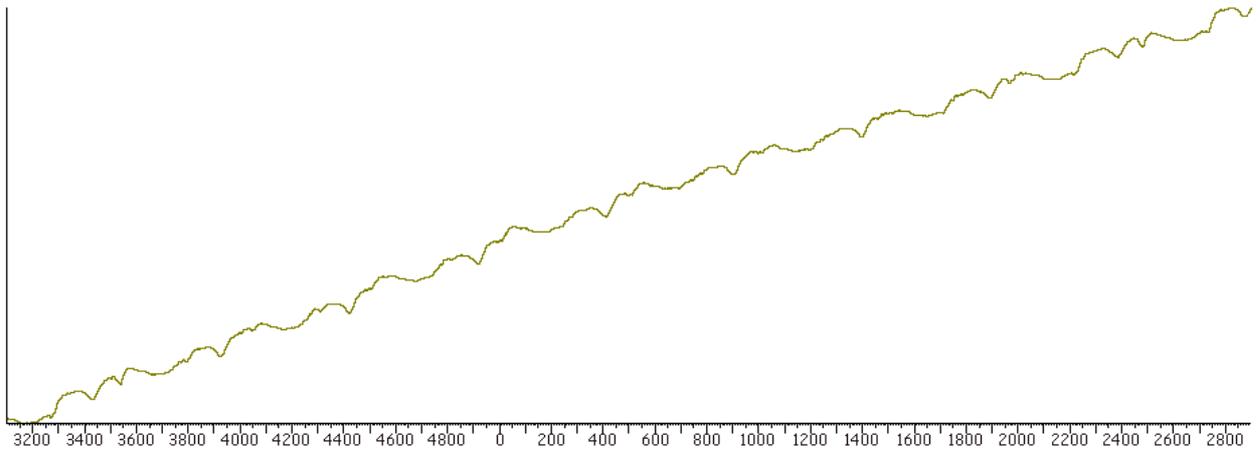

Figure 57 - Longer trend of angular momentum of Pluto relative to SSB.

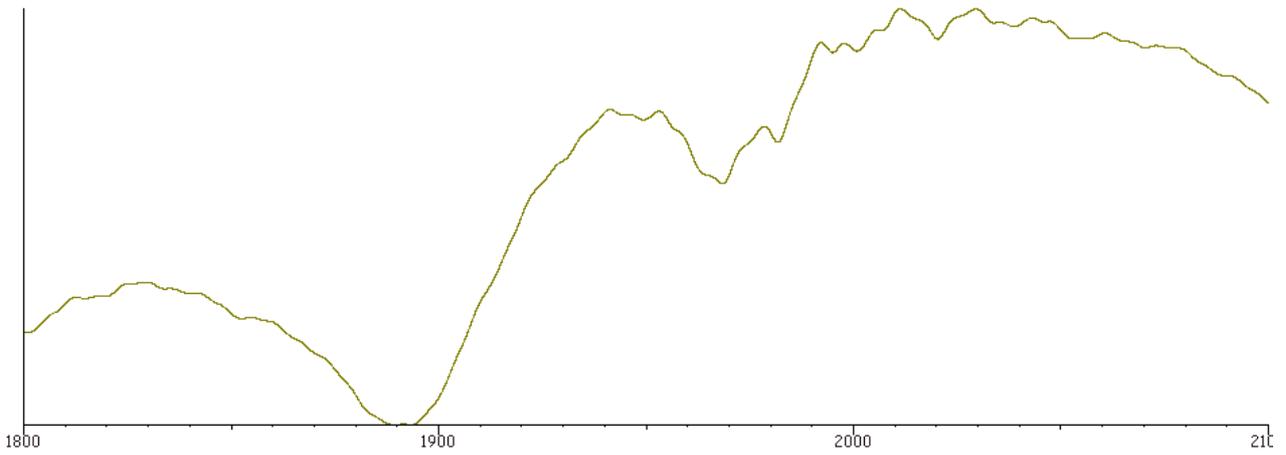

Figure 58 - detail of angular momentum of Pluto relative to SSB.

## Mercury

Mercury orbits on a heliocentric trajectory, it's angular momentum relative to Sun (fig. 59-62) is much more constant than relative to SSB (fig. 63-65).

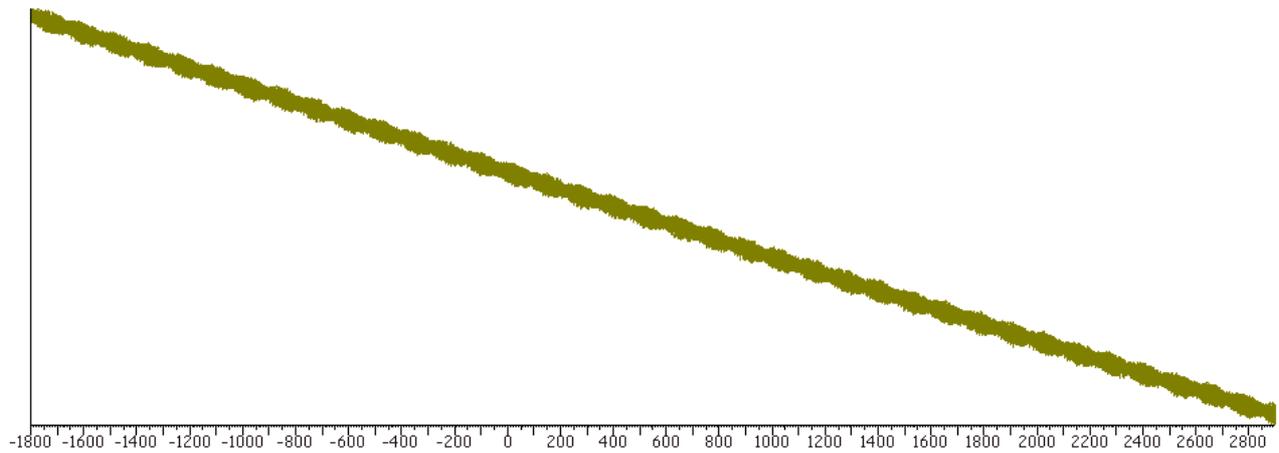
Figure 59 - Longer trend of angular momentum of Mercury relative to Sun

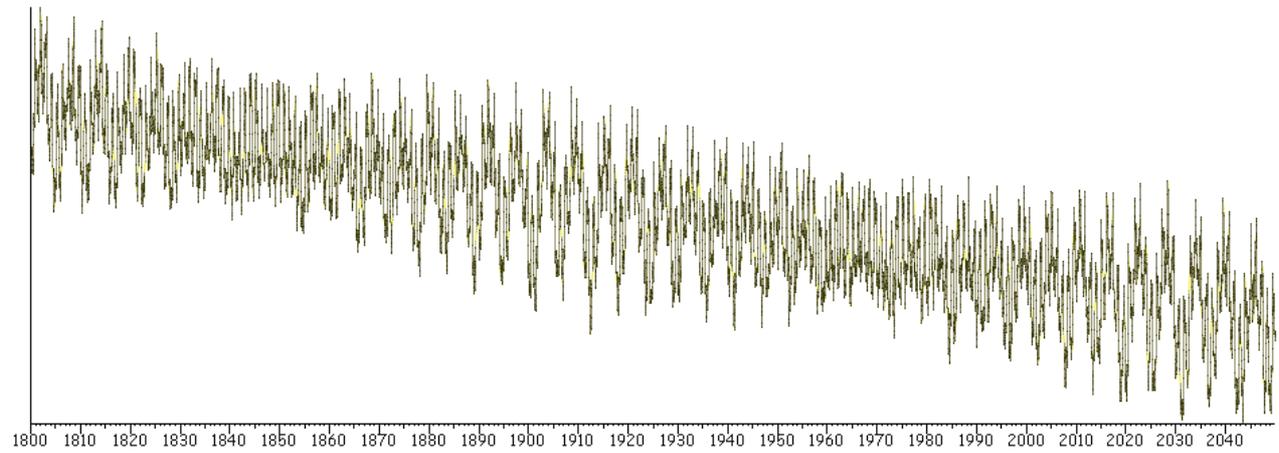
Figure 60 - Angular momentum of Mercury relative to Sun, in detail.

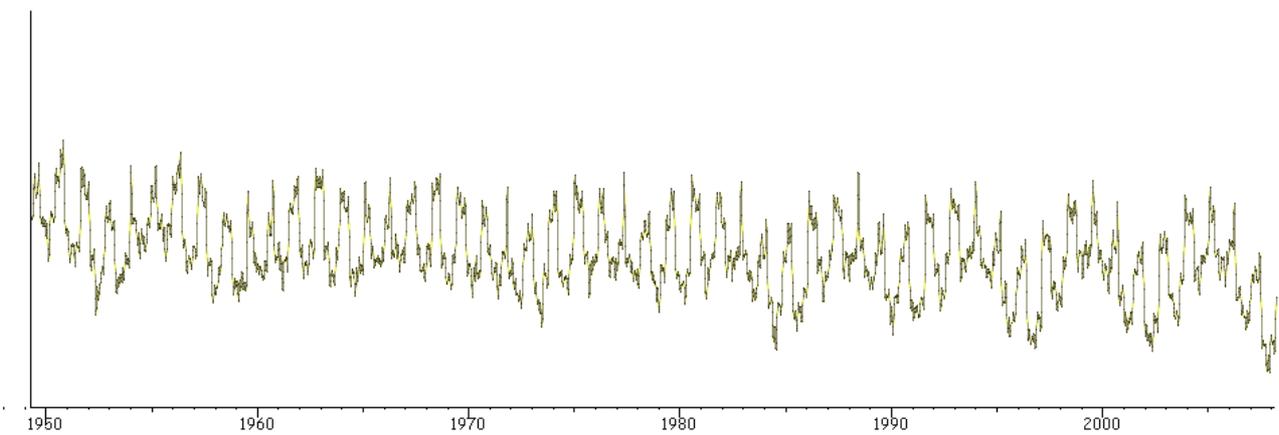
Figure 61 - Angular momentum of Mercury relative to Sun, in more detail.

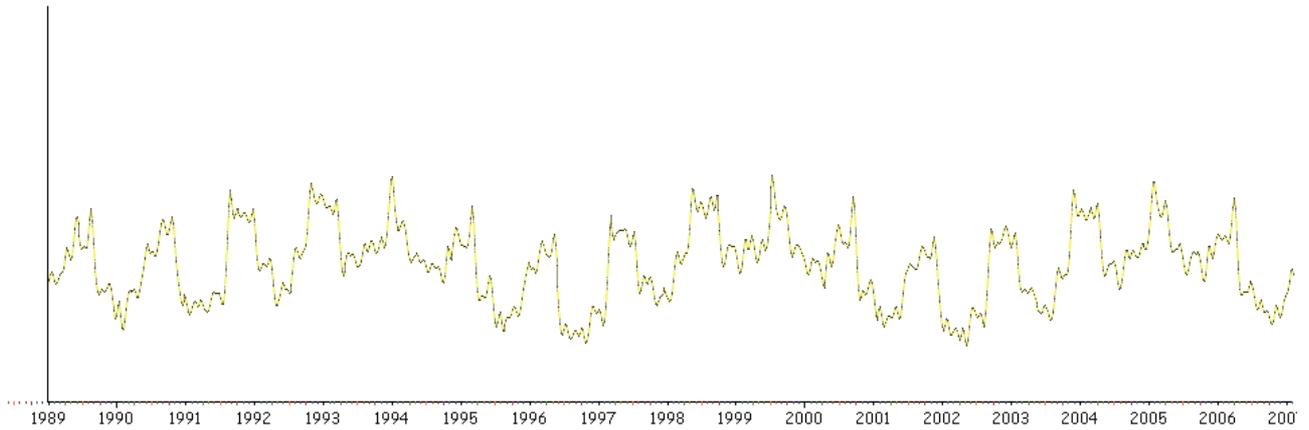
Figure 62 - Angular momentum of Mercury relative to Sun, in more detail.

Main frequency of angular momentum of Mercury relative to Sun is 1.1 years (403 days), other are 5.89 year, 1.37 year, 106.5 days, 72.2 days, 91.5 days, 169.6 days, 202.6 days, 11.23 year, 13.8 year and other, as determined by FFT analysis.

Angular momentum of Mercury relative to SSB shows the same cycle as that of the Sun relative to SSB (see fig. 64 and a chapter on the Sun below):

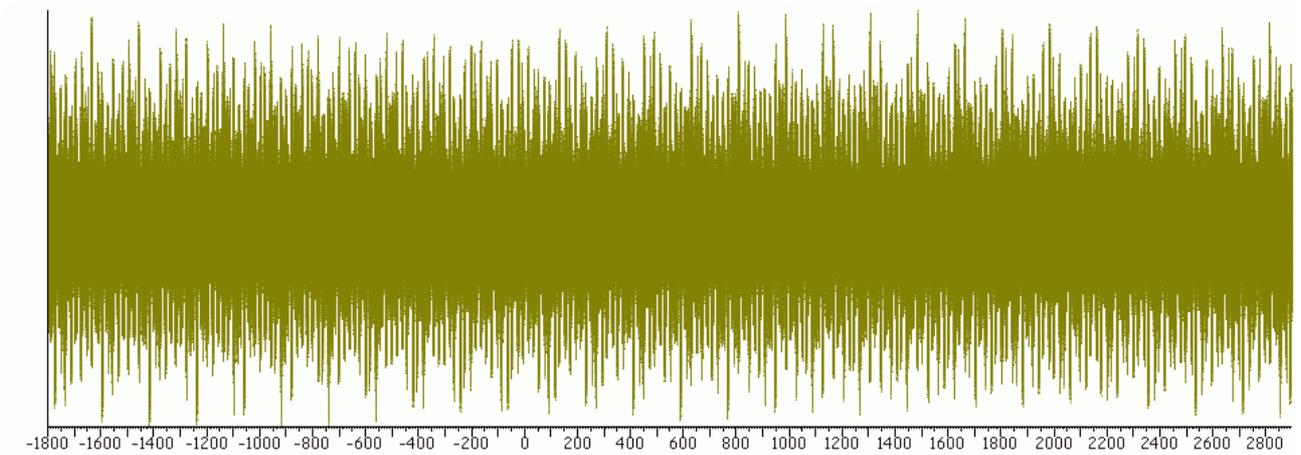
Figure 63 - Angular momentum of Mercury relative to SSB.

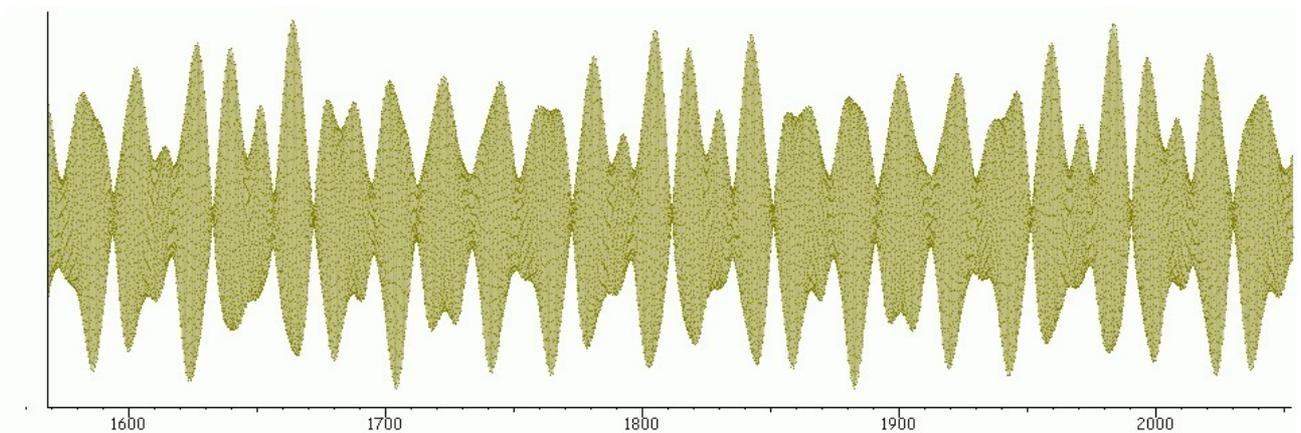
Figure 64 - Angular momentum of Mercury relative to SSB, in detail.

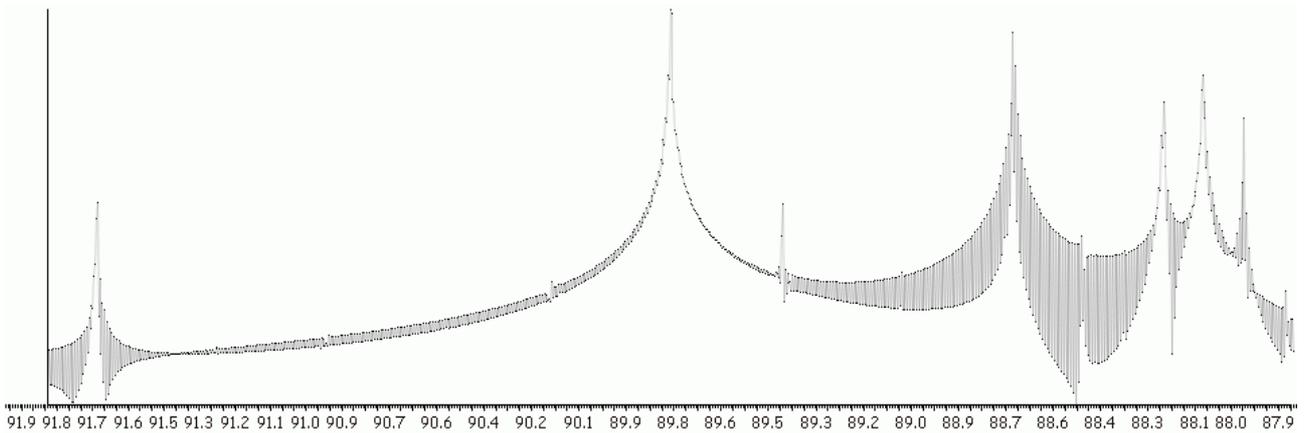

Figure 65 - part of FFT analysis of angular momentum of Mercury relative to SSB.

Main frequency is here 89.785 days, other are 88.691 days, 88.092 days, 88.214 days, 87.965 days, 91.687 days, 11.83 years, 115.86 days, 144.55 days, 1.11 years, 29.5 years and other harmonics, sorted by importance, as determined by FFT analysis.
The envelope follows the Sun move arround the solar-system barycenter, since the trajectory is heliocentric...

## Mars

Mars planet also orbits on a heliocentric trajectory and the envelope of its angular momentum relative to SSB shows the Sun's orbit cycle (fig. 67).

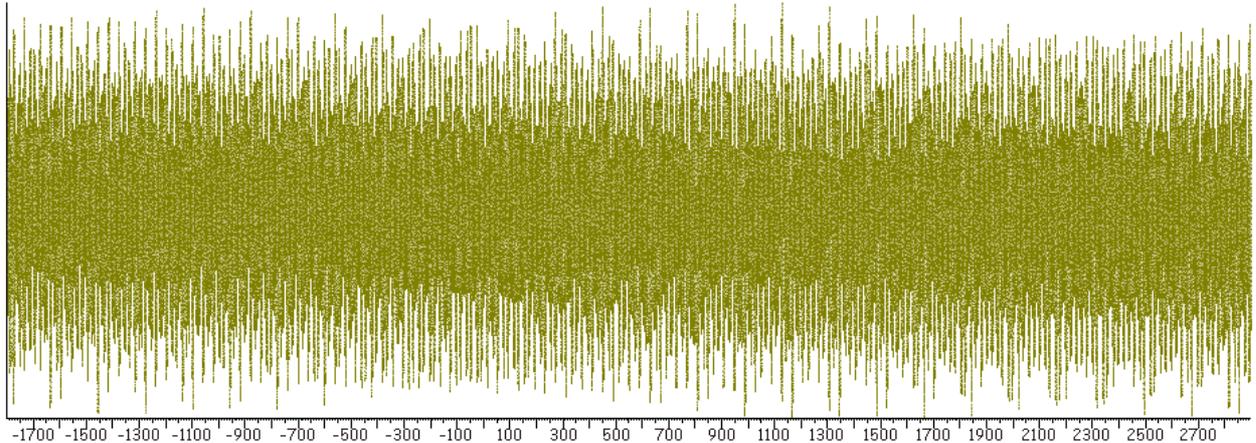
Figure 66 - Angular momentum of Mars planet relative to SSB

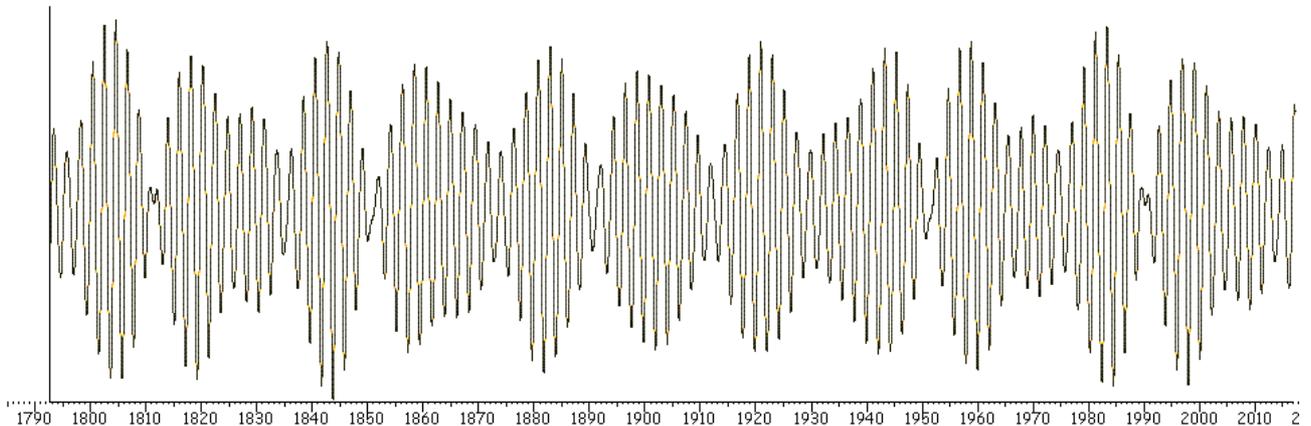
Figure 67 - Angular momentum of Mars planet relative to SSB, in detail.

Main frequency of angular momentum of Mars relative to SSB is 2.24 years, 2.01 years, 1.90 years, 1.92 years, 1.88 years, 2.75 years, all previous repeated as first harmonic, 11.82 years, 29.37 years, 18.96 years and few other, sorted by importance.
Again, the envelope follows the Sun move arround the solar system barycenter and shows a much higher swing than the heliocentric trajectory.

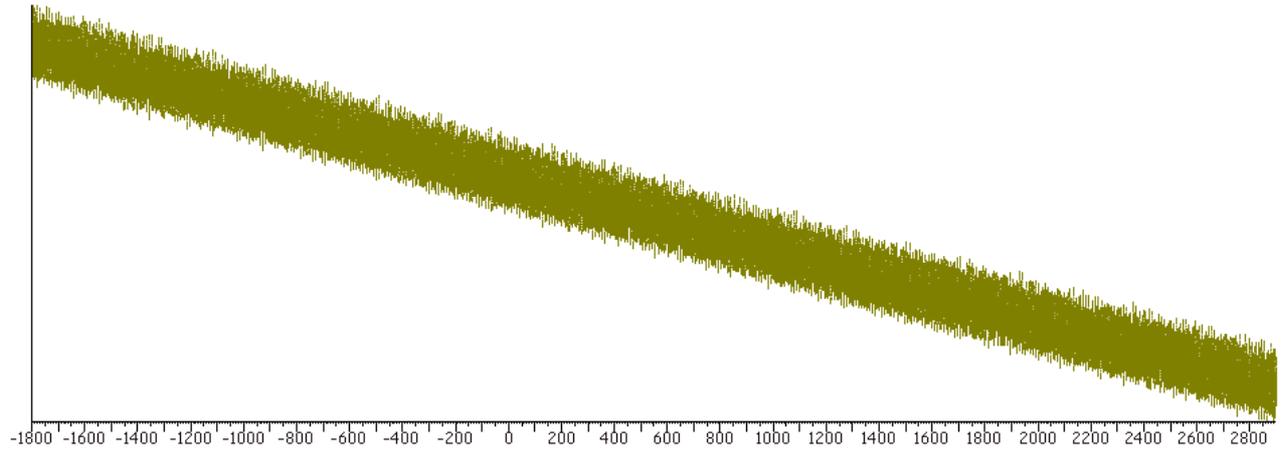
Figure 68 - Angular momentum of Mars planet relative to Sun, longer trend.

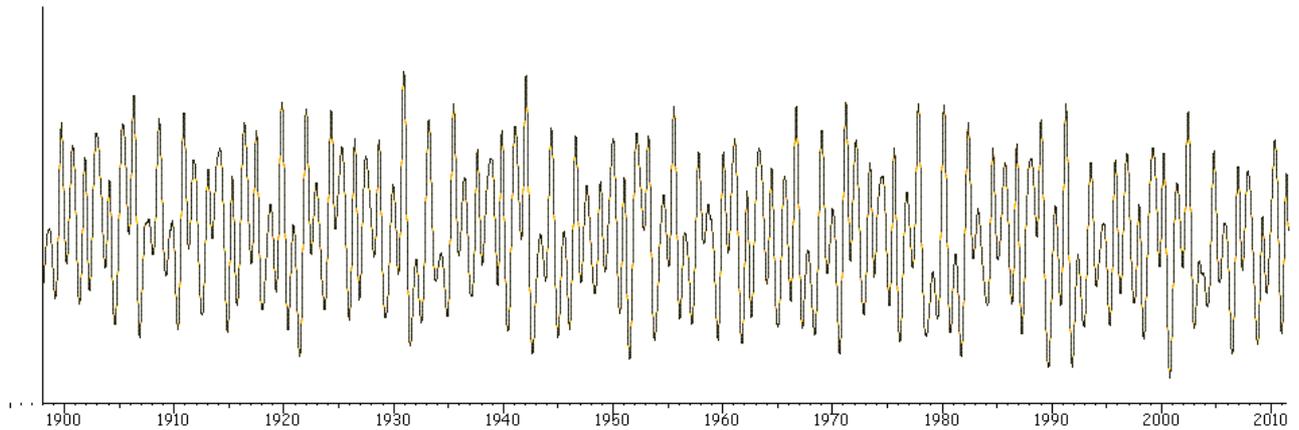
Figure 69 - Angular momentum of Mars planet relative to Sun, in detail.

Main frequency of angular momentum of Mars relative to Sun is 1.11 year, 2.75 year, 271.8 days, 11.7 years, 2.12 years, 333 days, 1.23 years, 203.8 days and other.

## Sun

Angular momentum of the Sun is mostly contrary to an angular momentum of Jupiter, which is its main counter-weight (figures 70,71,48c).

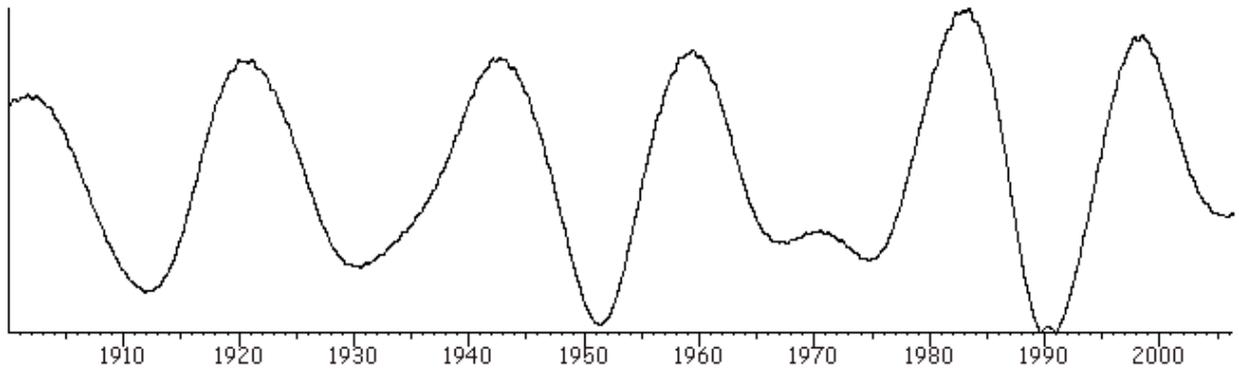
Figure 70 - Angular momentum of Sun relative to SSB.

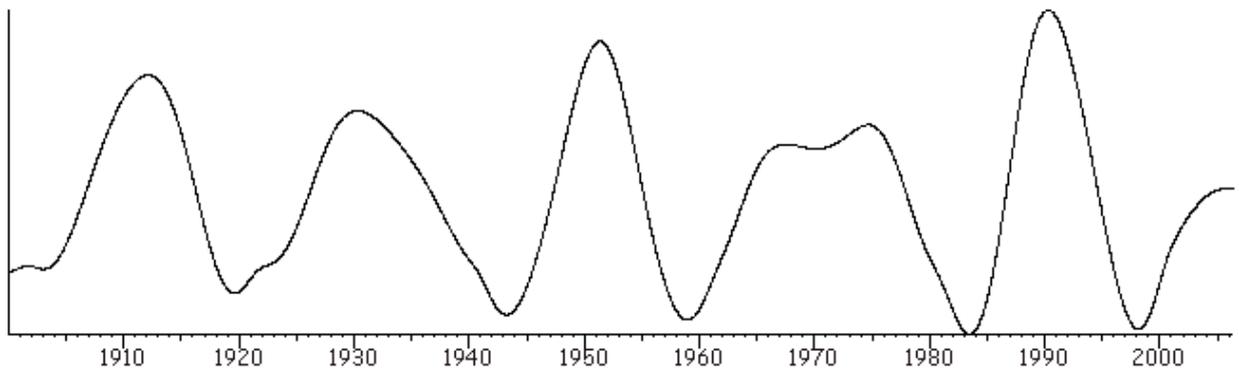
Figure 71 - Angular momentum of Jupiter relative to SSB.

It the Sun and Jupiter angular momentums are added, they could mostly cancel out, so that the rest, that is a "ripple" on the Sun's angular momentum chart, can reveal this shape (fig. 72):

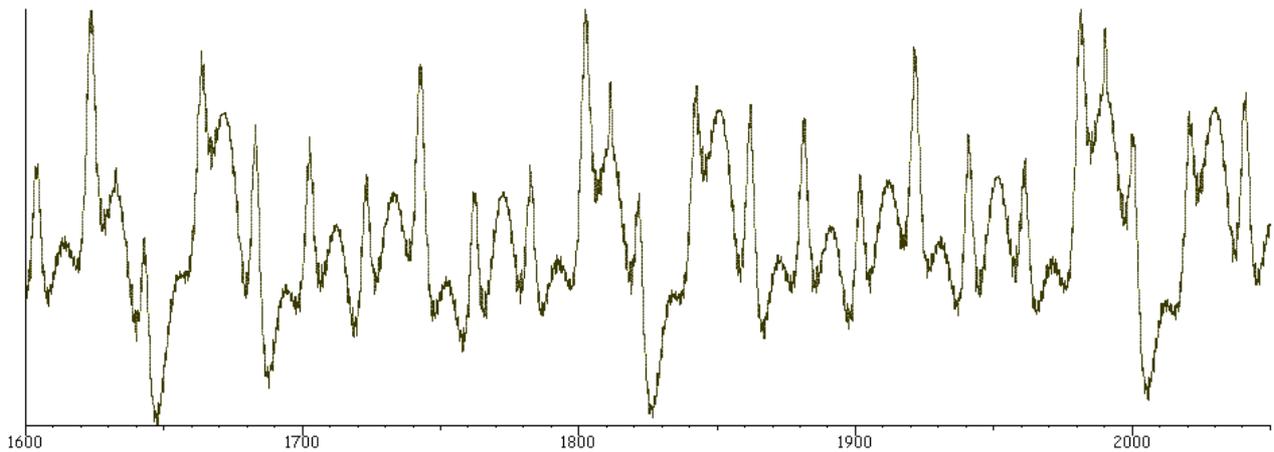
Figure 72 - Scalar sum of angular momentum of Sun and Jupiter.

Note the 178.8 year cycle of Uranus/Neptune, little similar to Gleissberg cycle (twice the Gleissberg cycle, which is actually an absolute value of otherwise sinusoidal function)...

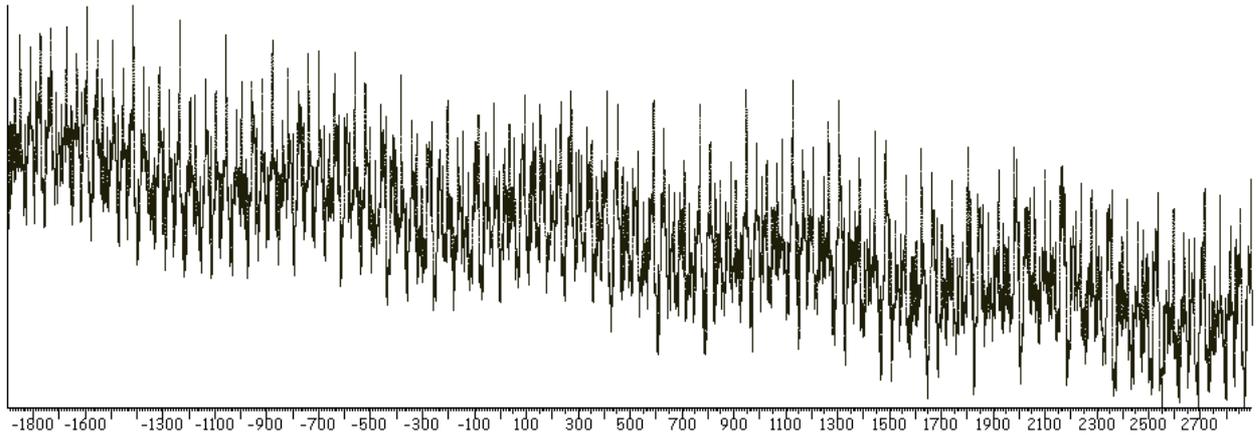
Figure 72b - Sum of Angular momentum of Sun and Jupiter relative to SSB, longer trend.

Note the 854-year cycle in sum of angular momentum of Sun and Jupiter relative to SSB (fig. 72b). Maxima of this cycle around years 1200 and 2050 correspond to Medieval optimum and Global warming periods, minima arround 700, 1650 correspond to Little ice age events (Maunder minimum and Wolf (???) minimum). The overall tendency (shrinking, corresponding to shrinking of angular momentum of Jupiter, see fig. 44, 48e) is balanced by the change of angular momentum of other planets (namely Saturn, see fig. 51)

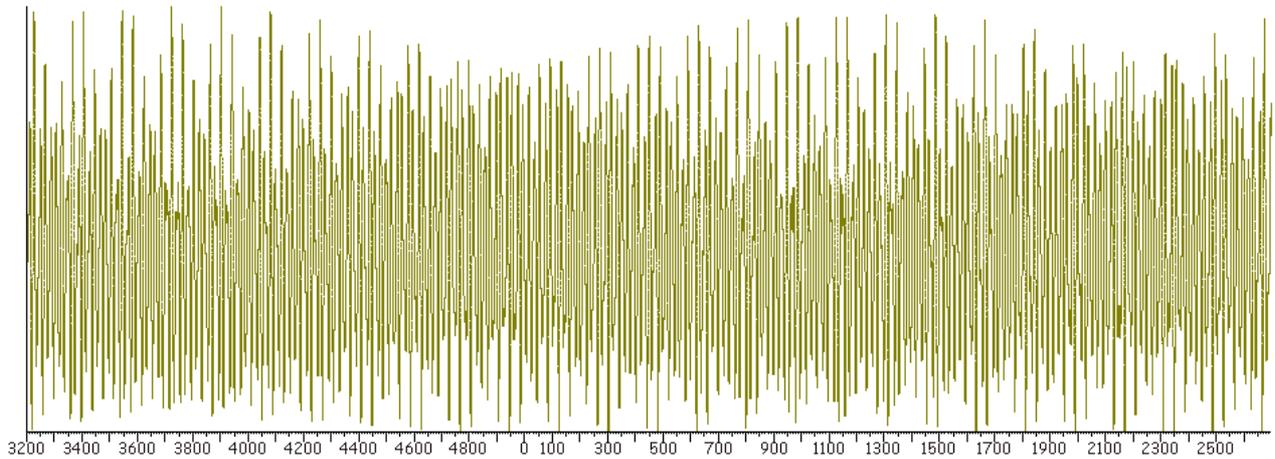
Figure 73 - Angular momentum of Sun relative to SSB.

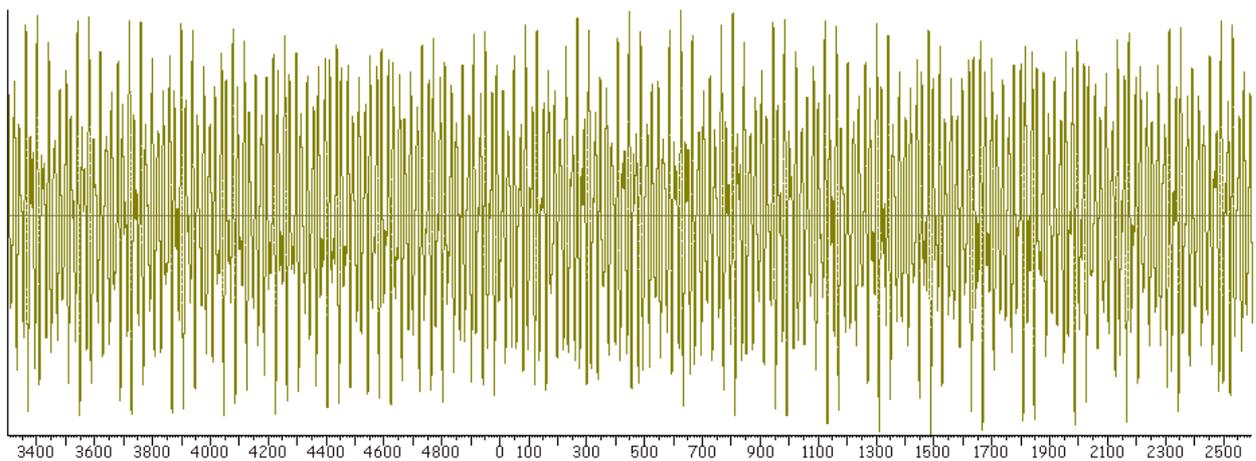
Figure 74 - First derivation of Angular momentum of Sun relative to SSB, showing the PTC cycle (note: give a reference to PTC) of roughly 2500 years.

For Perturbation of Torque Cycle (PTC), see ref. in Landscheidt et al., and fig. 75.

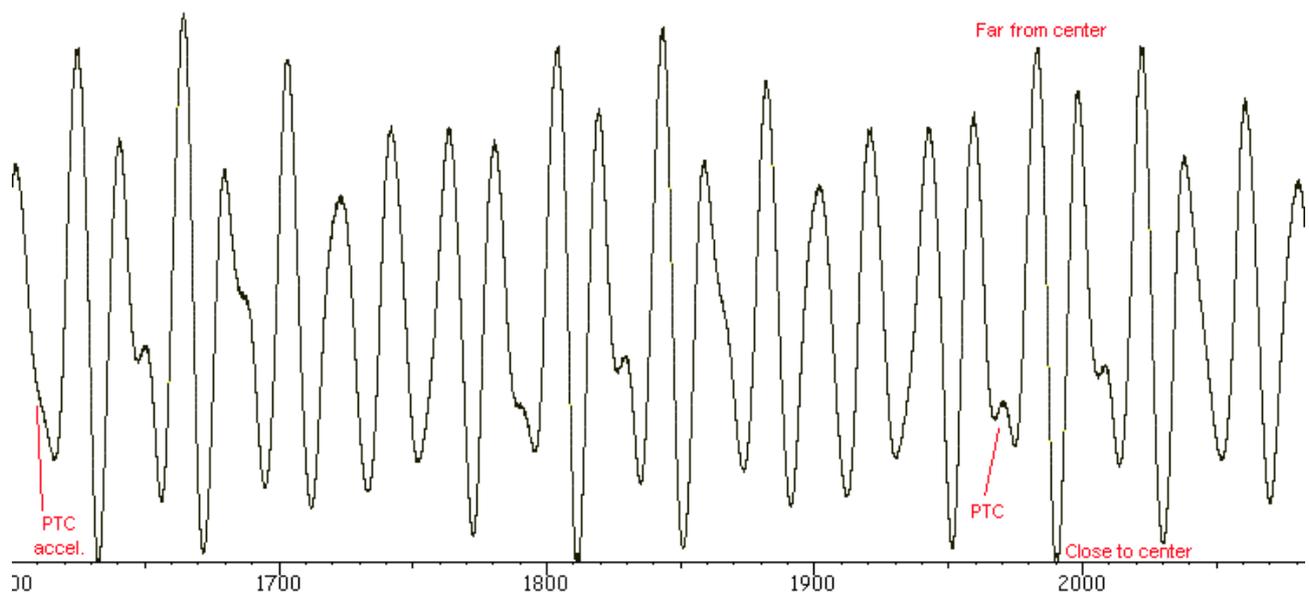
Figure 75 - Detail of angular momentum of Sun relative to SSB, annotated.

## Angular momentum sum of 9 planets and the Sun

### Scalar Sum

The "absolute value" chart (fig. 80) includes 0. It shows, that the sum seems very constant, and the "wave", seen on other charts, is only a "wave at the top of an ocean", on the order of $8.11*10^{-7}$, close to 1 in million...

The high peaks are during times, when the Sun approaches the solar-system barycenter. At these times, the Sun moves in the contrary (retrograde or highly inclined) direction for a short while, and has got a negative angular momentum (with respect to the invariant plane), so it should actually subtract in a vector sum of the system, but here it is added in the scalar sum (there is no negative scalar (absolute value or **vector length**) angular momentum)...

The first derivation of angular momentum sum (fig. 78) only little matches the sun-spot cycle, but the high-peak around 1990 could be correlated with a drop of solar-flare activity at the middle of preceeding Sunspot cycle 22, both peaks arround 1800 and 1990 having a damping effect on the Solar activity, possibly due to effects of exchange between Sun orbital angular momentum and spin angular momentum.

The "wave" of approximate period of 854 years, which could also probably be anti-correlated with Sun spin rate, seems to match the climatologic events of Medieval optimum and Global warming, and also the Little Ice age of Maunder minimum, and similar periods in earlier ages (fig. 81)...

If this is right, now the Solar activity could drop a little, but will approach a larger maximum arround year 2050, not disturbed by the peak anomaly, and then drop to a next little-ice-age arround 2400 AD.
The time-lag between the spin rate change and activity change is still uncertain...

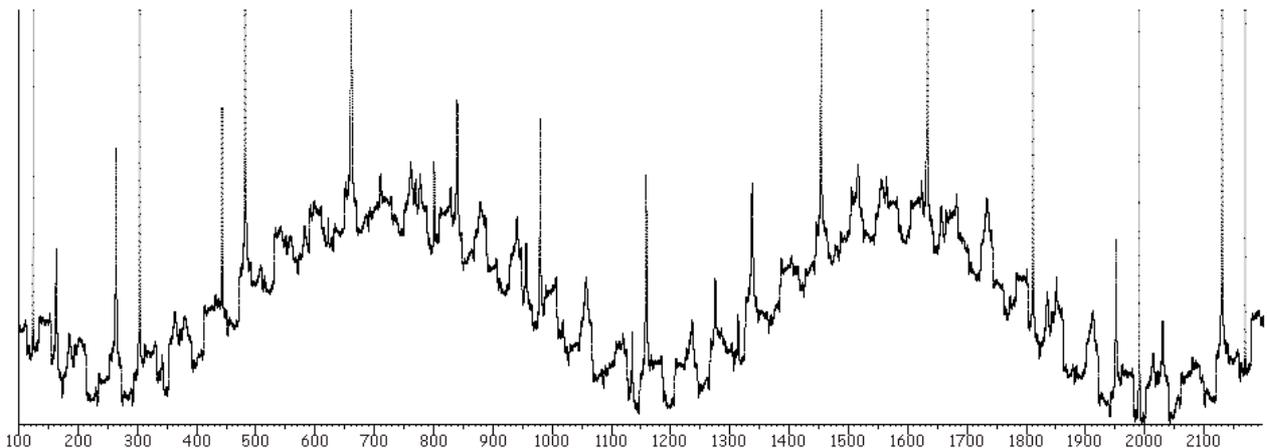
Figure 76 - Scalar sum of angular momentum of 9 planets and Sun, from 100 AD to 2200 AD, clipped at the top.

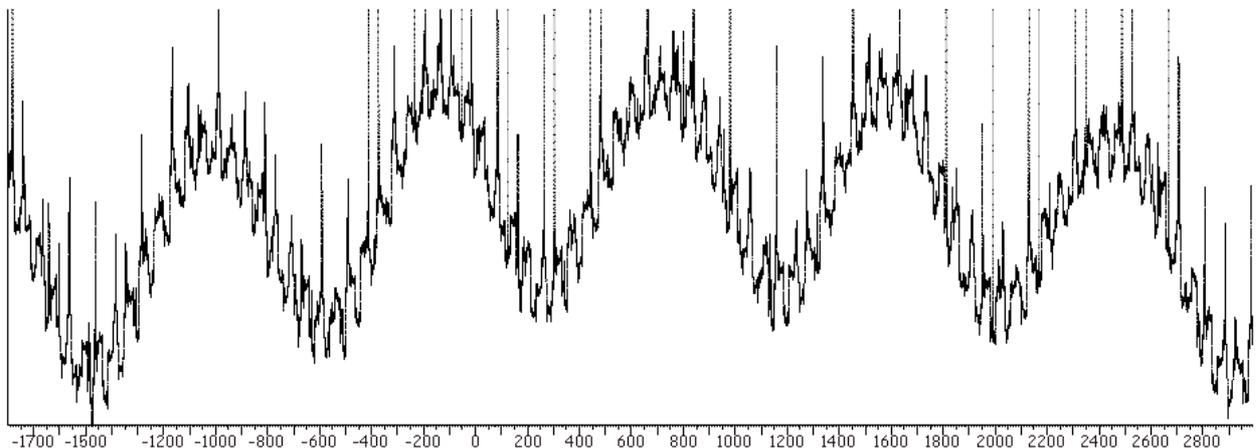
Figure 77 - Scalar sum of angular momentum of 9 planets and Sun, longer tendency (from 1800BC to 2700 AD), clipped at the top.

The periods of low scalar angular momentum (and higher Solar activity) roughly correspond to human civilization thriving: 1450BC Egypt, 600BC Greece, India and China, 200AD Rome and China, 1200 Medieval optimum (population growth in Europe), 2000AD (present "technical boom"). The periods of high scalar angular momentum (and lower Solar activity) correspond to crisis periods of human civilization.

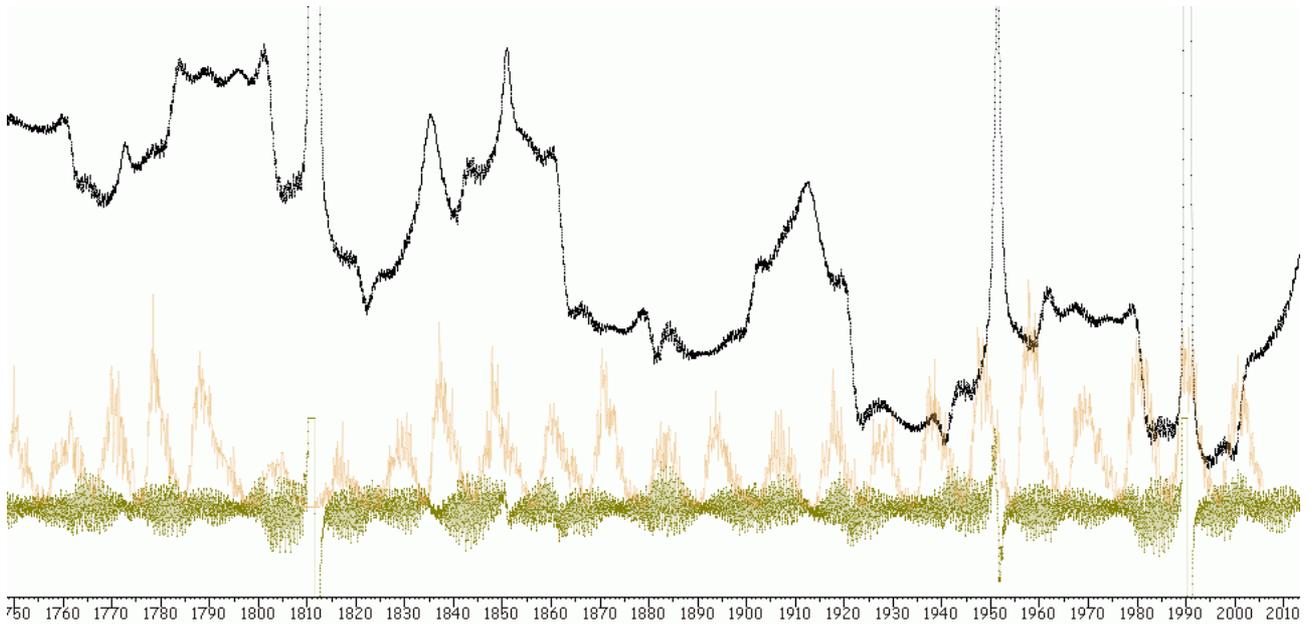

Figure 78 - Scalar sum of angular momentum of 9 planets and Sun (black), with first derivation (olive), compared with unsigned Sunspot cycle (orange), showing a different frequency.

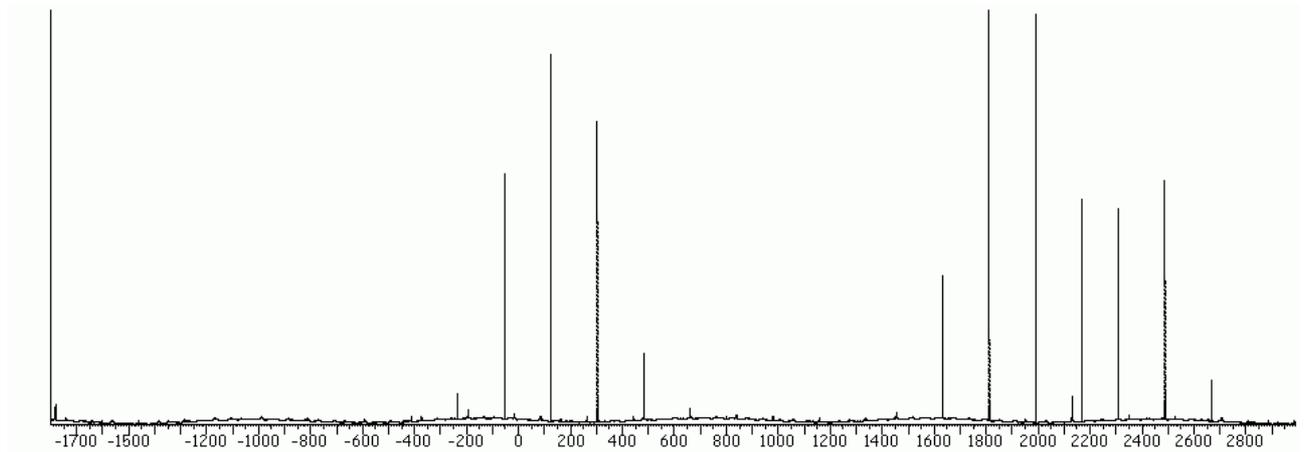

Figure 79 - Scalar sum of angular momentum of 9 planets and the Sun, not clipped.

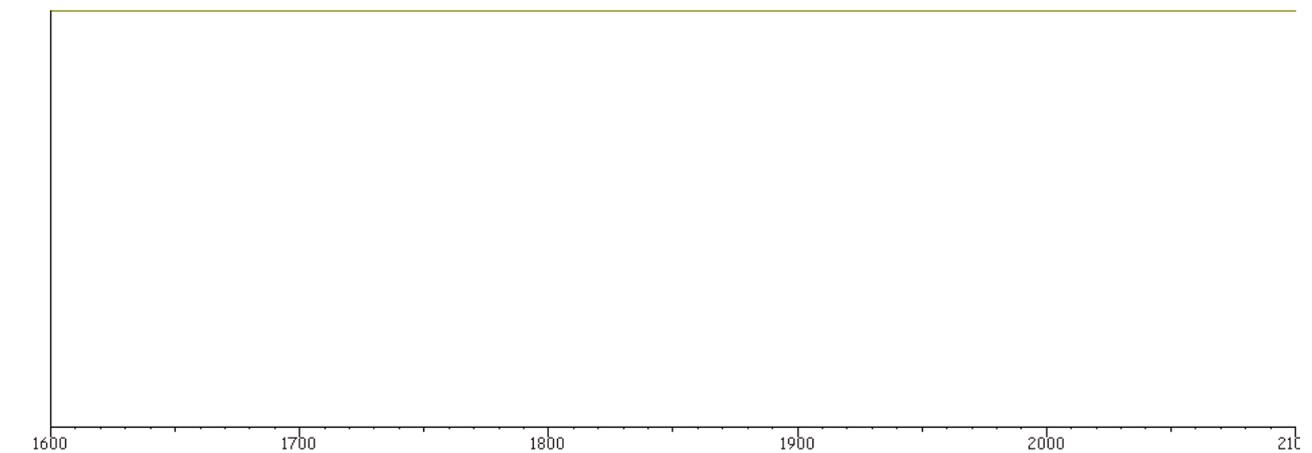

Figure 80 - Scalar sum of angular momentum of 9 planets and Sun, absolute value (olive line at the top)

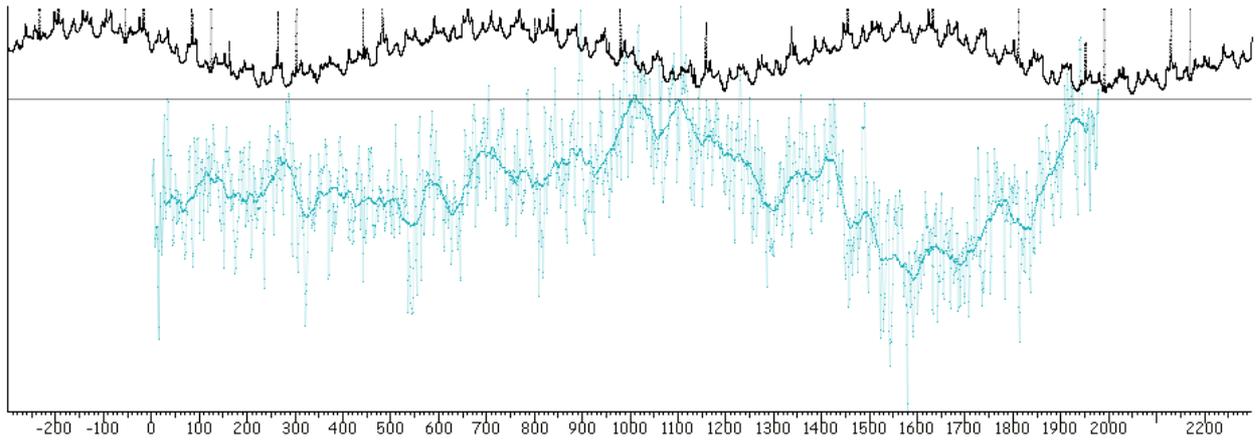

Figure 81 - Compared scalar sum of angular momentum of 9 planets and Sun with the climatologic data (Moberg at al. 2005, average temperature (light blue line), with gaussian filtering applied (bold blue line))

> Reference for the climatologic data:
> Moberg, A., D.M. Sonechkin, K. Holmgren, N.M. Datsenko and W. Karlén. 2005.
> Highly variable Northern Hemisphere temperatures reconstructed from low- and high-resolution proxy data.
> Nature, Vol. 433, No. 7026, pp. 613-617, 10 February 2005.

According to this connection, the current warming rate should slow down a little now, but will grow to local maximum arround year 2040, from which point it should drop to next little ice age arround year 2430 and to next warming arround year 2900.

## Vector Sum

The vector sum of angular momentum of 9 planets and the Sun is much more constant than the scalar sum.

The magnitude of vector-sum value is by 0.00005951 smaller than value of a scalar sum (difference is 1/16803.8 of the scalar sum), and does not vary much. Here (on fig. 82), the vector-sum value is offset by 12.46/209415 vertically to fit in the same chart with the scalar sum.

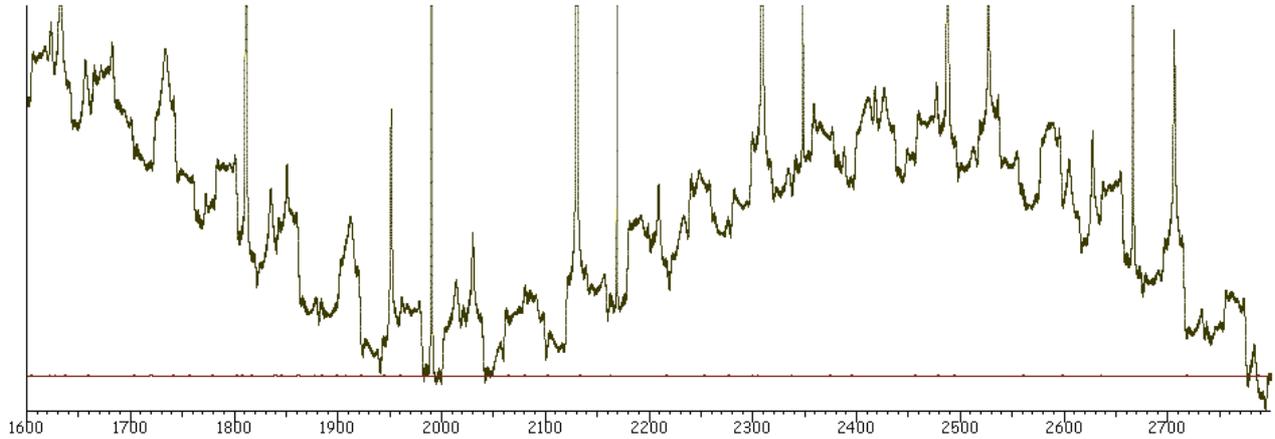

Figure 82 - Compared Vector sum of angular momentum of the Solar system (red line, offset vertically) with Scalar sum of angular momentum of the Solar system (olive line)

The Vector sum (in DE40x ephemerides) is still not exactly constant (fig. 83), but the rest is on the order of angular momentum of 4 big and 300 small asteroids, which were included in the ephemerides calculation.

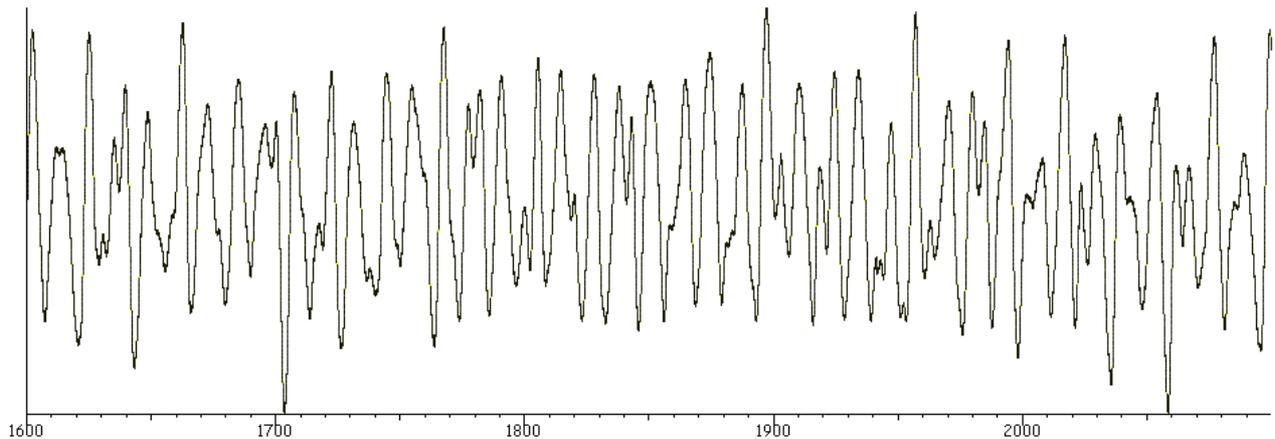

Figure 83 - Vector sum of angular momentum of the Solar system, in detail (contrary to angular momentum of Asteroids).

The vector sum of angular momentum is the conserved property of the system. Anyhow, if the angles of orbital planes incline, the individual absolute values grow or shrink to make the vector sum constant. Hence there is sometimes more energy stored in individual planet orbits, but working one against the other it nullifies. Still interesting is the connection of the scalar sum with the climatic cycle of little ice ages and warmings...

The vector angle of Angular momentum vector between Jupiter and Saturn shows a cycle similar to aforementioned scalar-sum cycle (fig. 84). This main frequency (854.021 years) of the Solar system cycles is 0.03710905 nano Hertz in average, or *almost* the tone **E**...

Distances between matching (854y) meet-points of Jupiter and Saturn during present 5 millenia (see fig. 13):
Minimum: 311860.3d (853.827y, F=0.03711301 nHz, tone=D# +82.82%)
Maximum: 311931.2d (854.021y, F=0.03710458 nHz, tone=D# +82.42%)
Average: 311893.6d (853.918y, F=0.03710905 nHz, tone=D# +82.63%)
The frequency oscilates (mainly due to distance of meet-point from perihelions) and slowly shrinks(?) (fig. 84b)
Distances between matching (60y) meet-points of Jupiter and Saturn (see fig. 84c):
Minimum: 21739.7d (59.520y, F=0.53239253 nHz, tone=C# +94.06%)
Maximum: 21780.0d (59.630y, F=0.53140845 nHz, tone=C# +90.78%)
Average: 21760.3d (59.577y, F=0.53188901 nHz, tone=C# +92.38%)

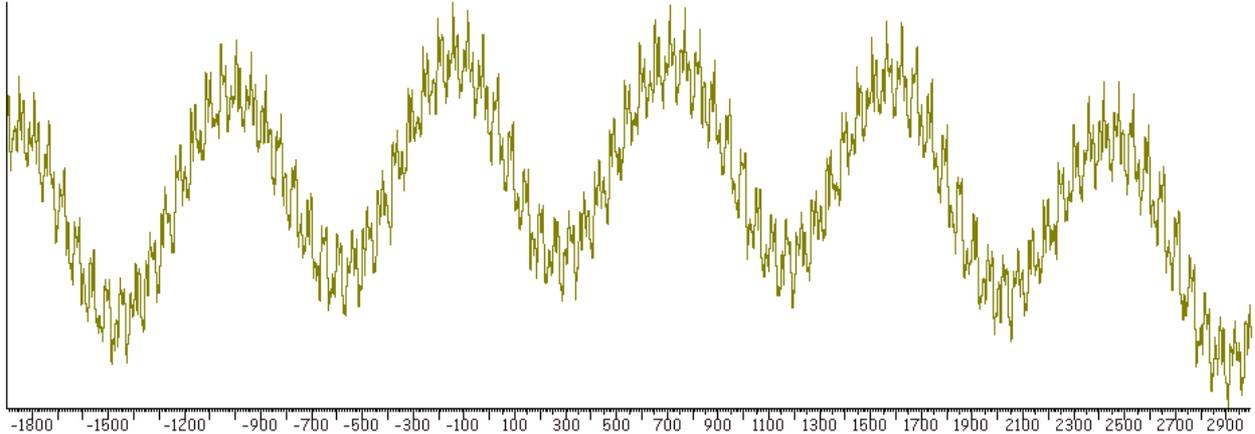
Figure 84 - The vector angle of angular momentum vectors between Jupiter and Saturn.

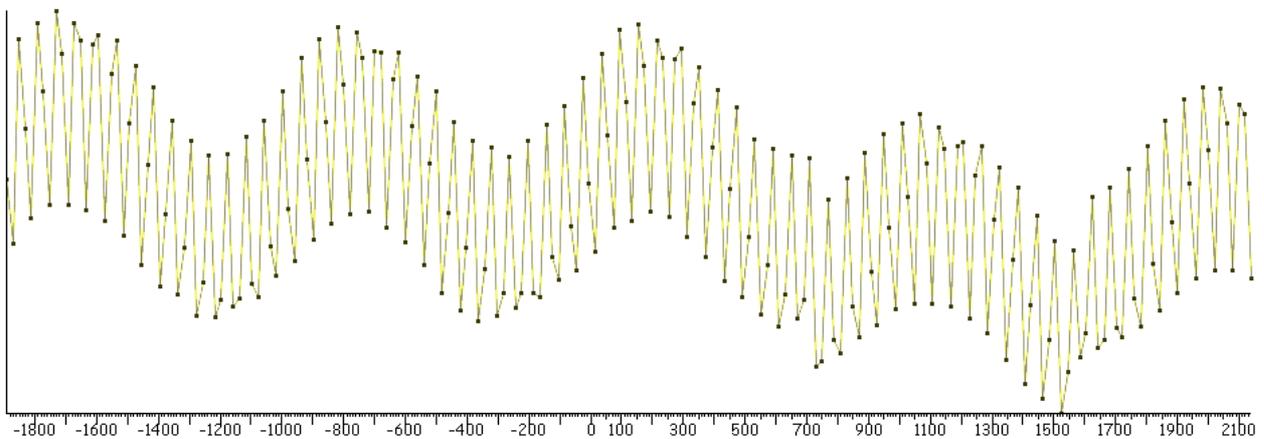
Figure 84b - Frequency of corresponding meet-points between Jupiter and Saturn (854 years), in nHz.

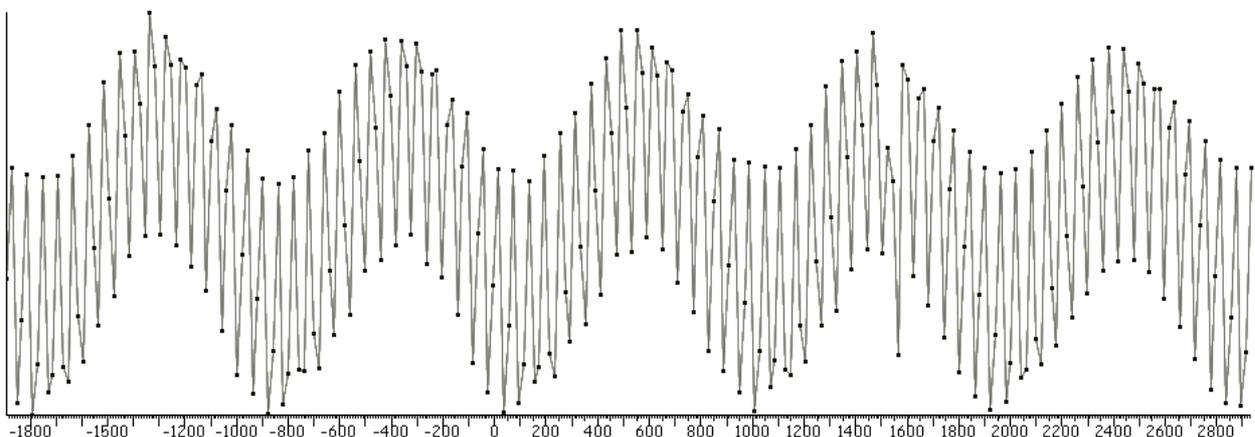
Figure 84c - Frequency of corresponding meet-points between Jupiter and Saturn (60 years), in nHz.

Now, lets call the vector sum of all angular moments the Normal to **invariant plane**. It should be (almost) constant, which it really is...

The angles between invariant plane normal and orbital plane normals of individual planets have these tendencies (see table 2), with range values for 5 millennia (1800BC-3000AD) :

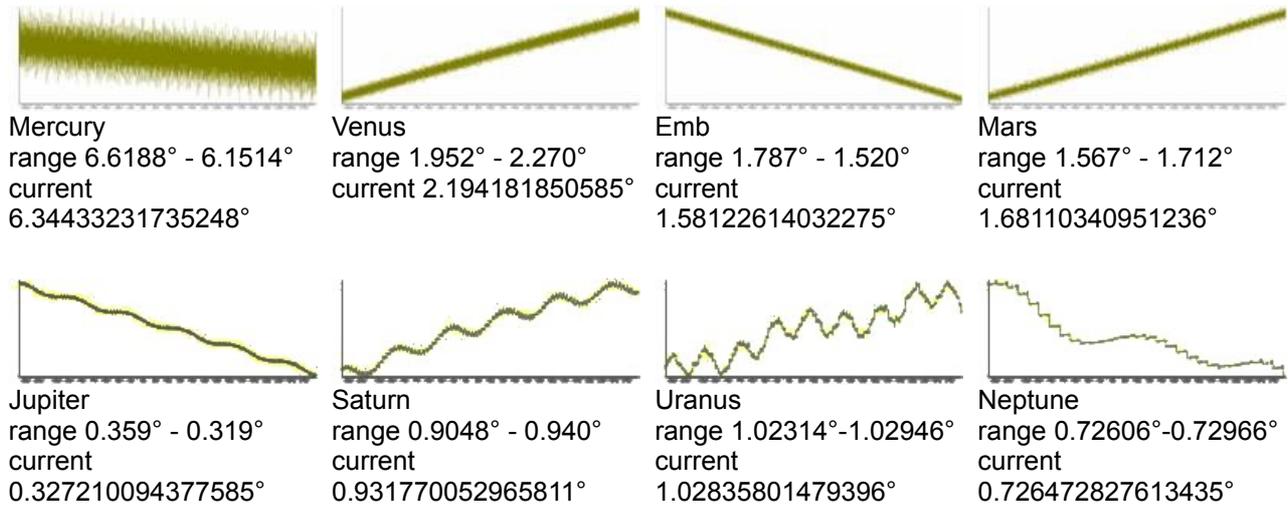

Mercury
range 6.6188° - 6.1514°
current 6.34433231735248°

Venus
range 1.952° - 2.270°
current 2.194181850585°

Emb
range 1.787° - 1.520°
current 1.58122614032275°

Mars
range 1.567° - 1.712°
current 1.68110340951236°

Jupiter
range 0.359° - 0.319°
current 0.327210094377585°

Saturn
range 0.9048° - 0.940°
current 0.931770052965811°

Uranus
range 1.02314°-1.02946°
current 1.02835801479396°

Neptune
range 0.72606°-0.72966°
current 0.726472827613435°

Table 2 - orbital inclination of individual planets to the invariant plane, during 5 millennia

The invariant plane is closest to the orbital plane of Jupiter. All other planets are more inclined from it.

When the planet shows decreesing tendency (as with Earth, Jupiter and others), the planet gets currently more aligned with the invariant plane. If the tendency is increasing (as with Venus, Saturn etc), the planet gets more diverted from the invariant plane. Actually, there are longer-scale cycles with sinus-like course, so that if the Earth is currently aligning with the invariant plane, it will be diverting later, but this time span is not included in DE4xx ephemerides...

The Sun orbital plane also varies from the invariant plane, sometimes even by more than 90° (in short times of retrograde sun) (see fig. 85-87)

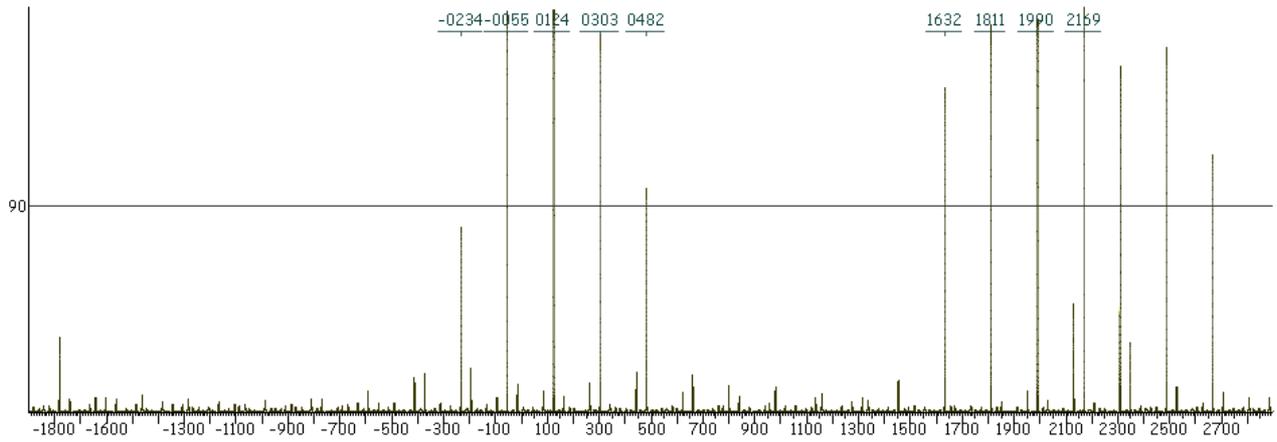
Figure 85 - Angle of Orbital plane of the Sun to the invariant plane

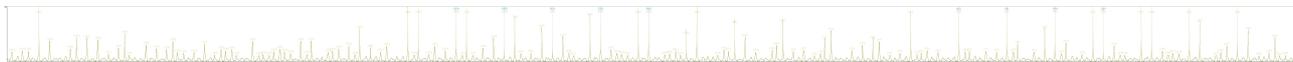
Figure 86 - Angle of Orbital plane of the Sun to the invariant plane, in very detail (see html version).

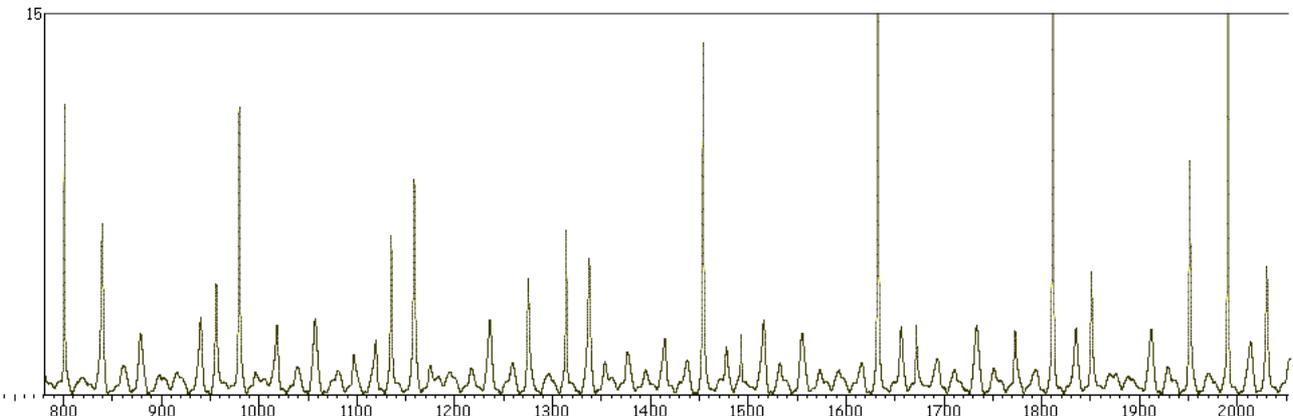
Figure 87 - Angle of Orbital plane of the Sun to the invariant plane, clipped at 15°.

# Sunspot cycle

## *Tidal force tables*

Scale used for values in these tables are:
- **Gravity force** - pull in $10^{18}$ Newtons on the whole planet, at its center:
  $F_g = G(m_1 \cdot m_2)/R^2 \cdot 1E{-}18$

- **Tidal force** - pull on surface, in 0.1 nano-newtons per 1 kilogram ($10^{-10}$ N/kg):
  $F_t = (G \cdot m_1 \cdot r)/R^3 \cdot 1E10$

Table 3 - Forces of planets onto **Sun**:

| Planet | Tidal force | | Gravity force onto Sun | |
|---|---|---|---|---|
| | Perihelium | Aphelium | Perihelium | Aphelium |
| Jupiter | 4.34907 | 3.24369 | 460 047.70085 | 378 354.10721 |
| Venus | 3.64316 | 3.49838 | 55 951.83789 | 54 459.40190 |
| Mercury | 3.15139 | 0.90141 | 20 715.25259 | 8 993.02017 |
| Earth/Moon bary. | 1.76497 | 1.59671 | 37 099.66726 | 34 702.61891 |
| Earth | 1.74341 | 1.57735 | 36 647.35500 | 34 281.60905 |
| Saturn | 0.21417 | 0.15468 | 41 346.85480 | 33 282.95551 |
| Mars | 0.06758 | 0.03852 | 1 995.55933 | 1 371.74508 |
| Moon | 0.02149 | 0.01927 | 451.47797 | 419.69332 |
| Uranus | 0.00394 | 0.00297 | 1 540.65205 | 1 275.09665 |
| Neptune | 0.00107 | 0.00102 | 683.92670 | 660.44589 |
| Pluto | 0.00000 | 0.00000 | 0.08802 | 0.03183 |
| *virtual points (resonance groups)* | | | | |
| Earth/Venus barycenter | **270.19721** | 4.65692 | 1 289 798.61569 | 86 059.82581 |
| Jupiter/Saturn barycenter | 168.07734 | 3.23097 | 5 738 340.93222 | 411 783.30751 |

Do you think, that Earth-Moon barycenter may interact and be computed as one planet, and Earth-Venus barycenter can not? Why?

Table 4 - For comparision, tidal forces onto **Earth** (in same units as above):

| Planet | Tidal force | | Gravity force onto planet | |
|---|---|---|---|---|
| | Perigeum | Apogeum | Perigeum | Apogeum |
| Moon | 13 020.96199 | 9 044.87272 | 221.66570 | 173.86203 |
| Sun | 5 319.97739 | 4 813.14218 | 36 647.57473 | 34 281.34414 |
| Venus | 0.67180 | 0.00238 | 1.24311 | 0.02887 |
| Jupiter | 0.05620 | 0.01806 | 1.73740 | 0.81499 |
| Mars | 0.01633 | 0.00009 | 0.05309 | 0.00168 |

Table 5 - For comparision, tidal forces onto **Moon** (in same units as above):

| Planet | Tidal force | | Gravity force onto planet | |
|---|---|---|---|---|
| | Perigeum | Apogeum | Perigeum | Apogeum |
| Earth | 292 736.27846 | 199 601.37614 | 232.14644 | 173.41398 |
| Sun | 1 458.38816 | 1 301.71176 | 452.58207 | 419.55781 |
| Venus | 0.14343 | 0.00067 | 0.01300 | 0.00036 |

Table 6 - For comparision, tidal forces onto **Venus**:

| Planet | Tidal force | | Gravity force onto planet | |
|---|---|---|---|---|
| | Maximum | Minimum | Maximum | Minimum |
| Sun | 12 942.66470 | 12 427.67159 | 55 952.46909 | 54 458.20237 |
| Earth/Moon barycenter | 0.79151 | 0.00280 | 1.25817 | 0.02922 |
| Earth | 0.78211 | 0.00277 | 1.24311 | 0.02887 |
| Jupiter | 0.04398 | 0.01952 | 1.24523 | 0.72461 |
| Mercury | 0.04316 | 0.00056 | 0.06864 | 0.00378 |

Whenever Earth, Jupiter or Mercury meets Venus, their tidal force is much higher than in other times, and they are all **retrogradating** at these times from Venus point of view. This is the most probable cause of Venus ill rotation (see Orbital resonance chapter above)... Sun performs a much bigger tidal force there, but it is stable, slowly progradating. The planets make *tidal shocks* instead...

## *Earth - Venus - Jupiter cycle*

As noted in chapter Earth and Venus, Orbital resonance, one of the most important events in Earth-Venus cycle (with respect to Sun), is the opposition, when the barycenter of the resonance group moves much closer to the Sun, temporarily stops orbiting, and after 2 weeks begins receding from the Sun to its more normal distance. Relative angle (or more exactly its half, as in tidal force formula and possibly others) between Sun to E/V-bary (same as angle to Earth from the Sun, since E/V barycenter is always on the Earth-side from the Sun) and between Sun to Jupiter at these times (fig. 88) plays somehow important role not only in planet orbits (it is remarkable in the angular momentums of both planets, see fig. 90-91), but also for Solar surface rotation rate and its damping, for the "wave" at the Solar surface, that slowly moves for 11 years toward Solar equator.

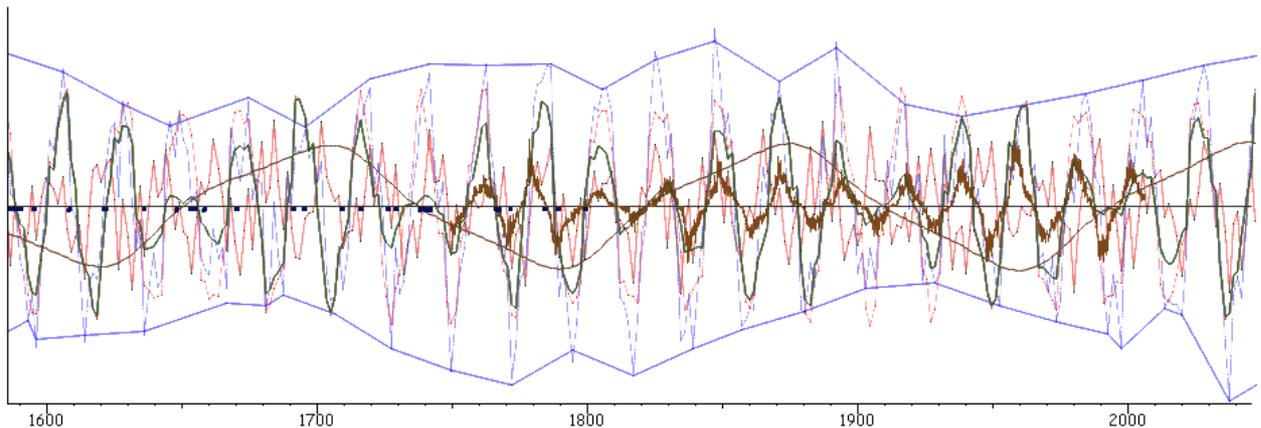

Figure 88 -Earth-Venus-Jupiter cycle compared to Signed Sunspot counts

Series in the chart on fig. 88 are:
- Orange - Real data - Signed sunspot counts (every other cycle is negative or positive)

All other series are computed (from ephemerides), at times of Earth-Venus opposition only:

- Bold green (which matches the Sunspot frequency) - Half of angle between Jupiter and Earth (or to EVB, with center in Sun) during Earth-Venus oppositions, multiplied (scaled vertically) by sinus of Uranus-Neptune angle at these times (to match cycle damping arround 1820 and 1910 of the Gleissberg cycle...The Uranus-Neptune cycle of 178.5 years seems to match the length of twice the Gleissberg cycle, observed in the Sunspot data.)
The damping arround 1650 (little ice age) and unexpectedly large values arround 1990 are due to another influences (matches cycle of overall angular momentum change, see relevant chapter)
- Purple serie - Uranus/Neptune cycle. To avoid a sinus-like symmetric appearance, the angle between those planets is multiplied by their relative velocity...
- Pink background serie - Tidal angle between Mercury and Jupiter at times of Earth-Venus oppositions.
- Outer blue serie - with connected maximums to show its envelope (see also fig. 89) - Tidal angle between Earth-Venus barycenter and Jupiter, with added or subtracted (with less importance) the Tidal angle between Jupiter and Mercury.
- Bold blue dots at X axis - historical record of "severe winters" in central Europe.

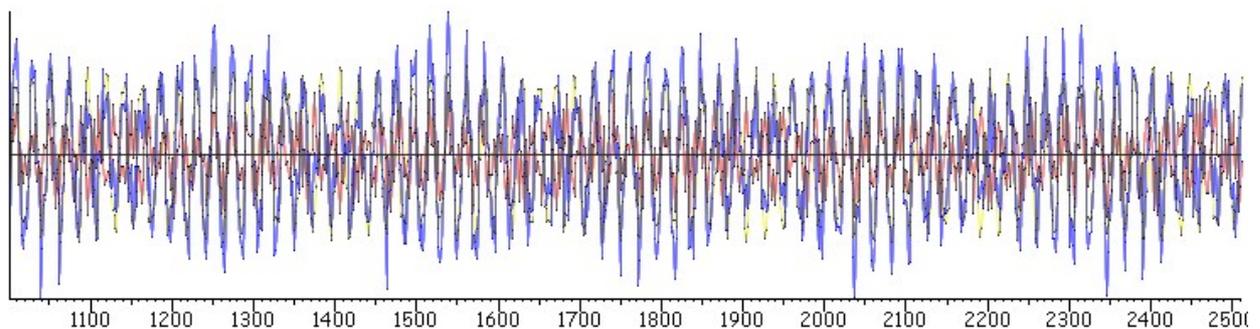

Figure 89 - Tidal angle between Jupiter and Mercury at times of Earth-Venus oppositions, longer trend.

Cycle of tidal angle between Jupiter and Mercury (fig. 89) during E-V oppositions seems to little match damping of the Sunspot cycle arround 1650 and 1910, but other parts of the envelope are only little matching real data or not at all.

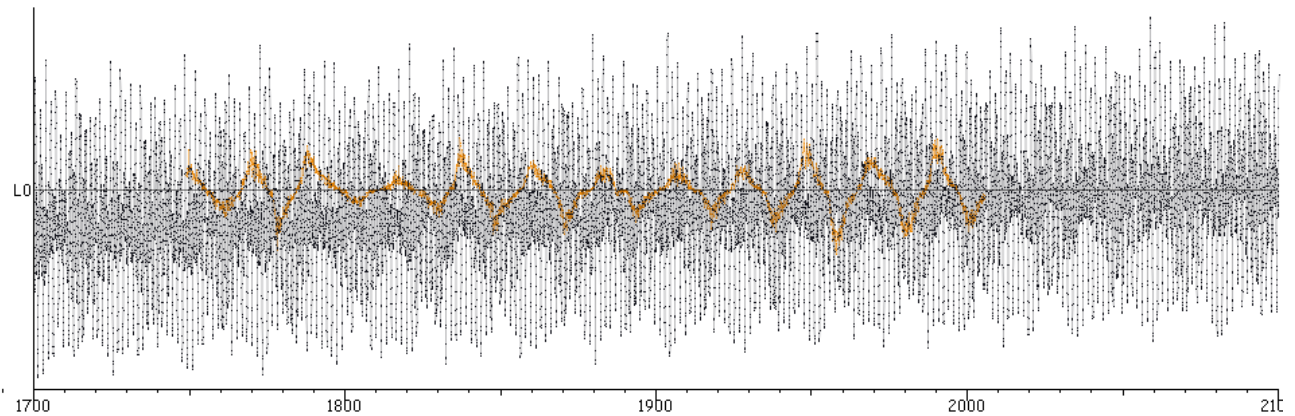

Figure 90 - Sum of angular momentum of Earth and Venus planets, compared with signed Sunspot cycle

Program formula for the chart in figure 90 is:
`((AngMoment(Emb,Sun)+AngMoment(Venus,Sun))*1e-24)-303.4095`
The chart is compressed horizontally to reveal the outer envelope of resonances cycles (PTCs in middle and high/low peaks). Positive phases of Solar cycle are matched with deep high-low peak mode and small central vibration, whereas negative phases of Solar cycle are matched with mode having more deep/high central parts of the serie.

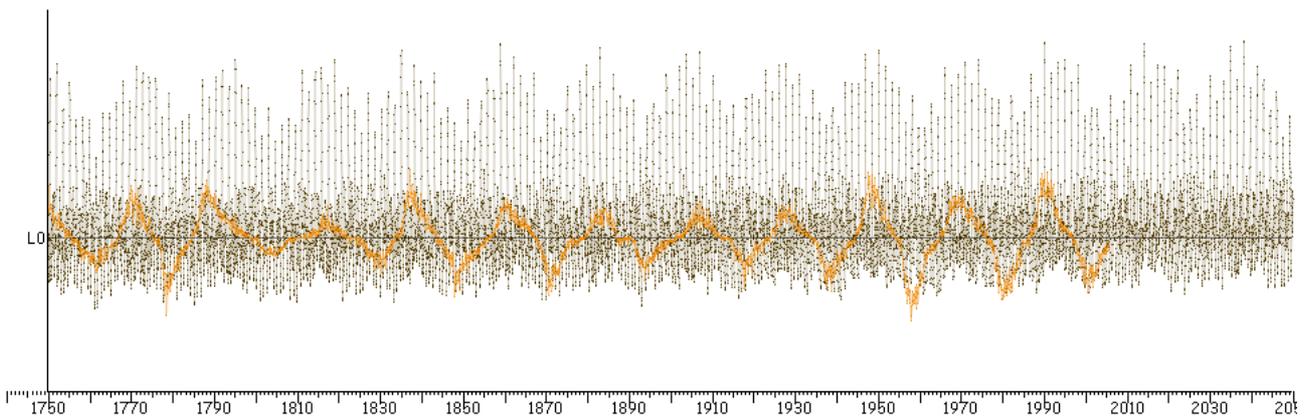

Figure 91 - Angular momentum of Venus only, compared with signed Sunspot cycle.

All high peaks on Venus angular momentum correspond with Earth-Venus heliocentric conjunctions. The Venus planet is attracted by Earth during each meating and its angular momentum grows. If the Jupiter planet "works with the Earth", the peaks are higher, than if Jupiter planet "works against the Earth". It again shows the same Earth-Venus-Jupiter cycle.

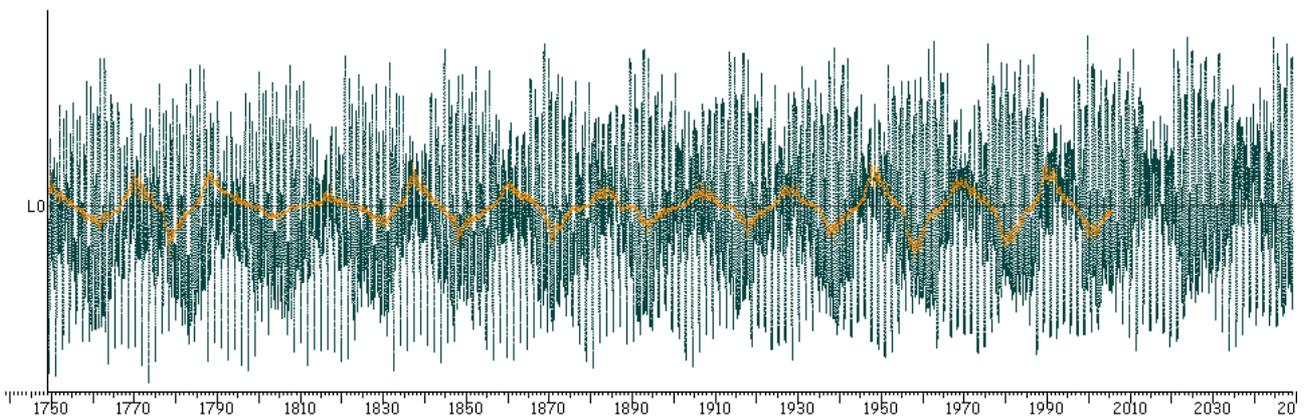

Figure 92 - Angular momentum of EMB only, compared with signed Sunspot cycle.

The same cycle may be seen in angular momentum of EMB with center in Sun (fig. 92-94)

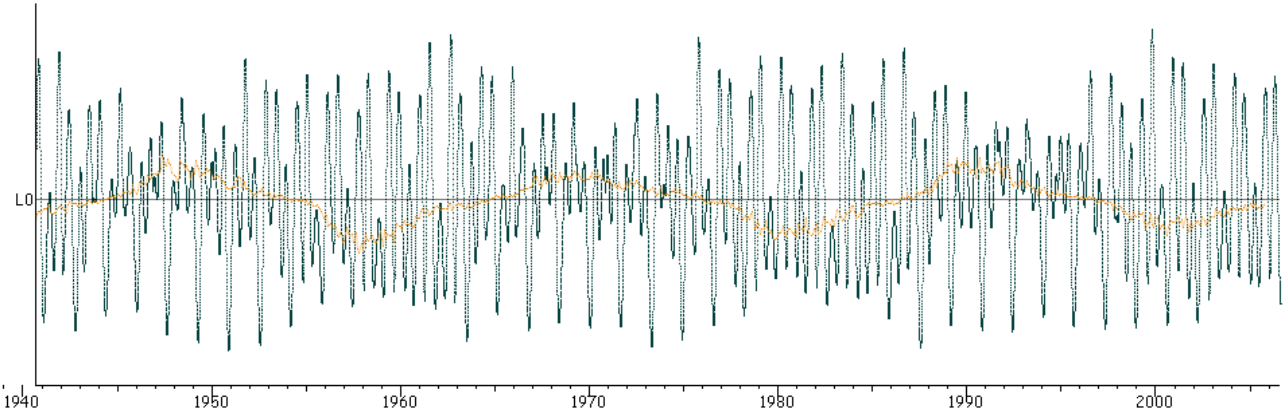

Figure 93 - Angular momentum of Emb relative to Sun, compared with signed Sunspot cycle, in detail.

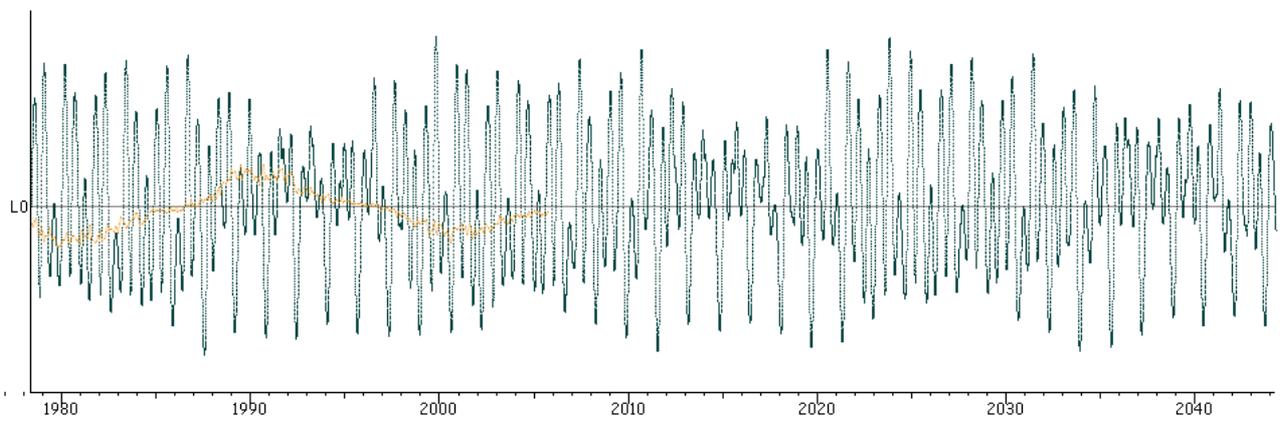

Figure 94 - Angular momentum of Emb relative to Sun, compared with signed Sunspot cycle, in detail, with "prediction" of the cycle into near future.

Figure 94 shows extrapolation of angular momentum of Emb relative to Sun, with a prediction of next Sunspot cycle maxima arround 2013,2022,2036 and minima arround 2020 and 2032, (with still a large level of uncertainty - since the cycle depends on more variables, among others by damping and exciting by overall (whole-system) angular momentum changes (with a main cycle of 854 years), by Uranus/Neptune cycle (178.5 year cycle corresponding with the Gleissberg cycle) and possibly other cycles and interferences).

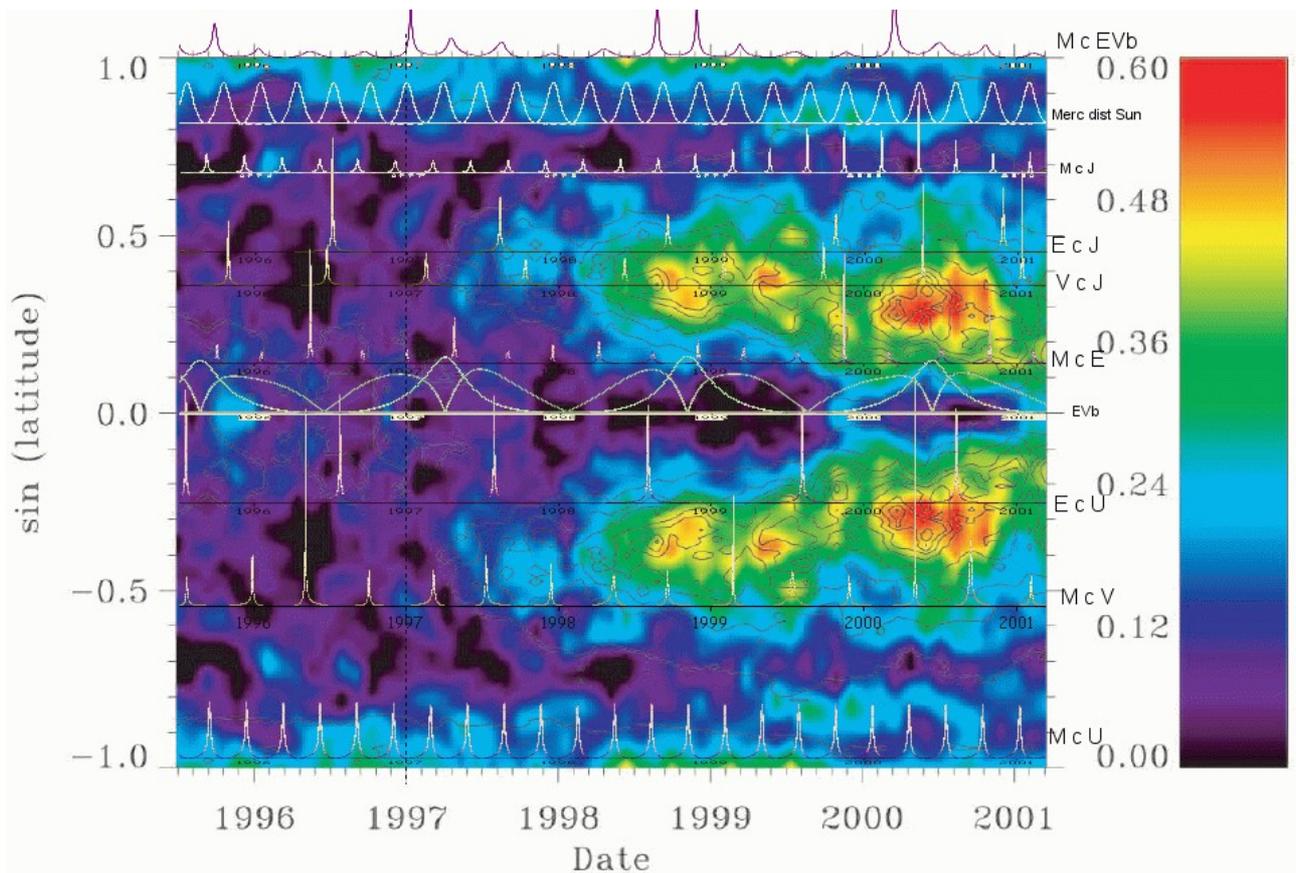

Figure 95 - Solar Cycle frequency shifts in Global p-Modes with overlay of planetary positions.

The image, from Localizing the Solar Cycle Frequency Shifts in Global p-Modes by R.Howe at al, 2002, shows change in p-mode frequencies, with my overlay comparing planetary conjunctions and Earth/Venus barycenter tidal events (fig. 95):

In the middle, the serie named "EVb" is Earth-Venus barycenter tidal force onto solar surface, and the same multiplied by its relative speed to the Sun.
The serie Mercury distance from Sun is inverted, so that high peaks are when Mercury is nearest to Sun.
Other series involve some conjunctions of planets E(arth),M(ercury),V(enus),J(upiter),U(ranus)...

The most prominent changes in Global p-Mode oscilations coincide with the Earth-Venus events (oppositions), but there are other changes, not explained yet... The influence seems to start with the onset of the event, not with its peak.

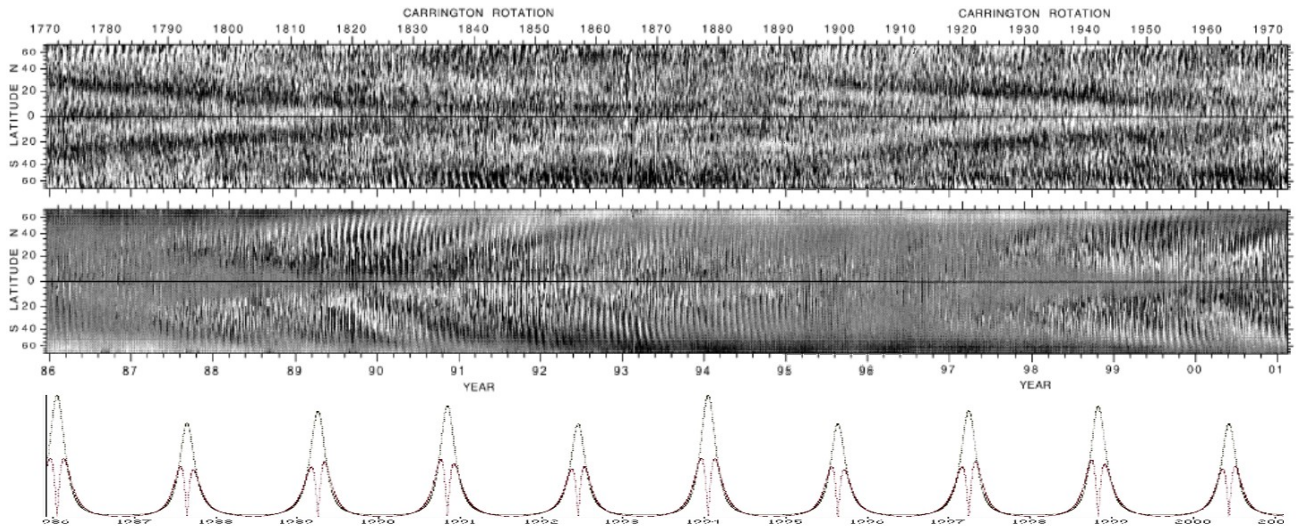

Figure 96 - Mt. Wilson synoptic map, from Very Long Lived Wave Patterns Detected in the Solar Surface Velocity Signal by Roger K. Ulrich, compared with Earth-Venus cycle.

Zonal velocity map together with a magnetic map for the period 1986 to 2001 (fig. 96), covering more than one Solar cycle. For the zonal velocity, the faster than average rotation is indicated by the black areas. The saturation velocity is 7.5 m s$^{-1}$, and for the magnetic map the saturation levels are 2 G.

The third row on fig. 96 shows tidal force of Earth-Venus barycenter on the Solar surface, with high peaks at Earth-Venus oppositions.

Origin of this "wave" (seen at top part of fig. 96) at Solar surface, which highly corresponds to Sunspot cycle, well may be caused by tidal effects of planets, and by partial exchange of orbital angular momentum of planets with surface spin angular momentum of the Sun.

The Solar surface rotation rate is quite strange, because it rotates faster on the equator (near the orbital plane of planets) than near the poles. One possible explanation of this is also the exchange of angular momentums of Solar surface levels with planets, the other explanation is in the similarity to mechanisms, responsible of accretion disks.

The partial waves of approximatelly 1.6 year period, seen on the magnetic synoptic map (bottom part of fig. 96), could correspond to the Earth-Venus cycle. The magnetic excitation, that happens near the equator, slowly travels toward poles, whereas the overall wave travels toward the equator (the Sunspots of the new Sunspot cycle begin at high latitudes and slowly travel toward equator during the cycle, where it diminishes, to start the new cycle at high latitudes).

The vertical (zebra-like) stripes (of length approximatelly 27 days) corresponds to Solar surface rotation - the active region on one side of the Sun periodically rotates into field of our view and out of it...

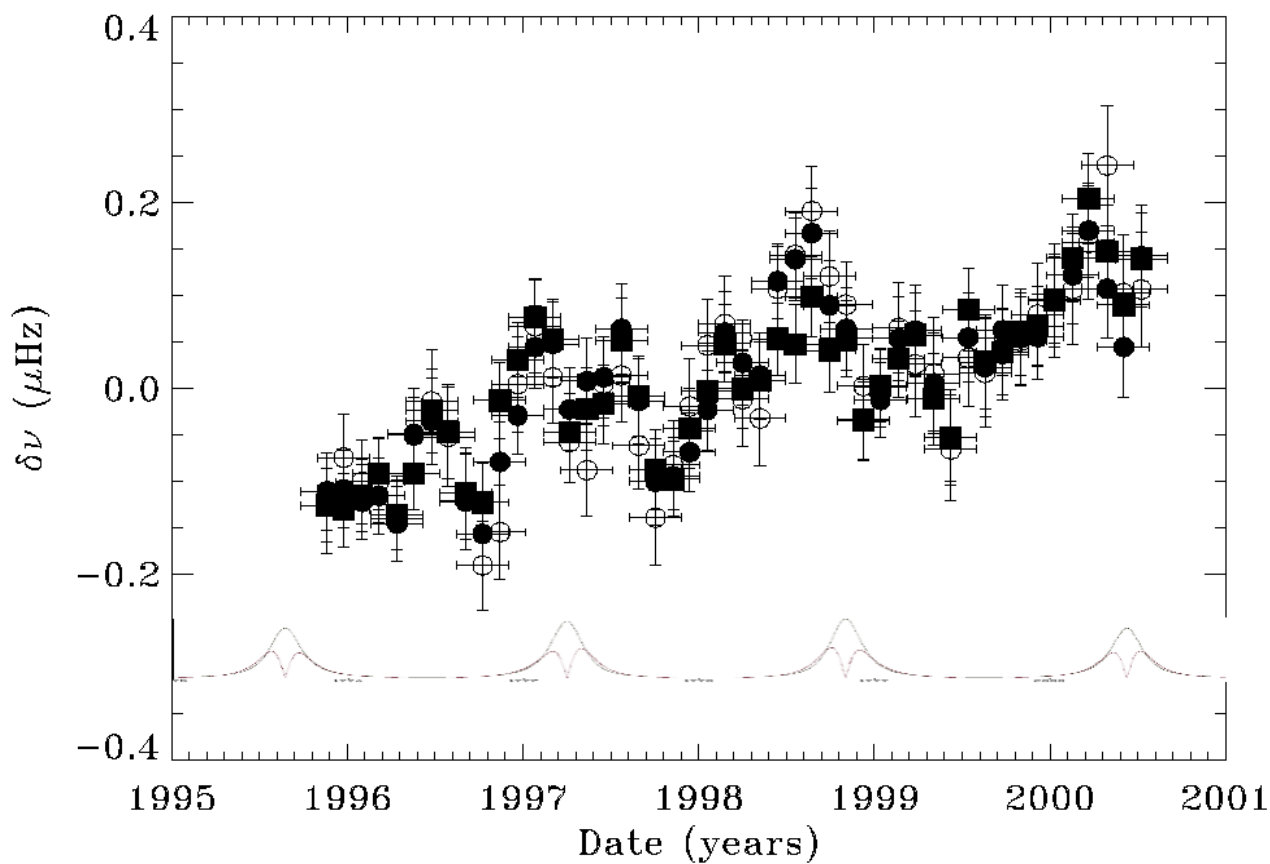

Figure 97 - Frequency shifts for GONG (filled circles) and BiSON (open circles) analyzed with the BiSON algorithms and for the GONG PEAKFIND results (squares), from "A Comparison of Low-Degree Solar p-Mode Parameters from BiSON and GONG" by Howe at al., compared with Earth-Venus barycenter force onto the Sun

The possible connection of Earth-Venus opposition events with the frequency shifts of Solar p-Modes is shown on fig. 97. Note, that the highest peaks of the p-Modes frequency shifts does not correspond with the maximum of the opposition events, but with the start of them, and the 1.6-year frequency approximatelly matches. It would need more observational data to prove this.

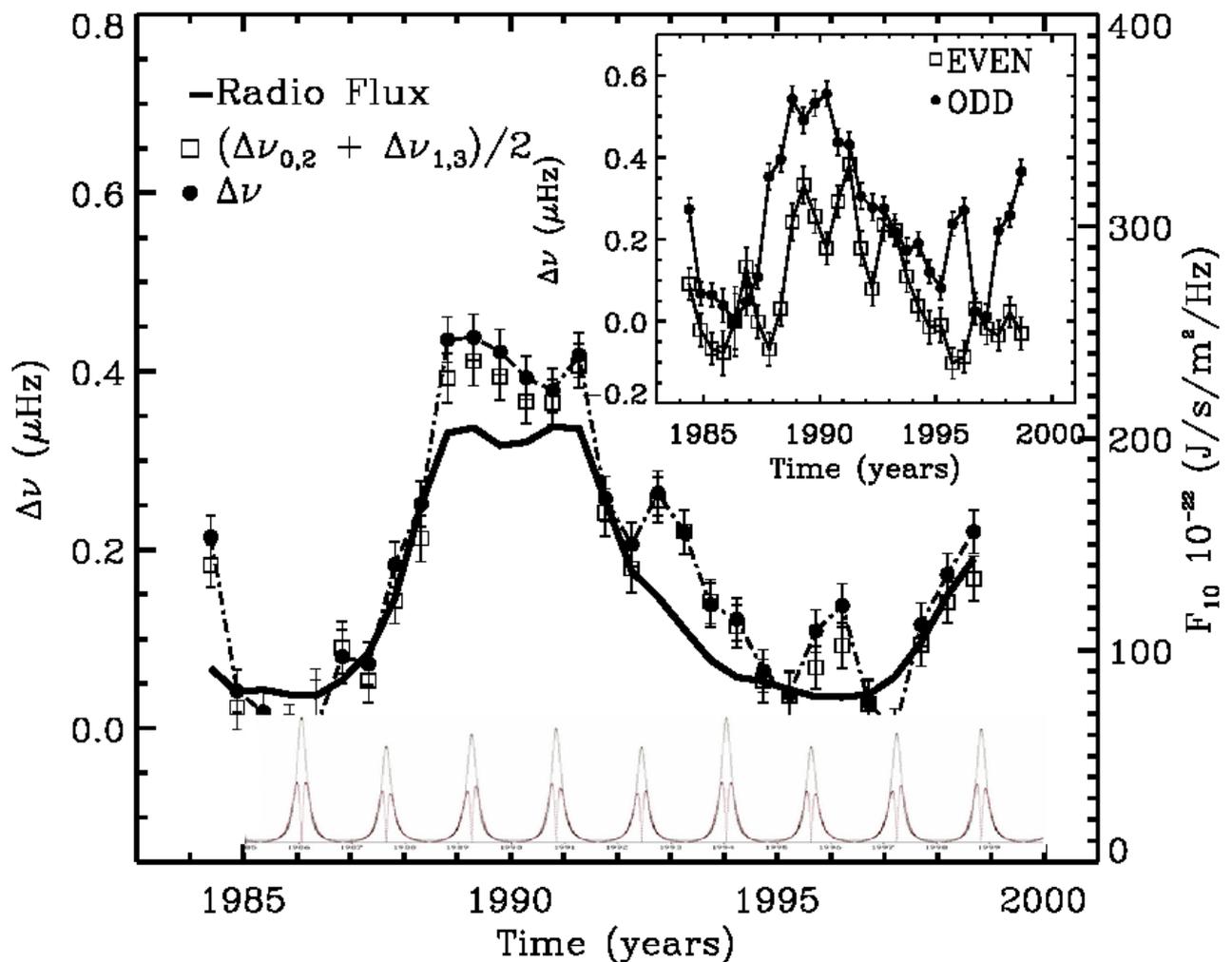

Figure 98 - Time variation of the integrated frequency shift for low-degree p-modes is plotted (black dots), where the radio flux at 10.7 cm is also shown (full line). In the sub plot, the results for the even and odd degrees separately are presented. The averages of these are plotted in the main panel (squares), from "Analysis of the solar cycle and core rotation using 15 years of Mark-I observations 1984-1999" by "S. J. Jiménez-Reyes et al.", compared with Earth-Venus tidal events (bottom serie).

The main rise of preceeding Solar cycle 22 at year 1987, the rises at 1992 and 1995 and the rise of Solar cycle 23 (fig. 98) may be correlated with Earth-Venus opposition events.

## *Conclusion*

Barycenter of the Earth-Venus resonance group seems to have a major impact on the Solar cycles, possibly through tidal forces, by stimulating and damping the widths, heights and frequencies of p-Modes. Its Tidal combination (half of the angle) with Jupiter planet shows a similar frequency to previously mysterious Sunspot cycle. The same cycle may be seen in the orbital angular momentum of both Earth and Venus planets (also caused mainly by the Jupiter planet), showing a possiblity of exchange of angular momentum of Earth or Venus planet(s) with the Solar surface levels (possibly magnetically, which would then select Earth from both planets, since there is little to none magnetic field on Venus planet).
Anyhow, if the influence was purely magnetical, then the Jupiter and Saturn planets would be more important, since their magnetic field influence is actually larger on the Solar surface. Anyhow, there is also a possible connection of 854-year climatic cycle (and possibly a Solar surface rotation rate and output power) with the large angular momentum cycle, mainly caused by Jupiter and Saturn cycle. The 178.5 year cycle of Uranus and Neptune resonance group is also seen in the Gleissberg cycle of Solar activity of similar length.

Appendix 1.

## *Mathematical formula for angular momentum calculation*

For calculating angular momentum of a body with respect to a center, we use this formula:

$$L = |\mathbf{r}||\mathbf{p}| \sin \theta_{r,p}$$

Angular momentum is a scalar multiplication of Momentum vector length ( Velocity * Mass ), multiplied by distance and multiplied by Sinus of angle between the line connecting the bodies and the relative velocity vector of the body with respect to the center...

The Scalar Sum of angular moments is just a scalar sum of moments of all individual bodies (9 planets and Sun)... The Solar-system barycenter is a coordinate center of these ephemerides and has got both vectors (distance and velocity) null... Anyhow - physically, the vector sum is more important, as it is a conserved property of the system!

Or, expressed as a pascal code:

```pascal
function PlanetMomentum(Planet,Center: TPlanet): Float;
var V: TVector3d;
begin
  // Relative velocity: V:=Planet.VelocityVector - Center.VelocityVector;
  Vector3dSub(Planet.VelocityVector, Center.VelocityVector, V);
  // ... multiplied by Mass:
  Result := Planet.Mass * Vector3dLength(V);
end;

function PlanetAngularMomentum(Planet,Center: TPlanet): Float;
var Momentum, Distance, SinAngle: Float;
    DistVector, RelativeVelocity: TVector3d;
begin
  // Relative momentum:
  Momentum := PlanetMomentum(Planet,Center);
  //
  // Distance:   DistVector:=Planet.PositionVector - Center.PositionVector;
  Vector3dSub(Planet.PositionVector, Center.PositionVector, DistVector);
  Distance := Vector3dLength(DistVector) * (1/km_to_AU); // Distance in AU...
  //
  // Angle of relative_velocity_vector and distance_vector:
  Vector3dSub(Planet.VelocityVector, Center.VelocityVector, RelativeVelocity);
  SinAngle := Sin(Vector3dAngle(DistVector, RelativeVelocity));
  //
  Result := Momentum * Distance * SinAngle  * 1e-24;
end;
```

All vectors are in km, masses are in kg, and it is just divided by AU at one place and by 1E24 at another to prevent precision lost and make it more readable...

Planet masses, used for the calculation, are fetched (recalculated) from the Ephemerides. The values, used with DE406 (most charts) are:

| Body | Mass (kg) | Note |
|---|---|---|
| **Sun** | 1.98843966345011E+30 | |
| **Mercury** | 3.30108185047166E+23 | |
| **Venus** | 4.86737884430283E+24 | |
| **Earth** | 5.97225784209252E+24 | |
| **Moon** | 7.34590000621462E+22 | |
| **EMB** | 6.04571684215467E+24 | (Earth-Moon Barycenter) |
| **Mars** | 6.41699593330543E+23 | |
| **Jupiter** | 1.89854616070534E+27 | |
| **Saturn** | 5.68467023180865E+26 | |
| **Uranus** | 8.68201283610302E+25 | |
| **Neptune** | 1.02432262501562E+26 | |
| **Pluto** | 1.47073939604298E+22 | |

Appendix 2

## *Orbital characteristics of planets*

Orbital characteristics of planets, collected from various sources (Wikipedia, NASA planetary tables, Explanatory Supplement to the Astronomical Almanac and more...)

| Planet | Average orbital period (days/years) | Spin (days/hours) | Distance from Sun (AU) (avg/min-max) | Distance from Sun (avg in light-hours) | Distance/Spin wave length (D/S, S/D) | Orbital speed (km/s) (avg/min-max) | Inclination to ecliptic (°) | Excentricity | Ascending node (RA) | Mass (kg) | Diameter (avg/min/max) | Equatorial radius | Surface Magnetism | Magnetic tilt | Density (g/cm^3) | Equatorial gravity | Spin tilt | Remarks |
|---|---|---|---|---|---|---|---|---|---|---|---|---|---|---|---|---|---|---|
| **Sun** | | 25.38 d 609.12 h | | | | 217 (galactic) 0.00851572 - 0.01609431 | | | | 1.9891E30 | 1392000 1391995 - 1392005 | 696000 | | | 1.408 | 273.95 | | rot.Obliquity 7.25° to eclip |
| **Mercury** | 87.96934 d 0.24084 y | 58.6462 d 1407.5088 h | 0.38709893 0.3074890352 - 0.4667065166 | 0.0536567 lh | 3.81217E-5 26231.74365 | 47.36 38.84376052 - 58.99057812 | 7.004986 | 0.205631752 | 48.33089° | 3.302E23 | 4879.4 | 2439 | 0.0033 | 169°, Lng 285°~115° | 5.427 | 3.701 | 0.0 | |
| **Venus** | 224.70069 d 0.61518 y | -243.0185 d -5832.444 h | 0.72333199 0.7183698239 - 0.7282889865 | 0.1002628 lh | 1.719E-5 58171.56512 | 35.02 34.76688994 - 35.27650343 | 3.394466 | 0.006771882 | 76.679920° | 4.8685E24 | 12103.7 | 6052 | 0 | | 5.204 | 8.87 | 177.3 | |
| **Earth** | 365.25641 d 1 y | 0.997258 d 23.9341 h | 1.0000000001 0.9831915115 - 1.0168061579 | 0.1386124 lh | 0.0057914 172.669256 | 29.783 29.26254785 - 30.31382552 | 0 | 0.016708617 | 0 | 5.9736E24 | 12745.591 12713.5 - 12756.27 | 6378.140 | 0.3076 | 11.4° (78.6°N) | 5.515 | 9.7801 | 23.45 | |
| **Mars** | 686.96 d 1.88076 y | 1.025957 d 24.6229 h | 1.52366231 1.3810512113 - 1.6662924910 | 0.2111986 lh | 0.0085773 116.586474 | 24.077 21.95392385 - 26.51824041 | 1.849726 | 0.093400620 | 49.558093° | 6.4185E23 | 6754.8 - 6804.9 | 3397.2 | | | 3.934 | 3.69 | 25.19 | |
| **Asteroids (Ceres)** | 1679.819 d 4.59901 y | 0.3781 ??? | 2.766 2.544 - 2.987 | | | 17.882 | 10.587 ? | 0.08 ? | 80.410° | 2.3E21 (sum) 9.5E20 (Ceres) | 950 | | | | | 0.27 | | dead planet between mar/jup, values for Ceres, mass 9.5e20 |
| **Jupiter** | 4333.2867 d 11.86368 y | 0.413538021 d 9.92491 h | 5.20336301 4.9473167719 - 5.4574965842 | 0.7212508 lh | 0.0726707 13.7606918 | 13.056 12.426479 - 13.711067 | 1.303269 | 0.048494851 | 100.46444° | 1.899E27 | 133709 - 142984 | 71398 | 4.28 | 9.6°, Lng 201.7° | 1.326 | 23.12 | 3.12 | |
| **Saturn** | 10756.1995 d 29.44835 y | 0.4440092592 d 10.65622 h | 9.53707032 8.9937104773 - 10.102270262 | 1.3219566 lh | 0.1240549 8.0609454 | 9.639 9.10081011 - 10.2229089 | 2.488878 | 0.055508622 | 113.66552° | 5.6846E26 | 108728 - 120536 | 60000 | 0.21 | <1° | 0.6873 | 8.96 | 26.73 | |
| **Uranus** | 30707.4896 d 84.07104 y | -0.718333333 d -17.23999 h | 19.19126393 18.280651144 - 20.107776505 | 2.6601479 lh | 0.15430101 6.4808389 | 6.795 6.48728023 - 7.13393835 | 0.773196 | 0.046295899 | 74.005947° | 8.6832E25 | 49946 - 51118 | 25400 | 0.228 | 58.6°, Lng ? | 1.318 | 8.69 | 97.86 | North pole 262° ecl, mag-pole 60° different! |
| **Neptune** | 60223.3528 d 164.87966 y | 0.6713 d 16.1112 h | 30.06896348 29.806522658 - 30.339172623 | 4.1679324 lh | 0.258697 3.8655137 | 5.432 5.38791362 - 5.48224695 | 1.769952 | 0.008988095 | 131.78406° | 1.0243E26 | 48681 - 49528 | 24300 | 0.142 | 46.9°, Lng 288° | 1.638 | 11.15 | 29.56 | |
| **Pluto** | 90613.3055 d 248.08135 y | 6.38723 d 153.29352 h | 39.48168677 29.649274404 - 49.318952713 | 5.4726529 lh | 0.0357005 28.010625 | 4.666 3.67761388 - 6.11692055 | 17.14216 | 0.249050 | 110.29714° | 1.305E22 | 2306 +-20 | 2500 | | | 2.03 | 0.58 | 118? | |
| **Kuiper Belt** | | | | 30-44 | | | | | | | | | | | | | | |

Notes to the table:
Minimum and maximum distances from Sun and orbital speeds were recalculated from the DE406 Ephemerides (whereas average values were kept from original sources), for period 1900-2100 for inner planets, for period 1600-2200 for outer planets..
Equatorial radius, Inclination to ecliptic, Excentricity and Spin tilt (originally called "Inclination of Equator to Orbit") are from the "Explanatory Supplement to the Astronomical Almanac" edition 2006, for epoch J2000.0, rounded to the precission specified here.
Distance/Spin wave length value (distance in light-hours and spin in hours) is particularly interesting for a Neptune planet, since Earth sometimes gets into the node of exactly 1/4 distance of the Neptune spin wave. More exact specification of the times during Earth year, when this happens, depends largely on an exact determination of the Neptune spin time.

## *References*

- E.M. Standish, **JPL Planetary and Lunar Ephemerides**, 1982-1995
- Theodor Landscheidt: **Solar Eruptions Linked to North Atlantic Oscillation**, 2001?
- Theodor Landscheidt: **Trends in Pacific Decadal Oscillation Subjected to Solar Forcing**, 2001?
- Theodor Landschedt: **Decadal-Scale Variations in El-Niňo Intensity**, 2003
- R.K.Ulrich: **Identification of Very Large Scale Velocity Structures on the Solar Surface Using Mt. Wilson Synoptic Observations**, 1998
- Roger K. Ulrich, **Very Long Lived Wave Patterns Detected in the Solar Surface Velocity Signal**, The A.P.J. 2001
- S.J.Jiménez Reyes, T. Corbard, P.L.Pallé, T. Roca Cortés, and S.Tomczyk: **Analysis of the Solar Cycle Core and Core Rotation Using 15 Years of Mark-I Observations: 1984-1999**, A&A 2001
- R.Howe, R.W.Komm, and F.Hill: **Localizing the Solar Cycle Frequency Shifts in Global p-Modes**, The A.P.J. 2002
- R.Howe, W.J.Chaplin, Y.P.Elsworth, F.Hill, R.Komm, G.R.Isaak, and R.New: **A Comparison of Low Degree Solar p-Mode Parameters From BiSON and GONG, Underlying Values and Temporal Variations**, The A.P.J. 2003
- Moberg, A., D.M. Sonechkin, K. Holmgren, N.M. Datsenko and W. Karlén.: **Highly variable Northern Hemisphere temperatures reconstructed from low- and high-resolution proxy data**, Nature 2005
- Jiří Svoboda, **Severe winters in Europe during the past millennium**, Vesmir 1997

Special thanks go to E.M.Standish for publishing the Ephemerides, to J.B.Gurman for requesting this work to be finished (by an almost automated email), to all the SOHO team for maintaining the station running through the whole cycle despite all complications, and to NASA A.D.S. for giving access to all the required scientific works...
Special thanks go to Radim S. for a hint with tone analysis of the resonances.

This work has been done from 2006 to 2009, first published on 2009, March 27 on the occasion of the Earth-Venus heliocentric conjunction.